%% file: ANA-FTAG-2019-07-PAPER.tex
\documentclass[cernpreprint, atlasdraft=false, texlive=2023, UKenglish, subfigure, texmf, orcidlogo]{atlasdoc}
 
\usepackage{atlaspackage}
\usepackage{atlasbiblatex}
 
\usepackage{atlasphysics}

\graphicspath{{logos/}{figures/}}
 
\usepackage{ANA-FTAG-2019-07-PAPER-defs}

\AtlasTitle{ATLAS flavour-tagging algorithms \\ for the LHC Run 2 $pp$ collision dataset}

\AtlasAbstract{
The flavour-tagging algorithms developed by the ATLAS Collaboration and used to
analyse its dataset of $\sqrt s = 13$~TeV $pp$ collisions from Run~2 of
the Large Hadron Collider are presented.
These new tagging algorithms are based on recurrent and deep neural networks,
and their performance is evaluated in simulated collision events.
These developments yield considerable improvements over previous jet-flavour
identification strategies.
At the 77\% $b$-jet identification efficiency operating point, light-jet
(charm-jet) rejection factors of 170 (5) are achieved in a sample of simulated Standard
Model $t\bar{t}$ events; similarly, at a $c$-jet identification efficiency of
30\%, a light-jet ($b$-jet) rejection factor of 70 (9) is obtained.
}

\AtlasRefCode{FTAG-2019-07}
 
\PreprintIdNumber{CERN-EP-2022-226}

\AtlasJournalRef{Eur.Phys.J.C 83 (2023) 7, 681}
\AtlasDOI{10.1140/epjc/s10052-023-11699-1}

\hypersetup{pdftitle={ATLAS document},pdfauthor={The ATLAS Collaboration}}
 
\begin{document}
 
\maketitle
 
\tableofcontents

\newpage

\section{Introduction}
 
The separation of jets containing $b$- and $c$-hadrons ($b$-jets and
$c$-jets, respectively) against jets containing neither $b$- or $c$-hadrons
(light-flavour jets) is of major importance in many areas of the physics
programme of the ATLAS experiment~\cite{PERF-2007-01} at the Large Hadron
Collider (LHC)~\cite{Evans:2008zzb}.
Flavour-tagging has been decisive in observations of the Higgs boson
decay into bottom quarks~\cite{HIGG-2018-04} and of its production in
association with a top-quark pair~\cite{HIGG-2018-13}, and plays a crucial role
in a large number of Standard Model (SM) precision measurements, studies of
Higgs boson properties, and searches for new phenomena.
 
The ATLAS Collaboration uses various algorithms to identify
$b$- and $c$-jets~\cite{PERF-2012-04}, referred to as flavour-tagging algorithms,
when analysing data from $pp$ collisions recorded during \RunTwo of the LHC (2015--2018)
at $\sqrt s = 13$~\TeV.
These algorithms exploit the long lifetime, high mass and high decay
multiplicity of $b$- and $c$-hadrons as well as the properties of heavy-quark
fragmentation.
Given a lifetime of the order of 1.5 ps ($\langle c\tau\rangle \approx 450~\mu$m),
energetic $b$-hadrons have a significant mean flight length, $\langle l\rangle=\beta \gamma c
\tau$, in the detector before decaying, generally leading to at least one vertex
displaced by a few~mm from the hard-scatter collision point.

The strategy developed by the ATLAS Collaboration is based on a two-stage
approach.
Low-level algorithms reconstruct the characteristic features of the
heavy-flavour jets via two complementary approaches: one that uses the
properties of individual charged-particle tracks (referred to as \enquote{tracks})
associated with a hadronic jet, and a second which combines the tracks to
explicitly reconstruct displaced vertices.
Then, in order to maximise performance, the results of low-level algorithms
are combined in high-level algorithms consisting of multivariate classifiers.
The analysis of the data from \RunTwo of the LHC is marked by improvements
and retuning of the low-level algorithms~\cite{FTAG-2018-01}, first introduced during \RunOne,
but also by the introduction of new low- and high-level algorithms respectively based on recurrent and deep
neural networks. This yields considerable improvements over previous work, which
was based on boosted decision trees or likelihood discriminants.
 
This paper is organised as follows. Section~\ref{sec:atlas} introduces the ATLAS
detector.
The simulated Monte Carlo events used in this work are described
in Section~\ref{sec:samples}. Section~\ref{sec:objects} contains the description
of the objects reconstructed in the detector which are key inputs to
flavour-tagging algorithms, while Sections~\ref{sec:lowlevel}
and~\ref{sec:highlevel} describe the low- and high-level tagging algorithms
respectively.
Finally, their performance, evaluated on simulated event samples, is presented in
Section~\ref{sec:performance}.


\section{The ATLAS detector}
\label{sec:atlas}

The ATLAS detector~\cite{PERF-2007-01} at the LHC covers nearly the entire solid
angle around the collision point.
It consists of an inner tracking detector (ID) surrounded by a superconducting
solenoid, electromagnetic and hadronic calorimeters and a muon spectrometer
incorporating three large superconducting toroid magnets.
 
The ID consists of a high-granularity silicon pixel detector which covers the
vertex region and typically provides four measurements per track. The innermost
layer, known as the insertable B-layer (IBL)~\cite{ATLAS-TDR-2010-19}, was added
in 2014 and provides high-resolution hits at small radius to improve the
tracking performance. For a fixed $b$-jet efficiency, the incorporation of the
IBL improves the light-flavour jet rejection of the $b$-tagging algorithms by up
to a factor of four~\cite{PIX-2018-001}.
The silicon pixel detector is followed by a silicon microstrip tracker
that typically provides eight measurements from four strip double layers.
These silicon detectors are complemented by a transition radiation tracker
(TRT), which enables radially extended track reconstruction up to a
pseudorapidity\footnote{ATLAS uses a right-handed coordinate system with its
origin at the nominal interaction point (IP)  in the centre of the detector and
the $z$-axis along the beam pipe.
The $x$-axis points from the IP to the centre of the LHC ring, and the $y$-axis
points upwards.
Cylindrical coordinates $(r,\phi)$ are used in the transverse plane, $\phi$
being the azimuthal angle around the $z$-axis.
The pseudorapidity is defined in terms of the polar angle $\theta$ as
$\eta = -\ln \tan(\theta/2)$.
Angular distance is measured in units of
$\Delta R \equiv \sqrt{(\Delta\eta)^{2} + (\Delta\phi)^{2}}$.}
of $|\eta| = 2.0$.
The TRT also provides electron identification information based on the fraction
of hits (typically 33 in the barrel and up to an average of 38 in the endcaps)
above a higher energy-deposit threshold corresponding to transition radiation.
The ID is immersed in a \SI{2}{\tesla} axial magnetic field and provides
charged-particle tracking in the pseudorapidity range $|\eta| < 2.5$.
 
The calorimeter system covers the pseudorapidity range $|\eta| < 4.9$.
Within the region $|\eta|< 3.2$, electromagnetic calorimetry is provided by
barrel and endcap high-granularity lead/liquid-argon (LAr) sampling
calorimeters,  with an additional thin LAr presampler covering $|\eta| < 1.8$ to
correct for energy loss in material upstream of the calorimeters.
Hadronic calorimetry is provided by a steel/scintillator-tile calorimeter,
segmented into three barrel structures within $|\eta| < 1.7$, and two copper/LAr
hadronic endcap calorimeters.
The solid angle coverage is completed with forward copper/LAr and tungsten/LAr
calorimeter modules optimised for electromagnetic and hadronic measurements,
respectively.
 
The muon spectrometer  comprises separate trigger and high-precision tracking
chambers measuring the deflection of muons in a magnetic field generated by the
superconducting air-core toroids.
The precision chamber system covers the region $|\eta| < 2.7$ with three layers
of monitored drift tubes, complemented by cathode-strip chambers in the forward
region.
The muon trigger system covers the range $|\eta| < 2.4$ with resistive-plate
chambers in the barrel and thin-gap chambers in the endcap regions.
 
A two-level trigger system~\cite{TRIG-2016-01} is used to select interesting
events.
The first level of the trigger is implemented in hardware and uses a subset of
detector information to reduce the event rate to a design value of at most
\SI{100}{\kHz}.
It is followed by a software-based trigger that reduces the event rate to a
maximum of around \SI{1}{\kHz} for offline storage.
 
An extensive software suite~\cite{ATL-SOFT-PUB-2021-001} is used in data simulation,
in the reconstruction and analysis of real and simulated data, in detector operations,
and in the trigger and data acquisition systems of the experiment.


\section{Monte Carlo Samples}
\label{sec:samples}
 
The optimisation of the ATLAS \RunTwo \btag algorithms is performed with jets from
a \enquote{hybrid} sample composed of a mixture of simulated SM \ttbar\ and high-mass
$Z^\prime \to q\bar q$ events.
The \Zprime events do not correspond to a single resonance but have a
broad \Zprime mass spectrum in order to optimise the $b$-tagging performance at
high jet momentum transverse to the beam-line (\pt).
The final hybrid sample for training is obtained by mixing all $b$-jets from
the available \ttbar\ events, if the corresponding $b$-hadron \pt\ is below
$250~\GeV$, with all jets containing a $b$-hadron with $\pt > 250$~\GeV\ from the
\Zprime sample.
A similar strategy, based on the jet \pt , is applied for $c$-jets and
light-flavour jets.
No attempt is made to distinguish quark-initiated jets from gluon-initiated jets in these samples.
 
The \ttbar simulation sample was produced using
\POWHEGBOX[v2]~\cite{Nason:2004rx,Frixione:2007vw,Frixione:2007nw,Alioli:2010xd},
which yields matrix elements at next-to-leading order (NLO) in the strong
coupling constant \alphas for top-quark pair production.
The first-gluon-emission cut-off scale parameter \hdamp was
set to $1.5m_t$, with $m_t=172.5~\GeV$ used for the top-quark
mass.
\POWHEGBOX was interfaced to \Pythia[8.230]~\cite{Sjostrand:2007gs} with
the A14 set of tuned parameters~\cite{ATL-PHYS-PUB-2014-021} and
\NNPDF[3.0nnlo] (\NNPDF[2.3lo])~\cite{Ball:2014uwa,Ball:2012cx}
parton distribution functions in the matrix elements (parton shower).
This set-up was found to produce the best modelling, out of a number of
available generator configurations, of the multiplicity of additional jets and
of both the individual top-quark and \ttbar system
\pt~\cite{ATL-PHYS-PUB-2016-020, ATL-PHYS-PUB-2020-023}.
The $\ttbar$ events with at least one leptonic $W$-boson decay are considered, which
ensures that a sufficiently large fraction of $b$-jets, $c$-jets, and light-flavour jets are
present in the jet population.
 
To train and to evaluate the performance of the \btag algorithms at high jet \pt ,
a \Zprime sample was generated using \Pythia[8.212] with the A14 set of
tuned parameters for the underlying event and the leading-order
\NNPDF[2.3lo]~\cite{Ball:2012cx} parton distribution functions.
The cross-section of the hard-scattering process was modified by applying an
event-by-event weighting factor to broaden the natural width of the resonance.
The branching fractions of the decays were set to be one-third each for the
$b\bar b$, $c\bar c$ and light-flavour quark pairs to obtain a sample
uniformly populated by jets of each flavour.
This results in a fairly flat jet \pt spectrum between 250~\GeV\ and 1.5~\TeV\ for
$b$-jets, $c$-jets, and light-flavour jets, with the falling tail of the \pt distribution
populated to 3~\TeV\ for each flavour.
 
The \EVTGEN~\cite{Lange:2001uf} package was used to simulate the decay
of heavy-flavour hadrons.
All simulated events have additional overlaid minimum-bias interactions
generated with \Pythia[8.160] with the A3 set of tuned
parameters~\cite{ATL-PHYS-PUB-2016-017} and \NNPDF[2.3lo] parton
distribution functions to simulate pile-up background.\footnote{Pile-up
interactions correspond to additional $pp$ collisions accompanying the
hard-scatter $pp$ interaction in proton bunch collisions at the LHC.}
These events are weighted to reproduce the observed distribution of the average
number of interactions per bunch crossing in the corresponding data sample.
The simulated events were processed through the full ATLAS detector
simulation~\cite{SOFT-2010-01} based on
\GEANT~\cite{Agostinelli:2002hh}.
Interactions of $b$-hadrons with the detector material were not
simulated; $c$-hadrons and $\tau$-leptons are similarly affected.
This may result in sub-optimal identification algorithm performance in data for
jets with very energetic particles that decay beyond the IBL.
ATLAS analyses apply dedicated correction factors and corresponding
uncertainties to account for the difference in performance between data and
simulation.


\section{Key ingredients for flavour-tagging}
\label{sec:objects}
 
Jet flavour identification relies upon a number of more fundamental objects
reconstructed in the ATLAS detector.
The most important of these are charged-particle tracks, primary
vertices (PVs), and hadronic jets (\enquote{jets}).
Limited information about the hadrons contained within hadronic jets from the
simulated event's record is also used during algorithm optimisation and
performance studies.
A short description of each of these key ingredients is given below, and more
detailed information is available in Section~3 of Ref.~\cite{FTAG-2018-01}.
 
Charged-particle tracks are reconstructed in the ID~\cite{PERF-2015-08}.
Only tracks with $\pt > 500~\MeV$ are used in jet flavour tagging, with further
selection criteria applied to reject fake and poorly measured
tracks~\cite{ATL-PHYS-PUB-2015-051}.
The total tracking efficiency for charged pions with $\pt > 4~\GeV$ ranges from
90\% for ${|\eta| < 1}$ to 70\% in the forward region (${2.3 < |\eta| < 2.5}$)
of the detector.
Additional selection criteria for reconstructed tracks are applied in the
low-level \btag algorithms described in Section~\ref{sec:lowlevel} to maintain a
high efficiency for charged particles from heavy-flavour hadron decays while
rejecting tracks originating from pile-up.
 
Primary vertex reconstruction~\cite{PERF-2015-01,ATL-PHYS-PUB-2015-026} on an
event-by-event basis is particularly important for \btag\ since it defines the
reference point from which track and vertex displacements are computed.
Reconstructed PVs are constrained to be within the luminous region of the
colliding LHC beams.
A longitudinal vertex position resolution of about $30~\mu$m is achieved for
events with a high multiplicity of reconstructed tracks, while the transverse
resolution ranges from 10 to 12~$\mu$m, depending on the LHC running conditions
that determine the beam-spot size.
Flavour tagging requires at least one PV in each event, and the PV with the
highest sum of squared transverse momenta of contributing tracks is selected
as the primary interaction point or \enquote{hard-scatter vertex}; all tracks
contributing to a vertex are used in the hard-scatter vertex determination --
they need not be associated to a hadronic jet.
Charged-particle tracks originating from $b$-hadron decays often have large
transverse and longitudinal impact parameters,
$d_0$ and $z_0$ respectively, where $d_0$ is the distance of closest approach of
the track to the PV in the transverse plane, and $z_0$ is the longitudinal
separation between the PV and the point on the track where $d_0$ is measured,
referred to below as the \enquote{perigee}.

Hadronic jets are built from \enquote{particle-flow objects}, which are
constructed from signals in the ATLAS calorimeters and ID; particle-flow objects
take advantage of the better resolution of particle tracking when low-\pt
charged hadrons result in geometrically-matched calorimeter deposits and a
charged-particle track~\cite{PERF-2015-09}.
The anti-$k_t$ algorithm with radius parameter $R=0.4$~\cite{Cacciari:2008gp},
implemented in \textsc{FastJet}~\cite{Fastjet}, is used for jet finding.
Jet transverse momenta are further corrected to the corresponding particle-level
jet \pt, based on the simulation~\cite{JETM-2018-05}.
Remaining differences between simulated events and observed data are
evaluated using in situ techniques, which exploit the transverse
momentum balance between a jet and a reference object such as a photon, $Z$
boson, or multi-jet system in data; corrections are applied to simulated jets to
bring them in line with data~\cite{JETM-2018-05}.
Jets with $\pt < 20~\GeV$ or $|\eta| \ge 2.5$ are not considered for jet flavour
identification, as low-\pt jets are outside the valid calibration range
and high-$|\eta|$ jets are outside the tracking fiducial volume.
In order to reduce the number of jets with large energy fractions from pile-up
collision vertices, the \enquote{jet vertex tagger} (JVT) algorithm is
used~\cite{ATLAS-CONF-2014-018}.
The JVT procedure builds a multivariate discriminant for each jet within
$|\eta| < 2.4$ based on the ID tracks ghost-associated\footnote{The
ghost-association algorithm collects tracks within a jet's geometric catchment
area, taking into account possibly overlapping jet areas and is
described in detail in Ref~\cite{Cacciari:2008gn}.} with the
jet~\cite{Cacciari:2008gn}; in particular, jets with a large fraction of
high-momentum tracks from pile-up vertices are less likely to pass the JVT
requirement.
The JVT efficiency for jets originating from the hard $pp$ scattering is 92\%
in the simulation, but the rate of pile-up jets with $\pt \ge 60~\GeV$ is
sufficiently small that the JVT requirement is removed above this threshold.
The reconstructed jet direction and transverse momentum are especially important
inputs to flavour-tagging, as they will determine which charged-particle tracks
should be considered for jet flavour identification.
 
Tracks are matched to jets by setting a maximum allowed angular separation
$\Delta R$ between the track momenta, defined at the perigee, and the jet axis.
Given that the decay products from higher-\pt\ $b$-hadrons are more collimated,
the $\Delta R$ requirement varies as a function of jet \pt, being wider for
low-\pt\ jets (0.45 for jet $\pt=20~\GeV$) and narrower for high-\pt\ jets (0.26
for jet $\pt=150~\GeV$); if more than one jet fulfils the matching criteria, the
closest jet is preferred.
The jet axis is also used to assign signed impact parameters to tracks, where
the sign is defined to be positive if the track intersects the jet axis in the
transverse plane in front of the primary vertex, and negative if the
intersection lies behind the primary vertex~\cite{PERF-2012-04}.
 
The flavour of a jet in simulation is determined by the nature of the hadrons it
contains.
Jets are labelled as $b$-jets if at least one weakly decaying $b$-hadron having
$\pt \ge 5~\GeV$ is found within a cone of size $\Delta R=0.3$ around the jet
axis.
If no $b$-hadrons are found, $c$-hadrons and then $\tau$-leptons are searched
for, based on the same selection criteria.
The jets matched to a $c$-hadron ($\tau$-lepton) are labelled as $c$-jets
($\tau$-jets).
The remaining jets are labelled as light-flavour jets.
For jets with more than one heavy-flavour hadron, e.g.\ from
gluon splitting to $b\bar b$ or $c\bar c$, the procedure above is still
followed, and $b\bar b$ ($c \bar c$) jets will receive a $b$ ($c$) label.


\section{Low-level $b$-taggers}
\label{sec:lowlevel}
 
This section describes the different low-level algorithms used for $b$-jet identification
in the ATLAS \RunTwo dataset.
These algorithms, designed to reconstruct the characteristic features of $b$-jets,
fall into two broad categories.
The first approach, implemented in the \iptwod and \ipthreed
algorithms~\cite{ATL-PHYS-PUB-2017-013} and described in Section~\ref{sec:IPTag},
is inclusive and based on exploiting the large impact
parameters of the tracks originating from the $b$-hadron decays.
The new \rnnip~\cite{ATL-PHYS-PUB-2017-003} algorithm, developed during \RunTwo
and described in Section~\ref{sec:rnn}, exploits a recurrent neural network~\cite{Graves:2012rnn} to learn
track impact-parameter correlations in order to further improve
the jet flavour discrimination.
The second approach explicitly reconstructs displaced vertices.
The \svone algorithm~\cite{ATL-PHYS-PUB-2017-011}, discussed in
Section~\ref{sec:SV1}, attempts to reconstruct an inclusive secondary vertex,
while the \jetfitter algorithm~\cite{ATL-PHYS-PUB-2018-025}, presented
in Section~\ref{sec:JetFitter}, aims to reconstruct the full $b$- to $c$-hadron
decay chain.

\subsection{Algorithms based on impact parameters}
\label{sec:IPTag}
 
\begin{figure}[htbp]
\centering
\subfigure[]{\includegraphics[width=0.45\textwidth]{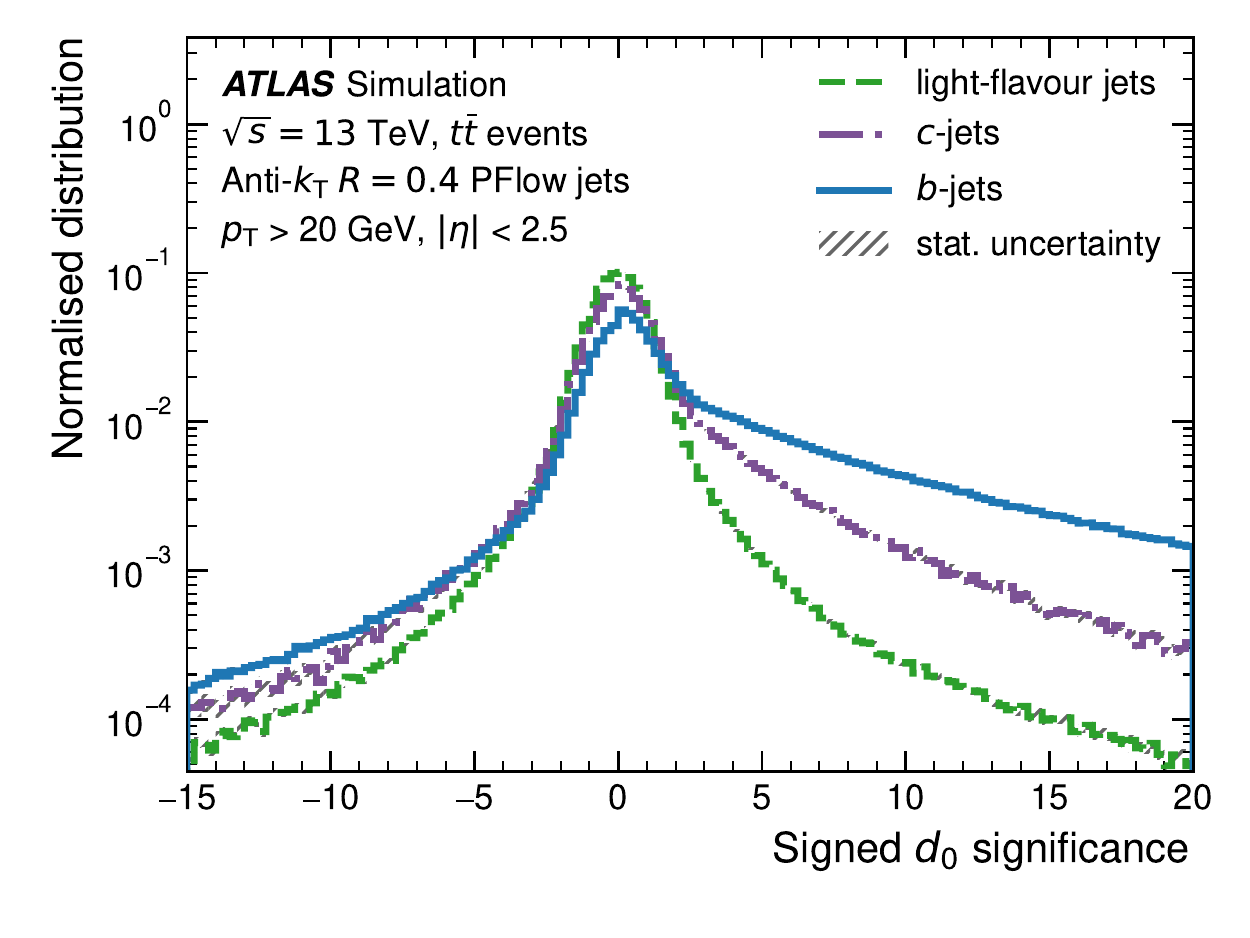}}
\subfigure[]{\includegraphics[width=0.45\textwidth]{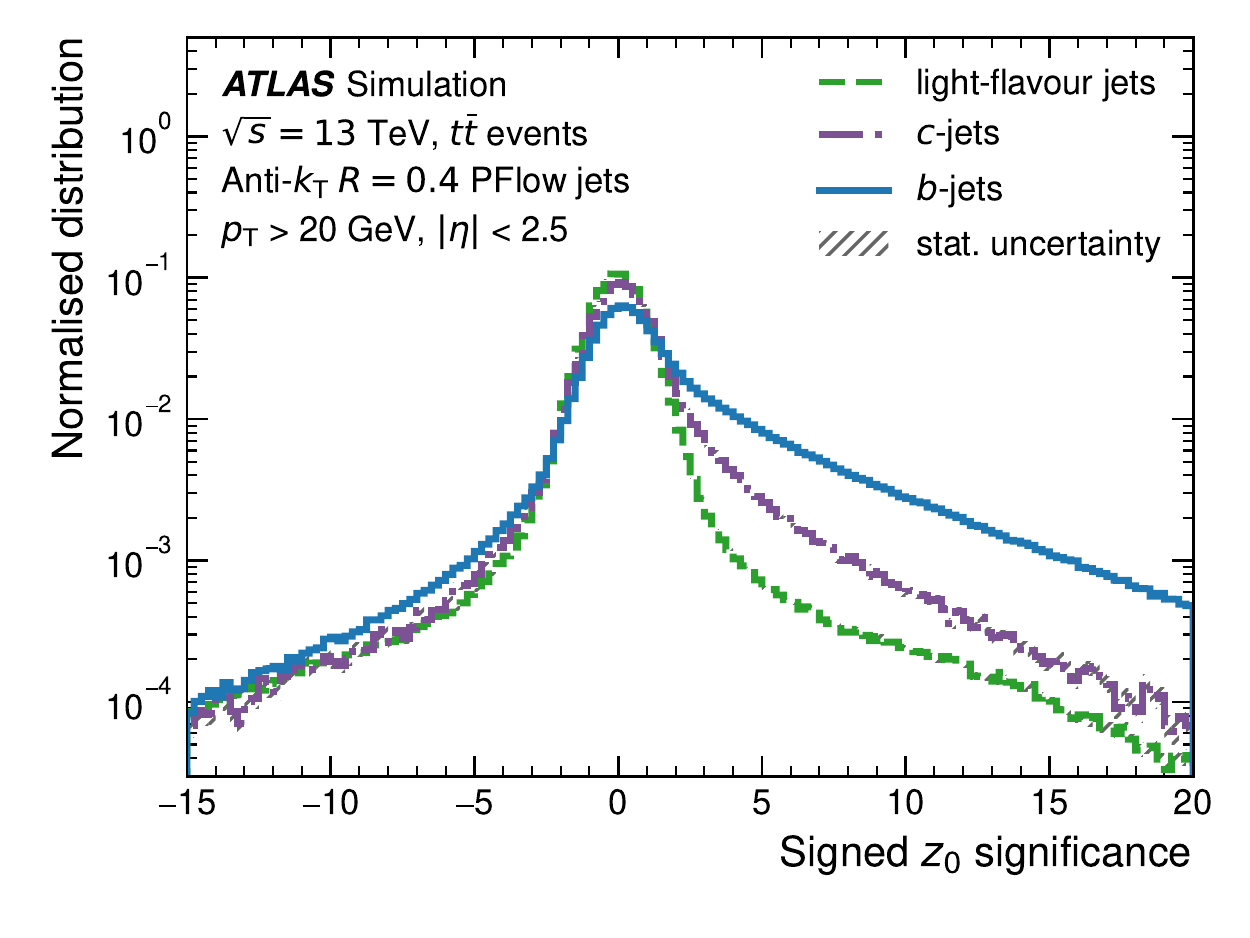}}
\caption{The (a) transverse and (b) longitudinal signed impact parameter significance of tracks for $b$-jets, $c$-jets and light-flavour jets in $t\bar{t}$ events. The first (last) bin in the distribution does not account for underflow (overflow).}
\label{fig:ipxd-significance}
\end{figure}

\begin{figure}[htbp]
\centering
\subfigure[]{\includegraphics[width=0.45\textwidth]{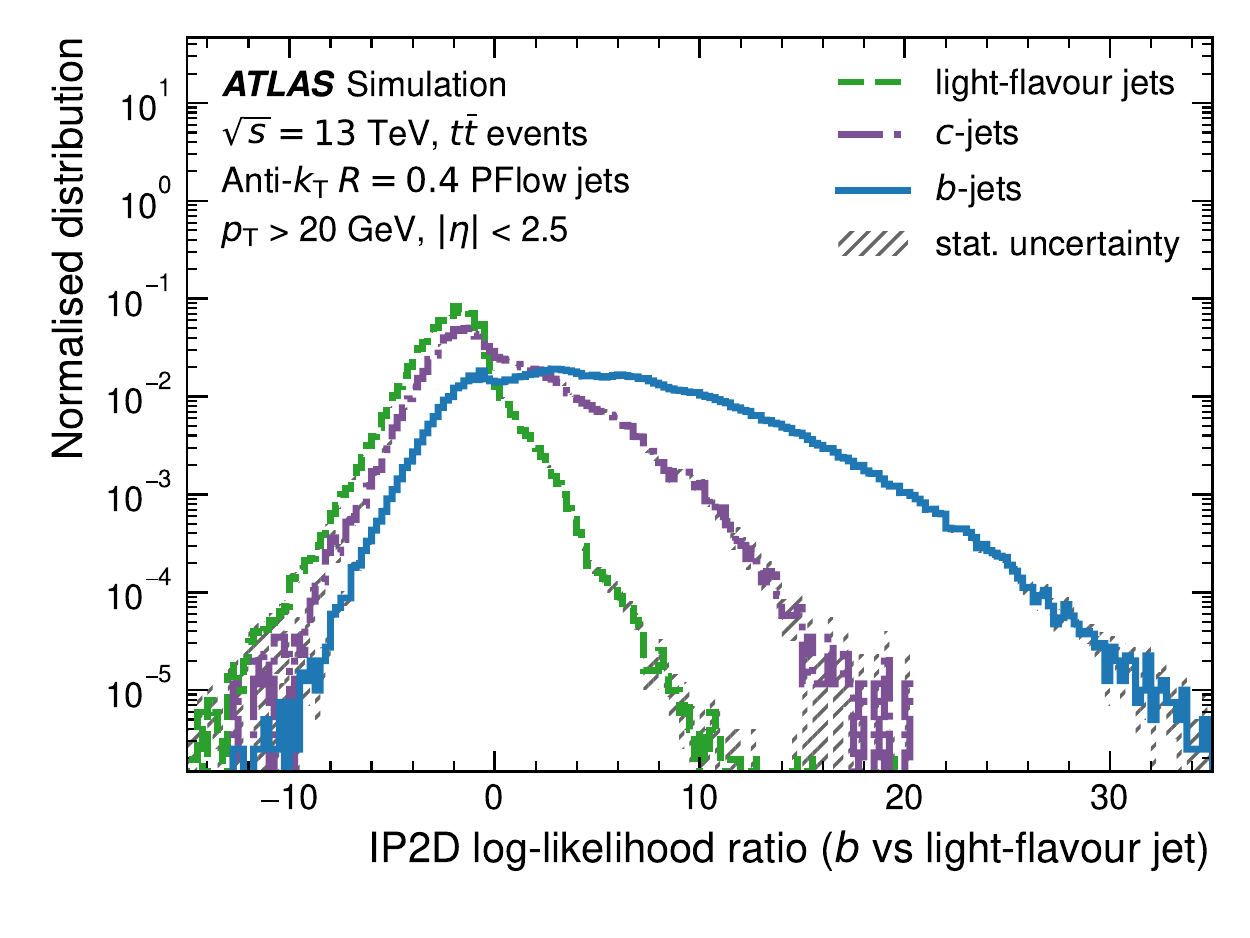}}
\subfigure[]{\includegraphics[width=0.45\textwidth]{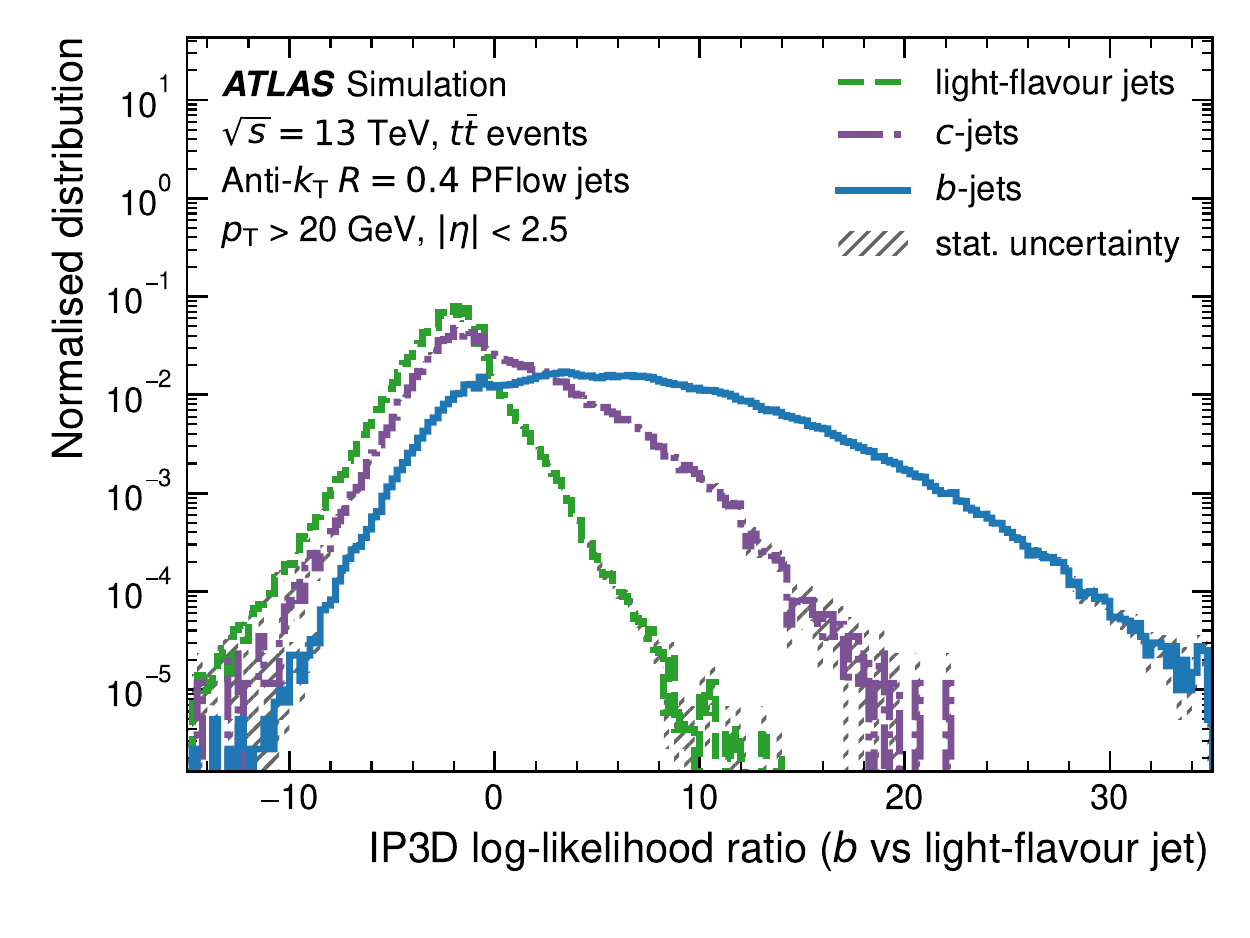}}
\caption{The log-likelihood ratio for the (a) \iptwod and (b) \ipthreed \btag algorithms for $b$-jets, $c$-jets and light-flavour jets in $t\bar{t}$ events. The log-likelihood ratio shown here is computed as the ratio of the $b$-jet to light-flavour jet hypothesis probabilities.
Jets with no tracks are not included in the plot, but assigned a large negative value.
The first (last) bin in the distribution does not account for underflow (overflow).}
\label{fig:ipxd-llr}
\end{figure}

Two impact-parameter-based (IP-based) algorithms, \iptwod and
\ipthreed~\cite{ATL-PHYS-PUB-2017-013}, are used by ATLAS.
The \iptwod tagger makes use of the signed transverse impact parameter
significance of tracks to construct a discriminating variable, whereas
\ipthreed uses both the signed transverse and signed longitudinal impact parameter
significances\footnote{The impact parameter significance $S_{i}$ of a track $i$
is defined as the ratio of the track impact parameter to its uncertainty.}
in a two-dimensional template to account for their correlation.
The signed transverse and longitudinal impact parameter significances
are shown in Figure~\ref{fig:ipxd-significance}.
Probability density functions (pdfs) obtained from reference histograms of the
signed transverse and longitudinal impact parameter significances of tracks
associated with $b$-jets, $c$-jets and light-flavour jets are derived from MC
simulation.
The pdfs are computed in exclusive categories that depend on the hit multiplicities of
the tracks in the different ID layers to increase the discriminating power.
In particular, the IBL hit pattern expectations and the second-innermost layer's
information are fully exploited to improve the \btag performance in LHC \RunTwo.
A set of template pdfs is produced using an equal
mix of simulated \ttbar and \Zprime events for track categories
with no hits in the first two layers, which are populated by long-lived
$b$-hadrons traversing the first layers before they decay.
The \ttbar\ sample is used to populate all remaining categories.
The pdfs are used to calculate ratios of the $b$-jet, $c$-jet and light-flavour
jet probabilities on a per-track basis.
Log-likelihood ratio (LLR) discriminants are then defined as the sum of the logarithms
of the per-track probability ratios for each jet-flavour hypothesis, e.g.\
$\sum_{i=1}^{N} \ln \left({\pb(S_{i})}/{\pu(S_{i})}\right)$ for the $b$-jet and light-flavour
jet hypotheses, where $N$ is the number of tracks associated with the jet and
$\pb(S_{i})$ $\left(\pu(S_{i})\right)$ is the template pdf for the $b$-jet (light-flavour jet)
hypothesis, with $S_{i}$ the impact parameter significance of the track $i$.
The flavour probabilities of the different tracks contributing to the sum are
assumed to be independent of each other.
The log-likelihood ratios separating $b$-jets from light-flavour jets
for the \iptwod and  \ipthreed \btag algorithms are shown in
Figure~\ref{fig:ipxd-llr}.
In addition to the LLR separating $b$-jets from light-flavour jets, two extra
LLR discriminants are defined to separate $b$-jets from $c$-jets and $c$-jets from
light-flavour jets, respectively.
These three likelihood discriminants for both the \iptwod and
\ipthreed algorithms are used as inputs to the high-level taggers.

\subsection{Track-based recurrent neural network tagger}
\label{sec:rnn}
 
\begin{figure}[htbp]
\centering
\includegraphics[width=0.95\textwidth]{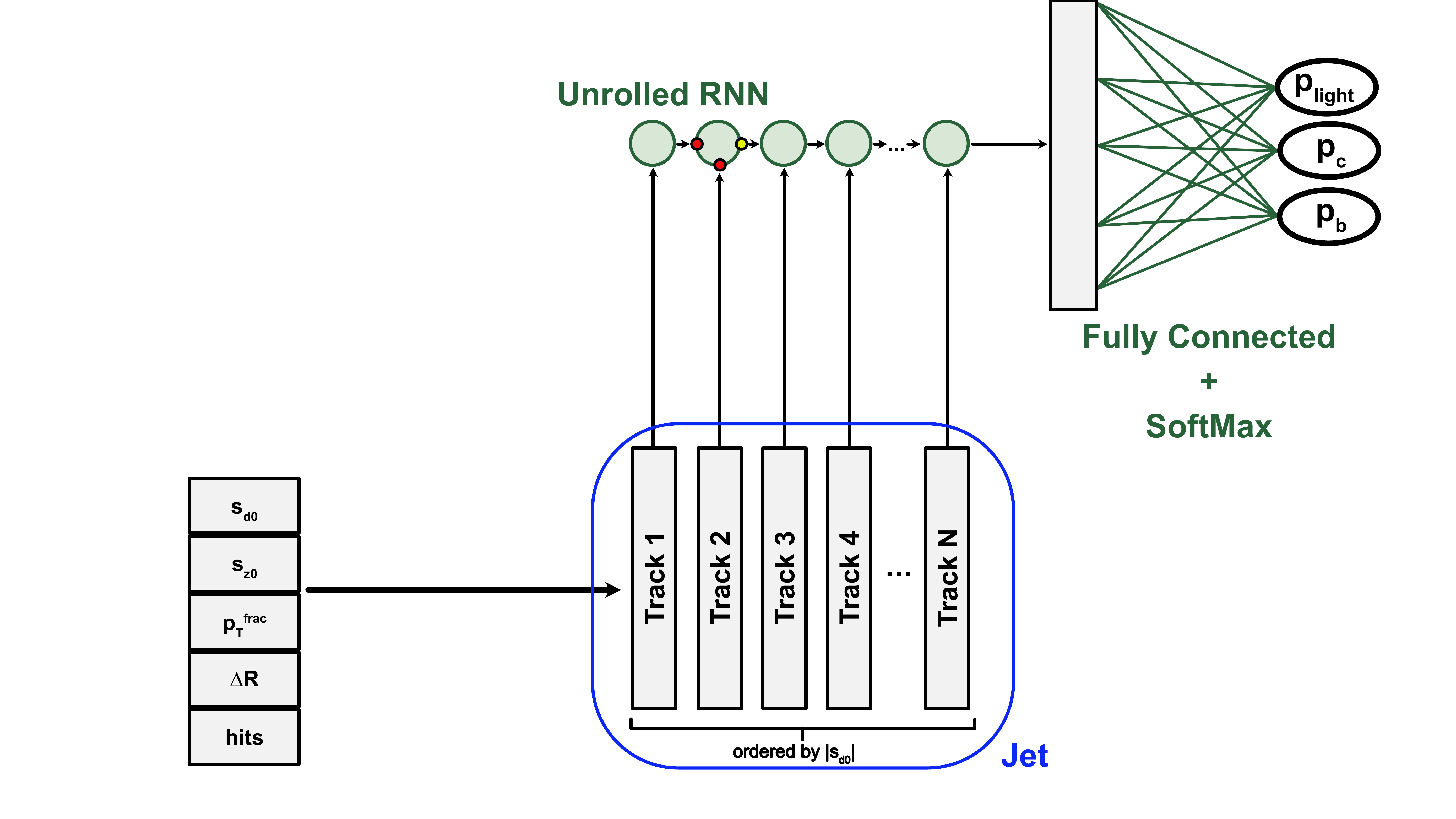}
\caption{Schematic drawing of the \rnnip neural network architecture. The $S_{d_0}$ and $S_{z_0}$ input variables correspond to the lifetime-correlated signed transverse and longitudinal impact parameter significances, while $\pT^{\textrm{frac}}$ and $\Delta R$ represent the fraction of transverse momentum carried by the track relative to the jet and the angular distance between the track and the jet axis, respectively.}
\label{fig:rnnip-diagram}
\end{figure}

In the case of a $b$-hadron decay, several charged particles can emerge from the
secondary (or tertiary) decay vertex with large impact parameters.
The impact parameters of these hadron-decay tracks are intrinsically correlated: if
one track is found with a large impact parameter then finding a second track with
a large impact parameter is more likely.
If no displaced decay is present, as in light-flavour jets, then such a
correlation for large impact parameter significance is not expected.
The baseline IP-based $b$-tagging algorithms, described in Section
\ref{sec:IPTag} uses likelihood templates to compute per-flavour conditional
likelihoods.
Due to the large event sample needed to compute such templates and therefore the
difficulty in exploiting more input variables as additional histogram axes,
IP-based $b$-tagging algorithms assume that the properties of each track in a
jet are independent of all other tracks, which limits their ability to fully model
the properties of $b$-jets.
Recurrent neural networks (RNN) can be used to overcome this challenge by directly learning
sequential dependencies between the variables in sequences of arbitrary length, as is the
case for the tracks in a jet.
For each selected track, the lifetime-correlated signed transverse ($S_{d_{0}}$) and longitudinal
($S_{z_{0}}$) impact parameter significances, the fraction of transverse
momentum carried by the track relative to the jet $\pT$
($\pT^{\textrm{frac}}$), the angular distance between the track and the
jet-axis ($\Delta R(\textrm{track, jet-axis})$) and the hit multiplicity of
the track in the different ID layers are fed into the
network~\cite{ATL-PHYS-PUB-2017-003}.
The tracks are ordered by the $|S_{d_{0}}|$ values to form a sequence emphasising the
particular importance this kinematic feature. This sequence is then passed to the neural
network cells as a vector of the ordered-track features.
During the training, the $b$-jet and $c$-jet $\pT$ spectra are separately reweighted to
the light-flavour jet spectrum so as to prevent the RNN from learning to discriminate directly
from sample- and flavour-specific momentum distributions.
The outputs provided by the network correspond to the $b$-jet, $c$-jet,
and light-flavour jet probabilities.
Figure~\ref{fig:rnnip-diagram} shows a schematic of the network architecture
used for this algorithm, known as \rnnip.
 
\begin{figure}[htbp]
\centering
\subfigure[]{\includegraphics[width=0.45\textwidth]{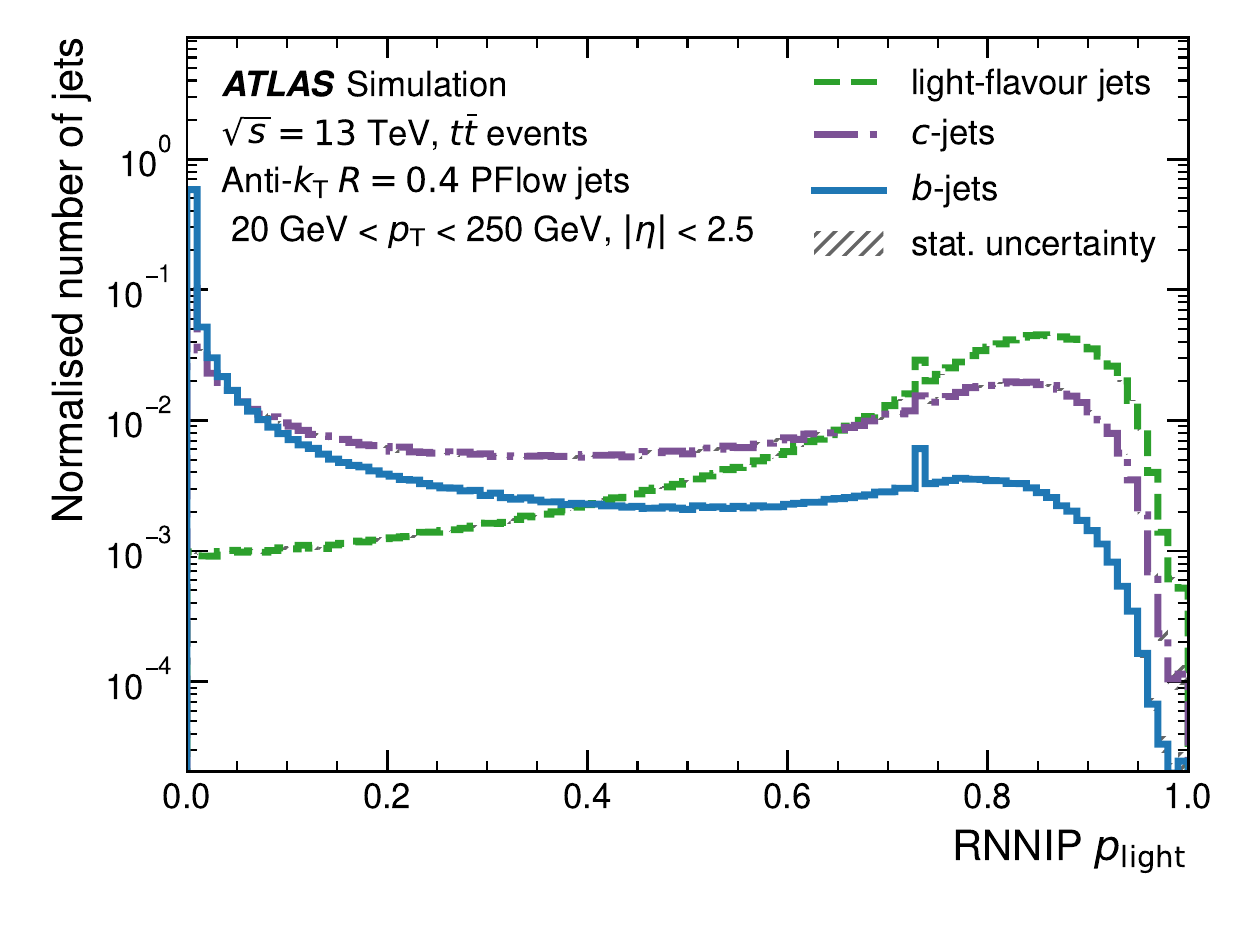} }
\subfigure[]{\includegraphics[width=0.45\textwidth]{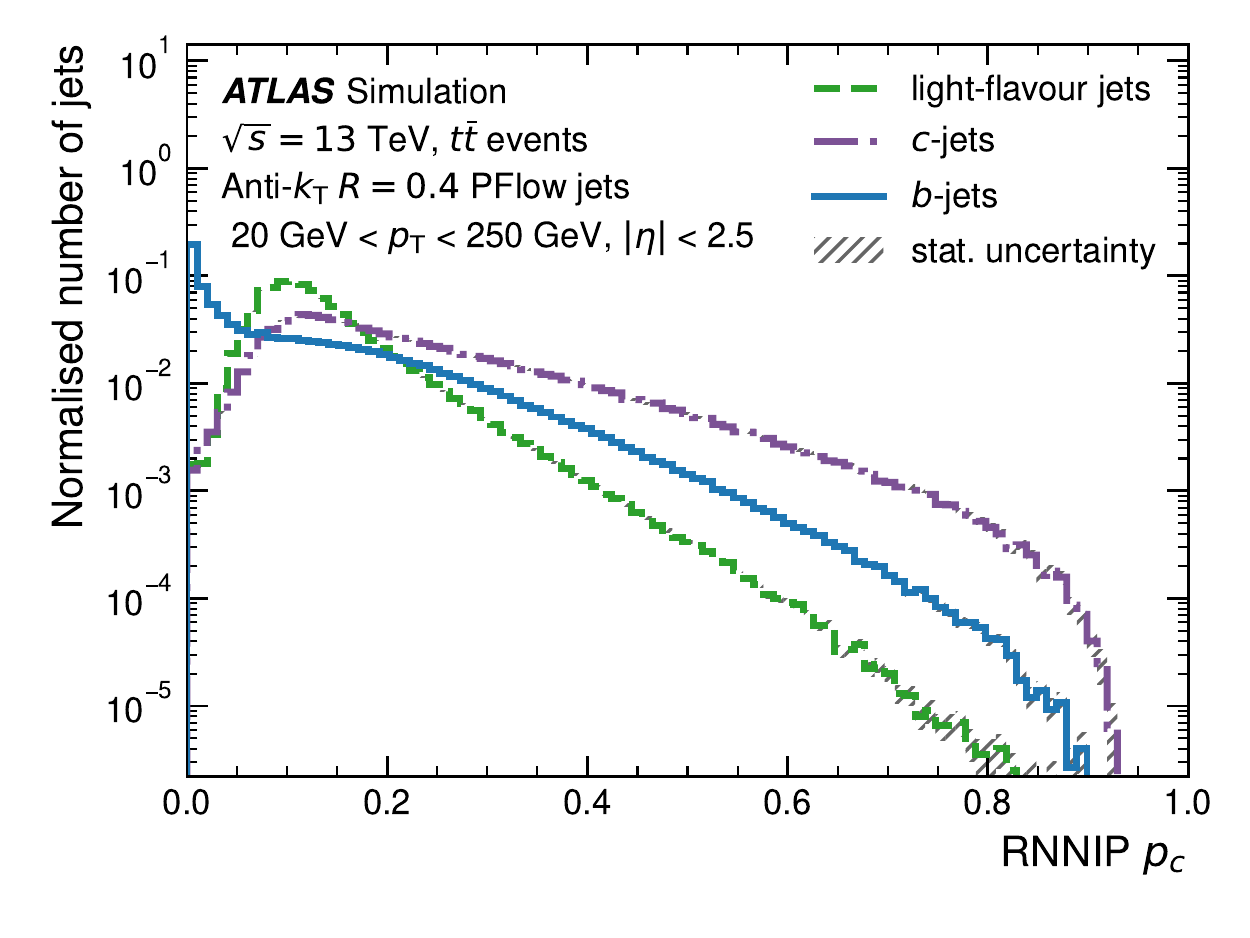} } \\
\subfigure[]{\includegraphics[width=0.45\textwidth]{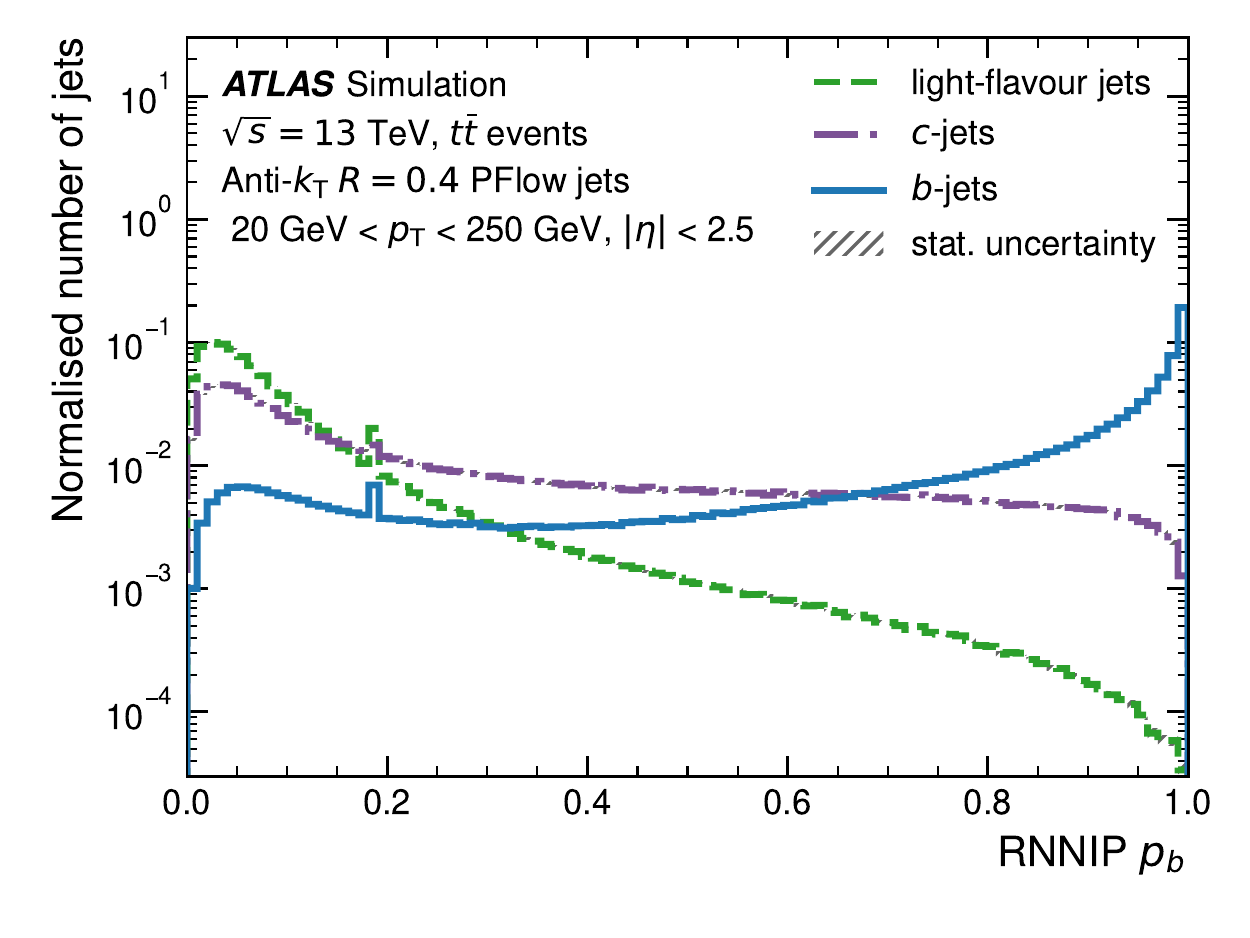} }
\subfigure[]{\includegraphics[width=0.45\textwidth]{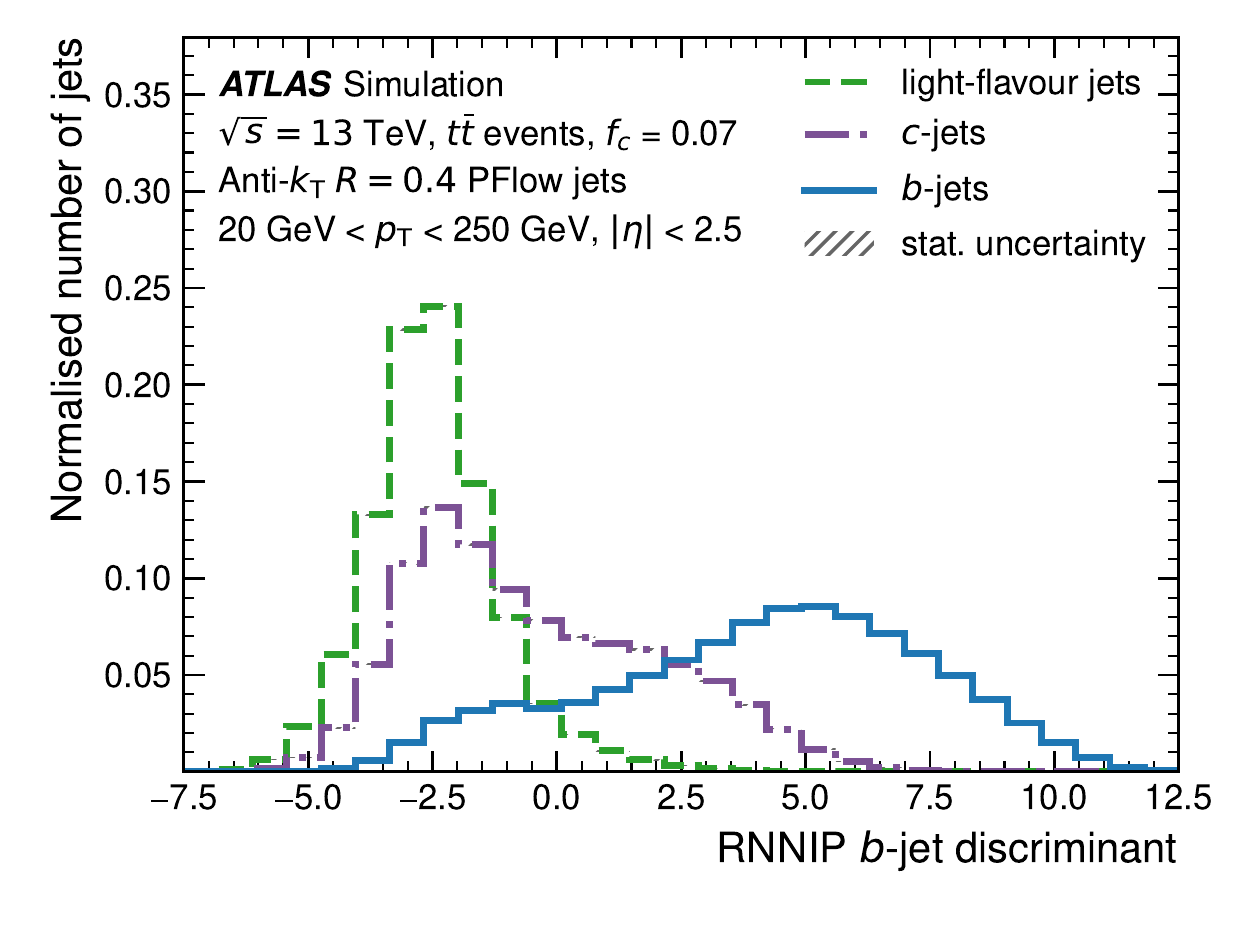} }
 
\caption{
Distributions of the outputs of the \rnnip \btag algorithm for
$b$-jets, $c$-jets and light-flavour jets in the baseline \ttbar\ simulated
events: (a) \pu, (b) \pc, (c) \pb, and (d) the \rnnip $b$-tagging discriminant.
The spikes at $\pu \approx 0.73$ and $\pb \approx 0.18$ originate from jets
with no tracks.
The location of these spikes is determined by the probability
that a jet of a given flavour has no associated tracks: since light-flavour
jets are more likely than $b$-jets to contain no tracks, $\pu > \pb$ in such
cases.
The corresponding spike at $\pc \approx 0.09$ is not easily visible since it
occurs within a steeply falling or rising region of the \pc distribution.
}
 
\label{fig:RNNIP}
\end{figure}

The outputs of the RNN are combined into the \btag discriminant function as defined as:
\begin{equation*}
D_{\mathrm{RNNIP}} = \ln \left( \frac{\pb}{f_c \cdot \pc + (1-f_c) \pu} \right),
\end{equation*}
where $p_{b}$, $p_{c}$ and $p_{\textrm{light}}$ are the three jet-flavour probabilities, and $f_c$ denotes the $c$-jet fraction.
The relative importance of $c$-jet and light-flavour-jet rejection
can be changed by varying $f_c$. In this paper an optimised $c$-jet fraction of 0.07 is used
in order evaluate the performance of the \rnnip \btag algorithm in $t\bar{t}$ and $Z'\rightarrow q\bar{q}$ events.
This value is chosen as a compromise to ensure
good rejection factors for both $c$-jets and light-flavour jets in a large \btag efficiency range.
The distribution of the \rnnip $b$-tagging discriminant for
$b$-jets, $c$-jets and light-flavour jets in the baseline \ttbar\ simulated
events is shown in Figure~\ref{fig:RNNIP}.
 
\subsection{Secondary-vertex-tagging algorithm}
\label{sec:SV1}
 
\begin{figure}[htbp]
\centering
\subfigure[]{\includegraphics[width=0.41\textwidth]{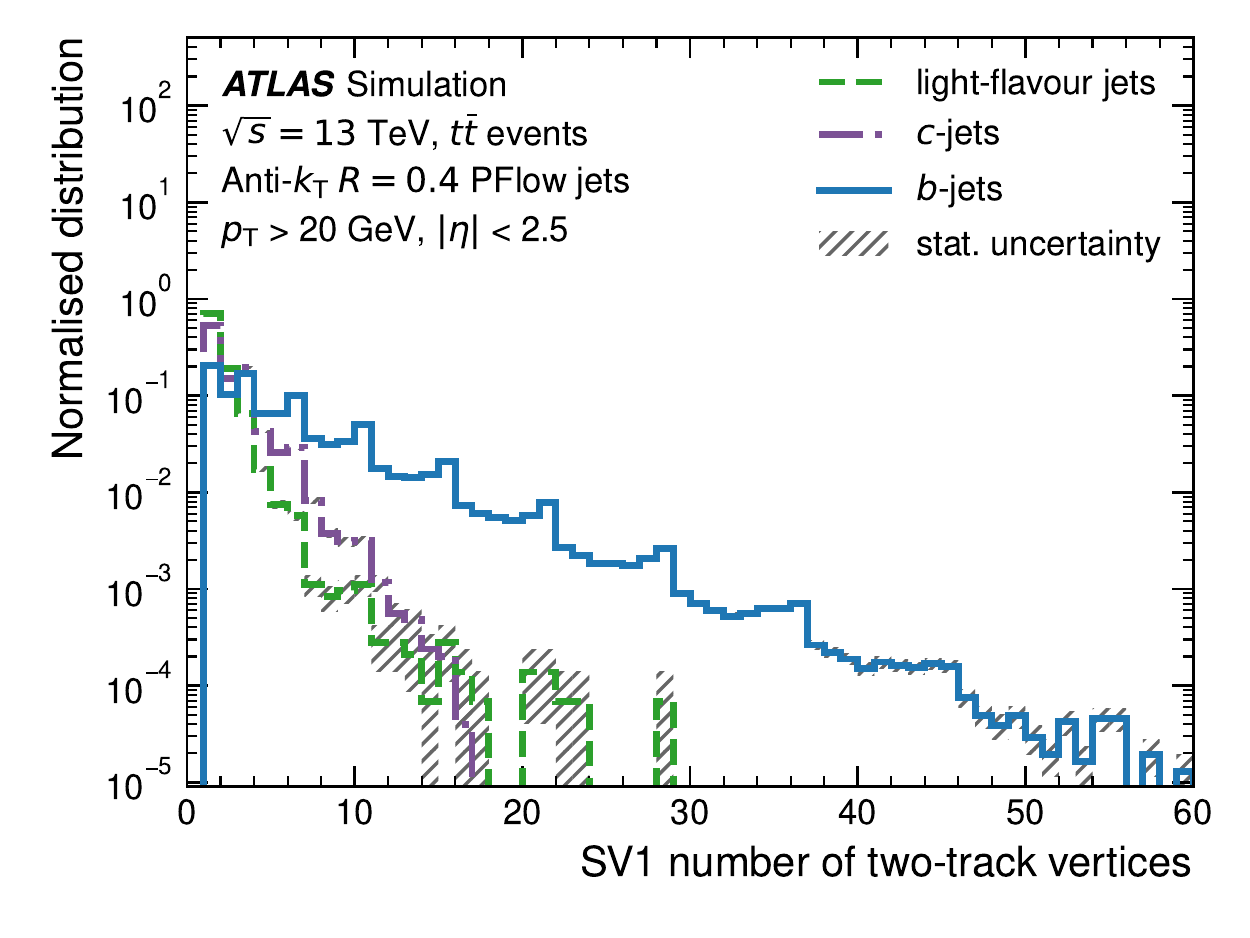}}
\subfigure[]{\includegraphics[width=0.41\textwidth]{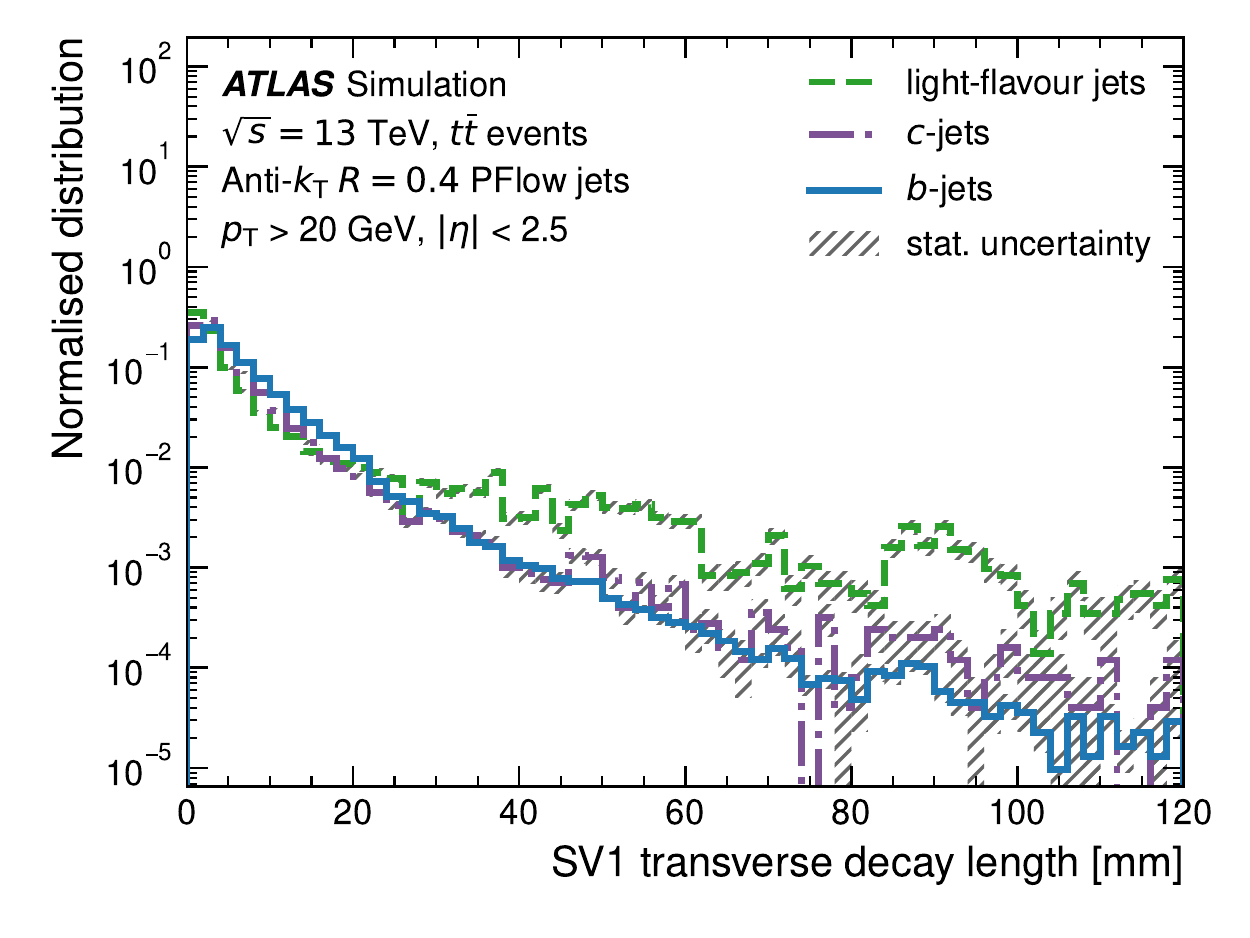}}
\subfigure[]{\includegraphics[width=0.41\textwidth]{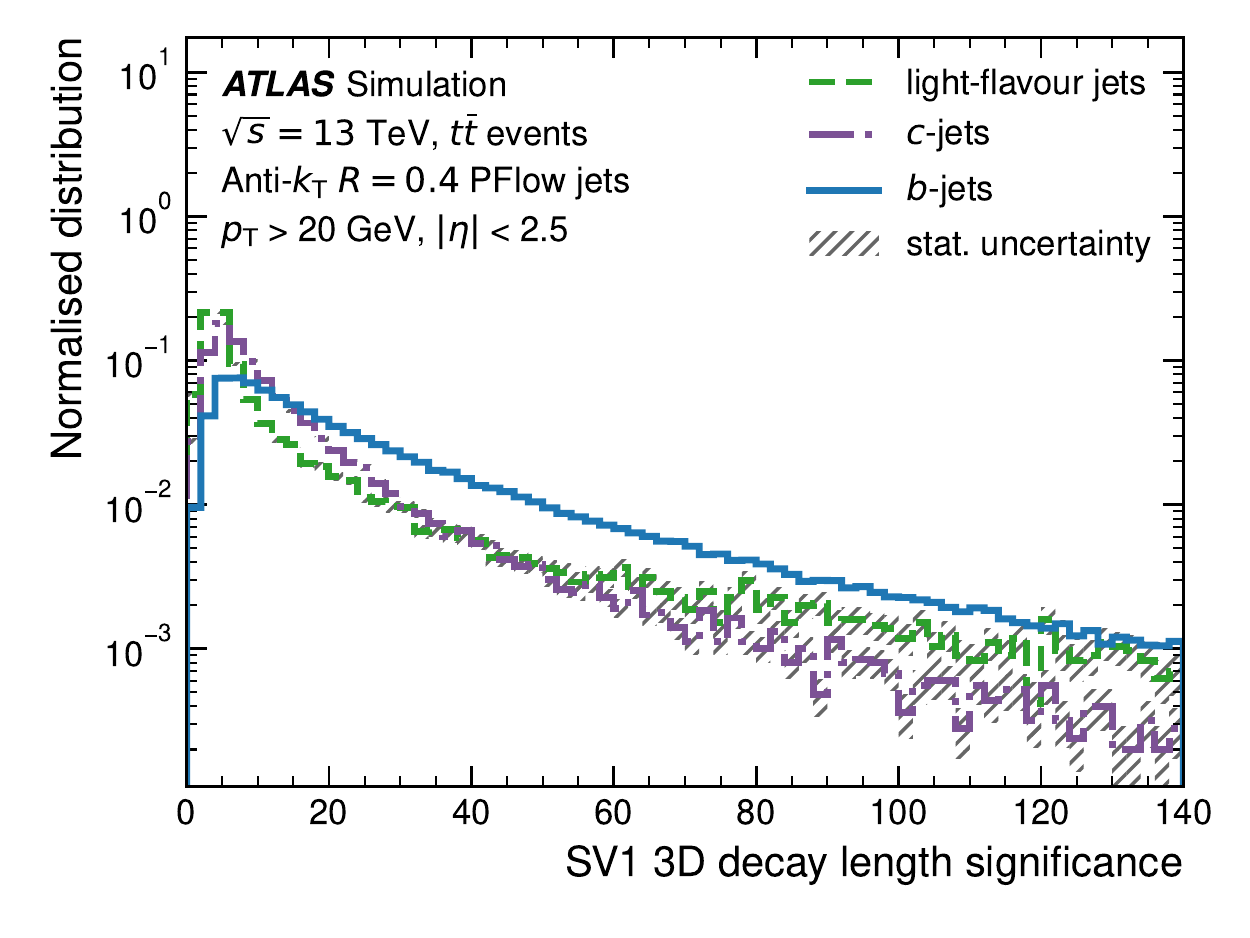}}
\subfigure[]{\includegraphics[width=0.41\textwidth]{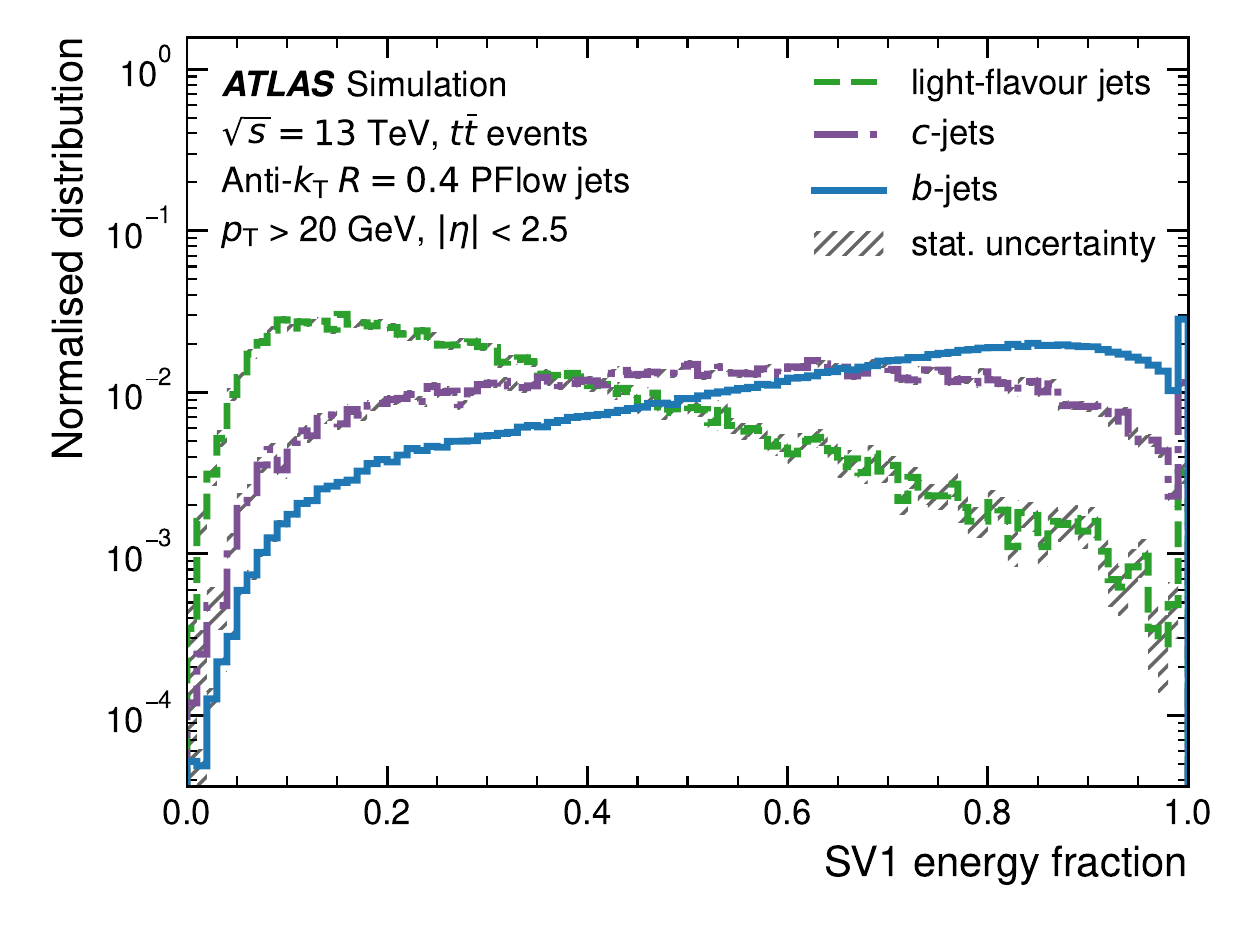}}
\subfigure[]{\includegraphics[width=0.41\textwidth]{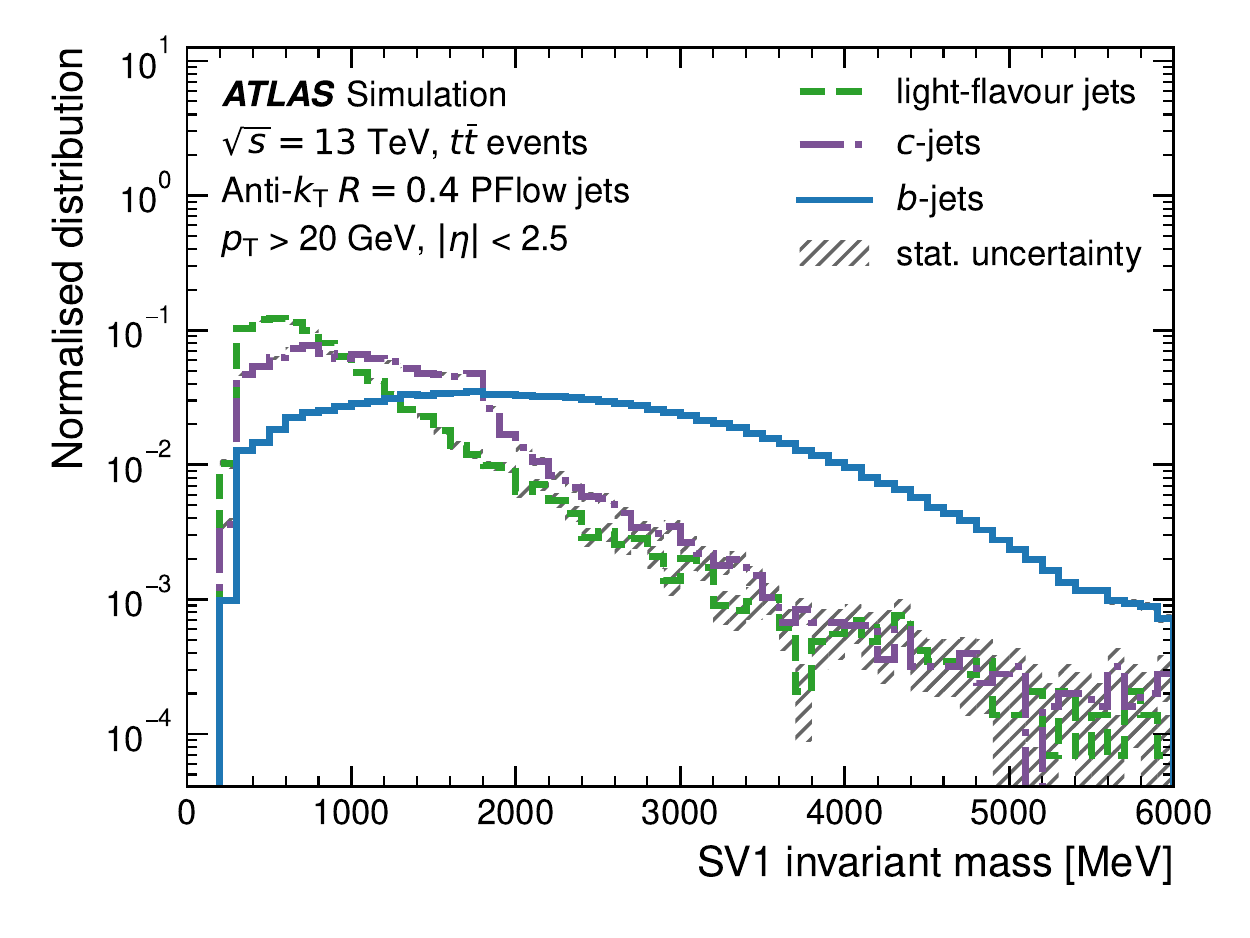}}
\subfigure[]{\includegraphics[width=0.41\textwidth]{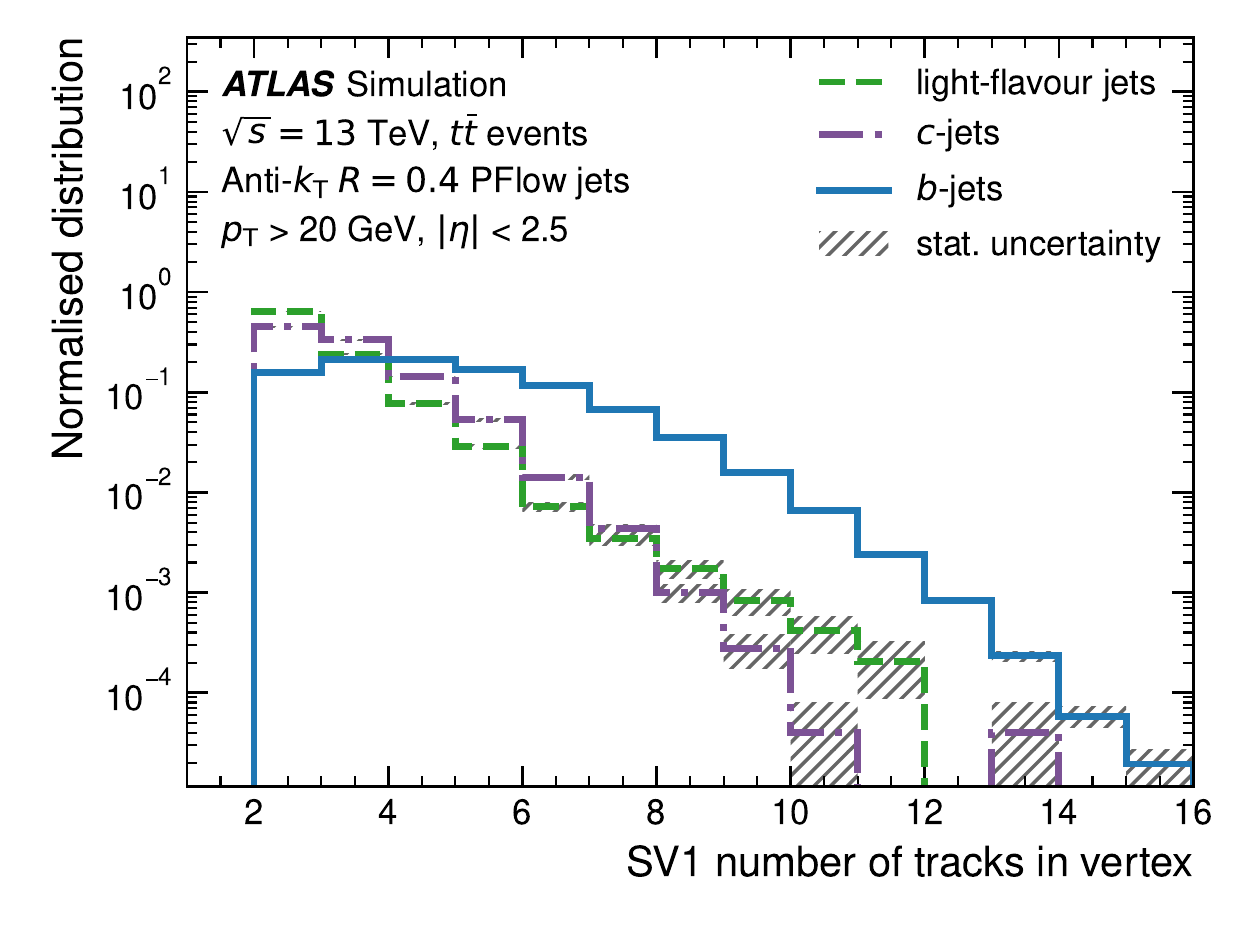}}
\caption{Properties of secondary vertices reconstructed by the \svone algorithm
for $b$-jets, $c$-jets and light-flavour jets in the baseline \ttbar\ simulated
events: (a) the number of two-track vertices reconstructed within the jet, (b)
the transverse decay length, (c) the 3D decay length significance defined as the
significance of the distance between the primary vertex and displaced vertex,
(d) the energy fraction, defined as the energy of the tracks in the displaced
vertex relative to the energy of all tracks reconstructed within the jet, (e)
the invariant mass and (f) the number of tracks associated with the vertex. The
increased rate of light-flavour jets at high transverse decay length values is
due to residual interactions with detector material.
The jumps in the frequency of reconstructed two-track vertices (a) originates
from combinatorics.
Expecting \((N\cdot(N-1))/2\) possible track pairs created from a set of \(N\)
tracks, this number is reduced due to track selection
criteria, resulting in low-side tails on each spike. The first (last) bin in the
distribution does not account for underflow (overflow).}
\label{fig:SV1var}
\end{figure}

The secondary-vertex-tagging algorithm,
\svone~\cite{ATL-PHYS-PUB-2017-011}, reconstructs a single displaced
secondary vertex in a jet.
The reconstruction starts by identifying the possible two-track vertices built
with all tracks associated with the jet, while rejecting any tracks making two-track
vertices compatible with \Kshort or $\Lambda$ decays,
photon conversions or hadronic interactions with the detector material.
The \svone algorithm runs iteratively on all tracks contributing to the
selected two-track vertices, trying to fit one secondary vertex.
In each iteration, the track-to-vertex matching is evaluated using a $\chi^2$
test.
The track with the largest $\chi^2$ is removed and the vertex fit is repeated
until an acceptable vertex $\chi^2$ and a vertex invariant mass less than 6~\GeV\
are obtained.
With this approach, the decay products from $b$- and $c$-hadrons are typically assigned to
a single common secondary vertex.
The \svone algorithm also benefits from several
improvements~\cite{ATL-PHYS-PUB-2017-011} introduced during \RunTwo, and resulting
in increased pile-up rejection and an overall enhancement of the
performance at high jet \pt.
Among the various algorithm improvements, additional track-cleaning requirements based on silicon detector hit multiplicity
are applied to jets in the high-pseudorapidity region ($|\eta |\ge1.5$) in order to improve the quality of the selected tracks.
The fake-vertex rate is also better controlled by limiting the algorithm to only
consider the 25 highest-\pt tracks in the jets.
Finally, eight discriminating variables, including the jet \pt and $\eta$, the
number of tracks associated with the \svone vertex, the invariant mass of the
secondary vertex, its energy fraction (defined as the total energy of all the
tracks associated with the secondary vertex divided by the energy of all the
tracks associated with the jet), and the three-dimensional decay length
significance, are used as inputs to the high-level taggers.
Six of these variables are illustrated in Figure~\ref{fig:SV1var}.
 
In order to quantify the flavour-tagging performance of the \svone algorithm, a
simple feed-forward neural network was trained by exclusively using the outputs
of the algorithm and the \pt and $\eta$ of the input jets.
The network was trained in the same way as the main high-level tagger and on
identical training samples, as explained in Section~\ref{sec:highlevel}.
This new low-level tagger, defined only to illustrate the performance of SV1 relative to other
algorithms described in this paper, is referred to as \svkine.

\subsection{Topological multi-vertex finding algorithm}
\label{sec:JetFitter}
 
\begin{figure}[htbp]
\centering
\subfigure[]{\includegraphics[width=0.41\textwidth]{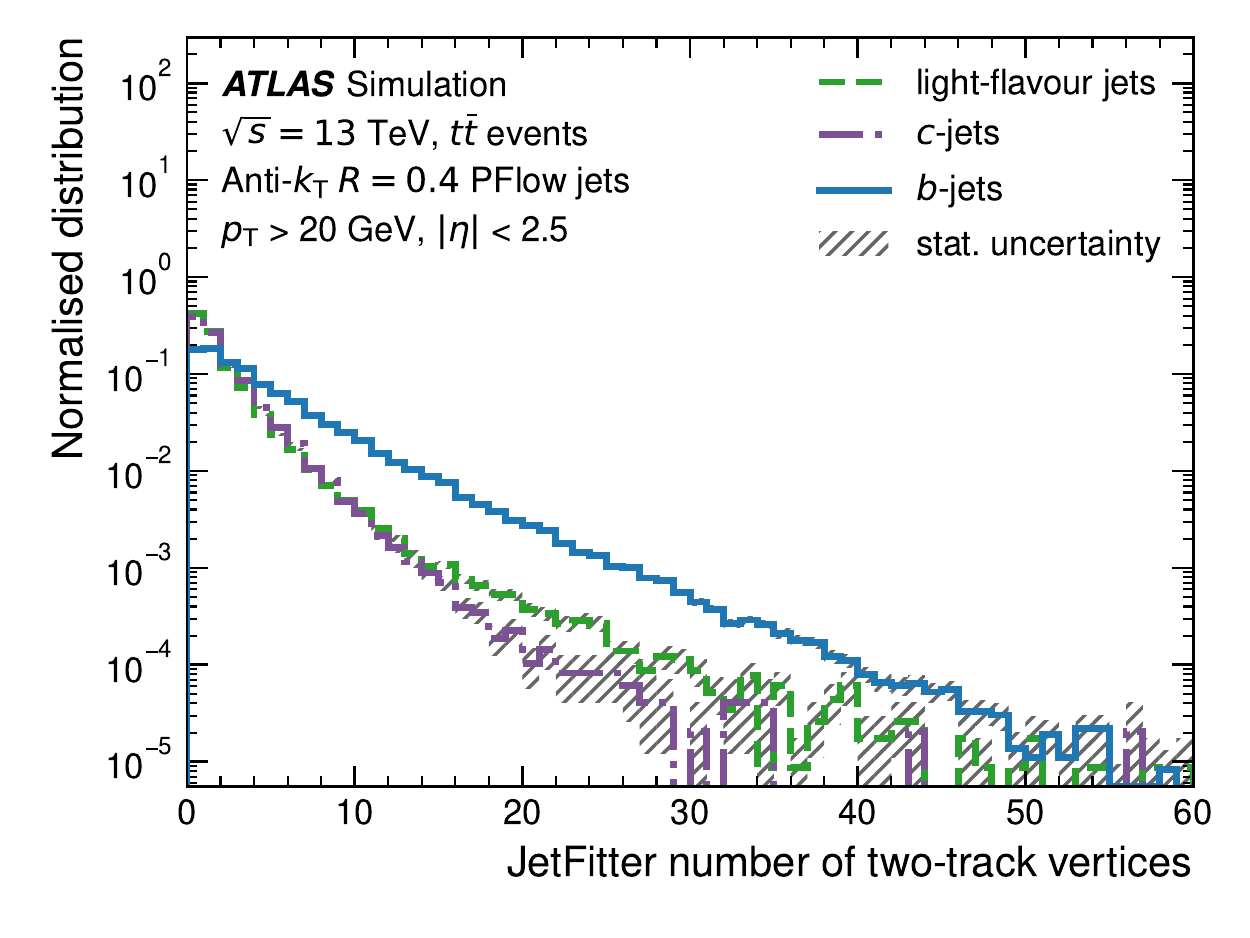}}
\subfigure[]{\includegraphics[width=0.41\textwidth]{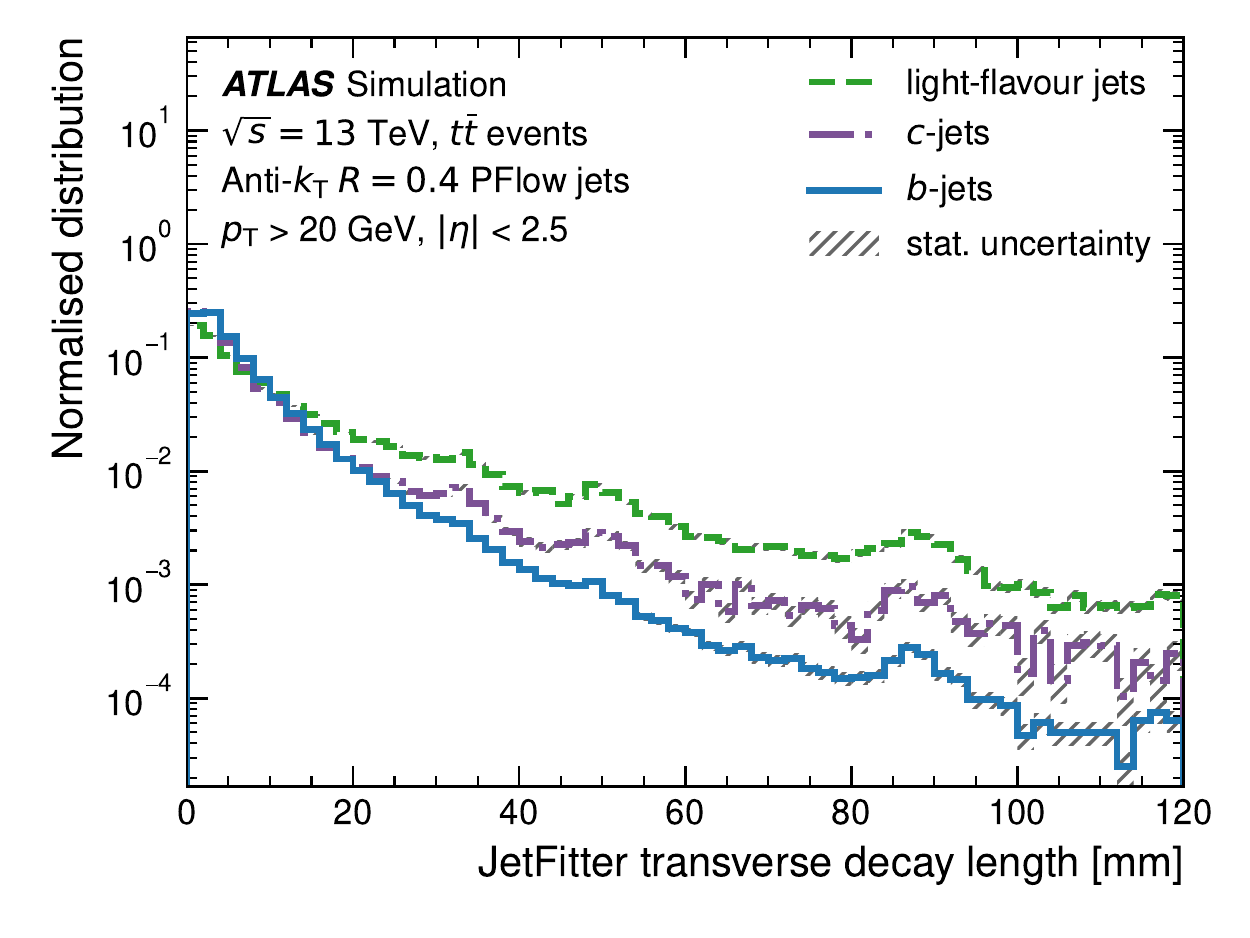}}
\subfigure[]{\includegraphics[width=0.41\textwidth]{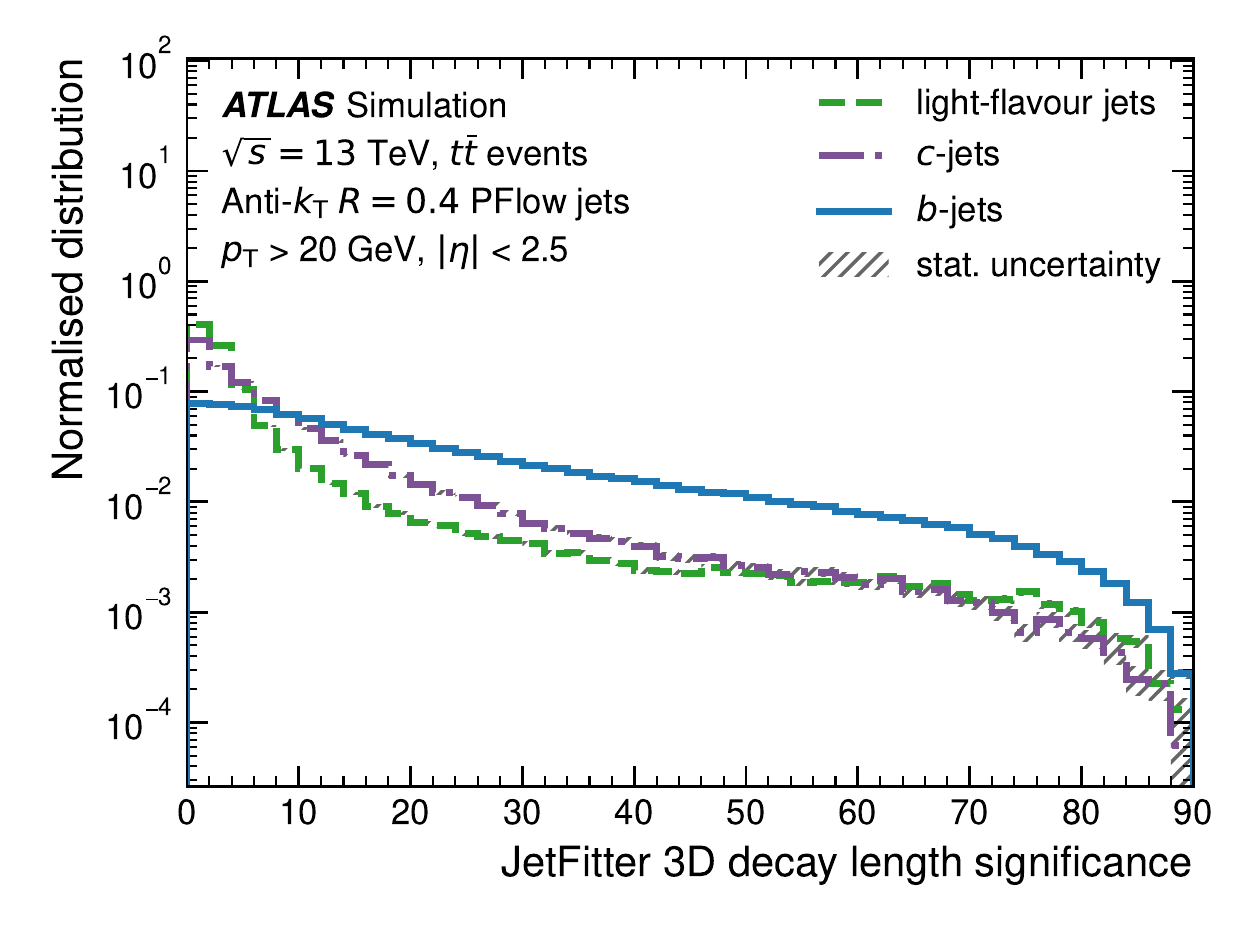}}
\subfigure[]{\includegraphics[width=0.41\textwidth]{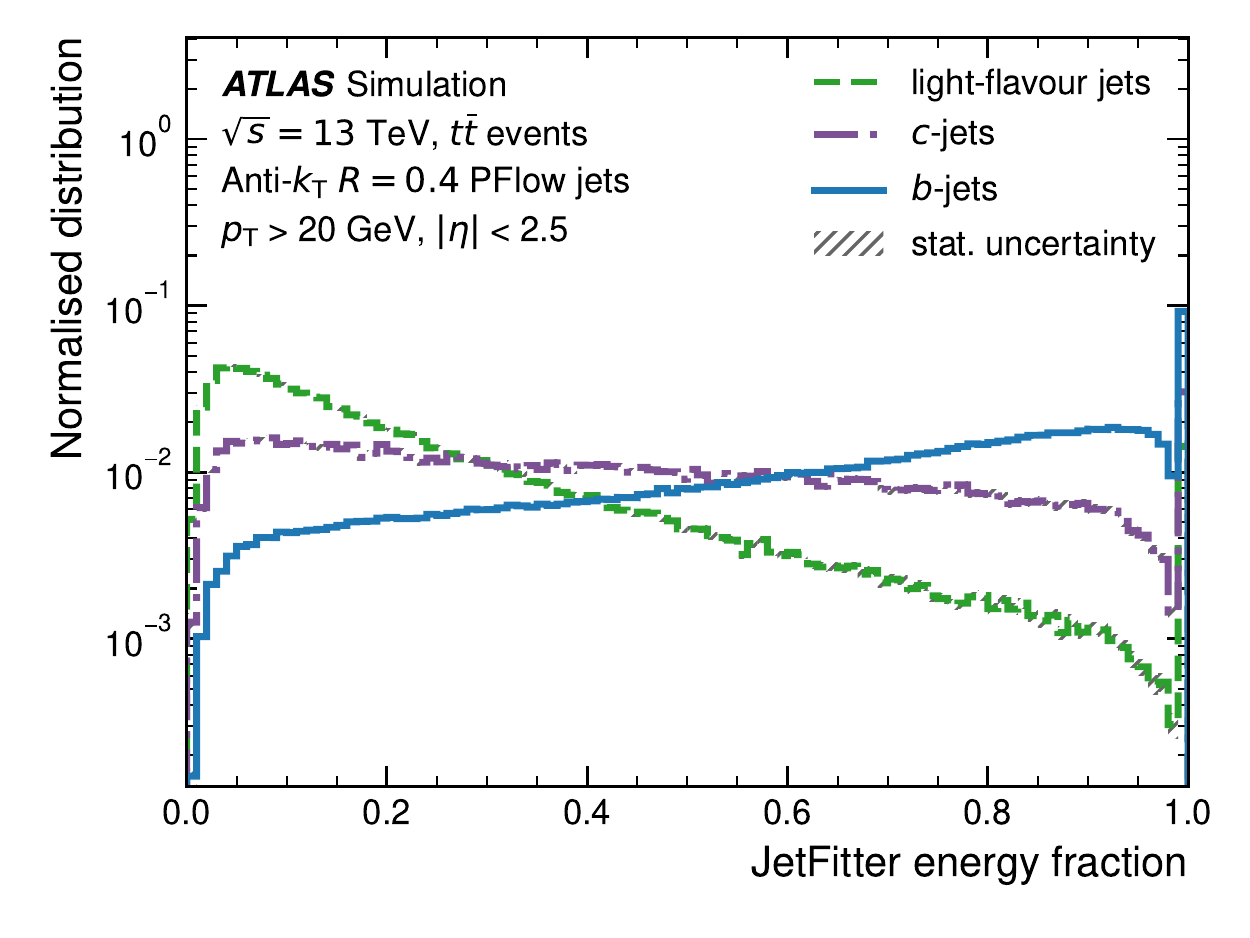}}
\subfigure[]{\includegraphics[width=0.41\textwidth]{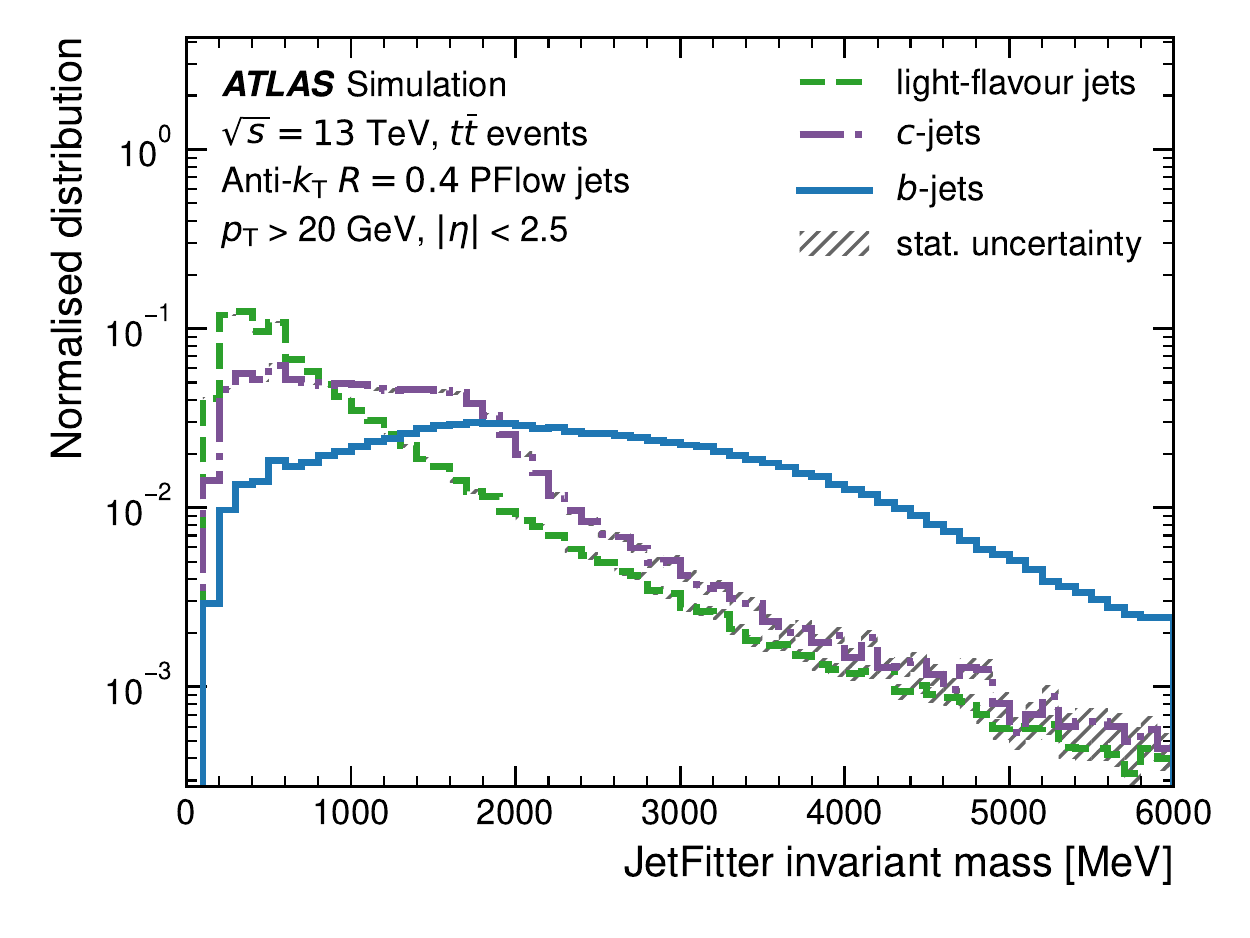}}
\subfigure[]{\includegraphics[width=0.41\textwidth]{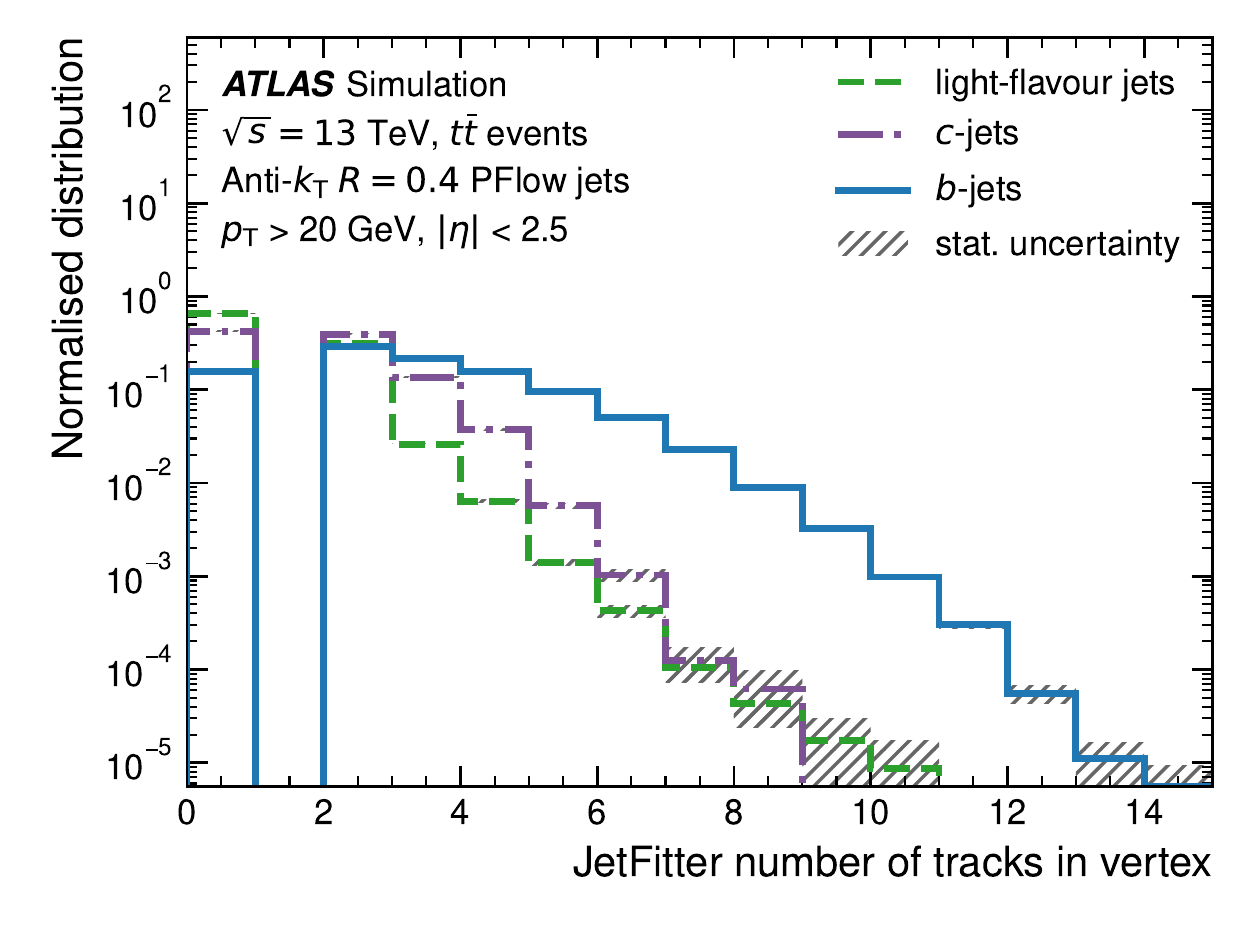}}
\caption{
Properties of secondary vertices reconstructed by the \jetfitter algorithm for
$b$-jets, $c$-jets and light-flavour jets in the baseline \ttbar\ simulated
events: (a) the number of two-track vertices reconstructed within the jet, (b)
the transverse decay length, (c) the average 3D decay length significance,
defined as the significance of the average distance between the primary vertex
and displaced vertices, (d) the energy fraction, defined as the energy of the
tracks in the displaced vertex relative to the energy of all tracks
reconstructed within the jet, (e) the invariant mass and (f) the number of
tracks associated with the vertex of at least two tracks. The increased rate of light-flavour jets at
high transverse decay length values is due to residual interactions with
detector material. Expecting jumps in the frequency of reconstructed two-track
vertices originating from combinatorics as in the case of \svone, the
\((N\cdot(N-1))/2\) possible track pairs created from a set of \(N\) tracks is
reduced due to track selection criteria, resulting in a smooth distribution. The
first (last) bin in the distribution does not account for underflow (overflow).
}
\label{fig:JetFittervar}
\end{figure}

The topological multi-vertex finding algorithm,
\jetfitter~\cite{ATL-PHYS-PUB-2018-025}, exploits the topological
structure of weak $b$- and $c$-hadron decays inside the jet and tries to
reconstruct the full $b$-hadron decay chain.
A modified Kalman filter~\cite{Fruhwirth:1987fm} is used to find a common line
on which the primary, $b$- and $c$-hadron decay vertices lie, approximating the
$b$-hadron flight path as well as the vertex positions.
With this approach, it is possible to resolve the $b$- or $c$-hadron decay vertex
even if only a single track is attached to it.
The \jetfitter algorithm also benefits from several
improvements~\cite{ATL-PHYS-PUB-2018-025} introduced during \RunTwo.
These include a reoptimisation of the track selection to better mitigate the
effect of pile-up tracks, an improvement in the rejection of interactions with
detector material, and the introduction of a vertex-mass-dependent selection during
the decay chain fit to increase the efficiency for tertiary vertex
reconstruction.
Eight discriminating variables, including the track multiplicity at the
\jetfitter displaced vertices, the invariant mass of tracks associated
with these vertices, their energy fraction and their average three-dimensional
decay length significance, are used as inputs to the high-level taggers.
Six of these variables are illustrated in Figure~\ref{fig:JetFittervar}.
Finally, the discrimination of $c$-jets from $b$-jets and light-flavour jets is further
improved by more specifically exploiting jets for which only a single secondary vertex is reconstructed
with intermediate charged decay multiplicity and comparable decay distance to $b$-hadrons in jets.
A set of nine additional variables~\cite{ATL-PHYS-PUB-2017-013}, among which
the number, the invariant mass and the energy of the tracks associated with
the secondary vertex as well as their rapidity, computed with respect to the
jet axis and the vector defined between the primary and secondary vertices,
are used as inputs to the high-level \btag algorithms.
 
Similarly to the \svone algorithm, the flavour-tagging performance of \jetfitter
is assessed from a simple feed-forward neural network trained by exclusively using
the outputs of \jetfitter and the input jets' \pt and $\eta$. This training was performed
in the same way as for the main high-level tagger and on identical training samples, as explained
in Section~\ref{sec:highlevel}. This new low-level tagger, defined only to illustrate the
performance of \jetfitter relative to other algorithms described in this paper, is referred
to as \jfkine.


\section{High-level flavour-taggers, the \dlone series}
\label{sec:highlevel}
 
To maximise the flavour-tagging performance for \RunTwo, the output quantities of the
low-level algorithms are combined using deep-learning classifiers, based on fully connected multi-layer
feed-forward neural networks (NN)~\cite{WITTEN2017417}, forming the so-called \dlone algorithm series.
 
These algorithms are trained with a hybrid training sample, for which 70\% of the jets
in the sample are from $t\bar{t}$ events and the remaining 30\% are from
$Z'\rightarrow q\bar{q}$ events, using  {\textsc TensorFlow}~\cite{chollet2015keras}
with the {\textsc Keras}~\cite{chollet2015keras} front-end
and the {\textsc Adam} optimiser~\cite{Kingma:2014aa}. The \dlone algorithm,
introduced at the beginning of \RunTwo in Ref.~\cite{FTAG-2018-01}, exploits as
input the \iptwod, \ipthreed, \svone and \jetfitter algorithm outputs, while the \dloner
algorithm also includes the jet \rnnip output probabilities.
 
The level of correlation between the different low-level algorithm outputs varies as a function
of the jet flavour and the kinematic range. In general, large correlations between the \iptwod,
\ipthreed, \svone and \jetfitter algorithms are observed for heavy-flavour jets. However, these correlations
are significantly reduced in the case of light-flavour jets. In addition, such correlations are further reduced
in high-$p_T$ regimes. On the other hand, the \rnnip algorithm contributes a set of input variables which are not strongly correlated.
The DL1r algorithm training exploits these correlation differences to reach the best tagging performance.

In addition, the kinematic properties of the jets, namely \pt\ and $|\eta|$, are included in the
training in order to take advantage of the correlations with the other input variables.
However, to avoid differences between the kinematic distributions of signal and background jets
being used to discriminate between the different jet flavours, the input training dataset is resampled.
The resampling procedure ensures that jets in the training sample are uniformly distributed in jet $\pT$
and $\eta$ for each flavour class.
No kinematic resampling is applied at the evaluation stage of the algorithms.
Table~\ref{tab:HLTaggerInputs} presents a detailed list of input variables used by each algorithm.
 
\newenvironment{DIFnomarkup}{}{}
 
\begin{DIFnomarkup}
\begin{table}[htbp]
\caption{Input variables used by the \svkine and \jfkine, \dlone and \dloner
algorithms.
}
\label{tab:HLTaggerInputs}
\begin{center}
\scriptsize
\begin{tabular}{|l|c|p{0.45\textwidth}| c | c | c | c |}
\hline
Input & Variable & Description & \svkine & \jfkine & \dlone & \dloner\\
\hline
\multirow{2}{*}{Kinematics}
& \pt & Jet \pt & \checkmark & \checkmark & \checkmark  & \checkmark\\
& $\eta$ & Jet $|\eta|$ & \checkmark & \checkmark & \checkmark  & \checkmark\\
\hline
\multirow{3}{*}{\iptwod, \ipthreed}
& $\log(P_{\mathrm{b}}/P_{\mathrm{light}})$  & Likelihood ratio of the $b$-jet to light-flavour jet hypotheses  & & & \checkmark & \checkmark  \\
& $\log(P_{\mathrm{b}}/P_{\mathrm{c}})$  & Likelihood ratio of the $b$-jet to $c$-jet hypotheses  & & & \checkmark & \checkmark\\
& $\log(P_{\mathrm{c}}/P_{\mathrm{light}})$  & Likelihood ratio of the $c$-jet to light-flavour jet hypotheses  & & & \checkmark & \checkmark\\
\hline
\multirow{3}{*}{\rnnip}
& $P_{\mathrm{b}}$  & $b$-jet probability  & & & & \checkmark \\
& $P_{\mathrm{c}}$  & $c$-jet probability  & & & & \checkmark \\
& $P_{\mathrm{light}}$ &  light-flavour jet probability & & & & \checkmark \\
\hline
\multirow{8}{*}{\svone}
& $m$(SV) & Invariant mass of tracks at the secondary vertex assuming pion mass & \checkmark & & \checkmark & \checkmark\\
& $f_E$(SV) & Jet energy fraction of the tracks associated with the secondary vertex  & \checkmark & & \checkmark & \checkmark\\
& $N_{{\mathrm{TrkAtVtx}}}$(SV) & Number of tracks used in the secondary vertex  & \checkmark & & \checkmark & \checkmark\\
& $N_{{\mathrm{2TrkVtx}}}$(SV) & Number of two-track vertex candidates  & \checkmark & & \checkmark & \checkmark\\
& $L_{xy}$(SV) & Transverse distance between the primary and secondary vertices  & \checkmark & & \checkmark & \checkmark\\
& $L_{xyz}$(SV)& Distance between the primary and secondary vertices   & \checkmark & & \checkmark & \checkmark \\
& $S_{xyz}$(SV)& Distance between the primary and secondary vertices divided by its uncertainty  & \checkmark & & \checkmark & \checkmark \\
& $\Delta R(\vec p_{\mathrm{jet}}, \vec p_{\mathrm{vtx}})$(SV) & $\Delta R$ between the jet axis and the direction of the secondary vertex relative to the primary vertex.  & \checkmark & & \checkmark & \checkmark\\
\hline
\multirow{8}{*}{\jetfitter}
& $m$(JF) & Invariant mass of tracks from displaced vertices  & & \checkmark & \checkmark & \checkmark\\
& $f_E$(JF) & Jet energy fraction of the tracks associated with the displaced vertices  & & \checkmark & \checkmark & \checkmark\\
& $\Delta R(\vec p_{\mathrm{jet}}, \vec p_{\mathrm{vtx}})$(JF) & $\Delta R$ between the jet axis and the vectorial sum of momenta of all tracks attached to displaced vertices  & & \checkmark & \checkmark & \checkmark\\
& $S_{xyz}$(JF) & Significance of the average distance between PV and displaced vertices  & & \checkmark & \checkmark & \checkmark\\
& $N_{{\mathrm{TrkAtVtx}}}$(JF) & Number of tracks from multi-prong displaced vertices  &  & \checkmark & \checkmark & \checkmark\\
& $N_{{\mathrm{2TrkVtx}}}$(JF) & Number of two-track vertex candidates (prior to decay chain fit)  &  & \checkmark & \checkmark & \checkmark\\
& $N_{{\mathrm{1\mbox{-}trk\ vertices}}}$(JF) & Number of single-prong displaced vertices &  & \checkmark & \checkmark & \checkmark\\
& $N_{{\geq \mathrm{2\mbox{-}trk\ vertices}}}$(JF) & Number of multi-prong displaced vertices  &  & \checkmark & \checkmark & \checkmark\\
& $L_{xyz}(2^{{\mathrm{nd}}})$(JF) & Distance of $2^{{\mathrm{nd}}}$ vertex from PV  & & \checkmark & \checkmark & \checkmark\\
& $L_{xy}(2^{{\mathrm{nd}}})$(JF) & Transverse displacement of the $2^{{\mathrm{nd}}}$ vertex  & & \checkmark & \checkmark & \checkmark\\
& $m_{\mathrm{Trk}}(2^{{\mathrm{nd}}})$(JF) & Invariant mass of tracks associated with the $2^{{\mathrm{nd}}}$ vertex  & & \checkmark & \checkmark & \checkmark\\
& $E (2^{{\mathrm{nd}}})$(JF) & Energy of the tracks associated with the $2^{{\mathrm{nd}}}$ vertex  & & \checkmark & \checkmark & \checkmark \\
& $f_E (2^{{\mathrm{nd}}})$(JF) & Jet energy fraction of the tracks associated with the $2^{{\mathrm{nd}}}$ vertex  & & \checkmark & \checkmark & \checkmark \\
& $N_{{\mathrm{TrkAtVtx}}}(2^{{\mathrm{nd}}}) $(JF) & Number of tracks associated with the $2^{{\mathrm{nd}}}$ vertex  & & \checkmark & \checkmark & \checkmark \\
& $ \eta_{\mathrm{trk}}^{\mathrm{min, max, avg}}(2^{{\mathrm{nd}}}) $(JF) & Min., max.\ and avg.\ pseudorapidity of tracks at the $2^{{\mathrm{nd}}}$ vertex  & & \checkmark & \checkmark & \checkmark \\ [0.1cm]
\hline
\end{tabular}
\end{center}
\end{table}
\end{DIFnomarkup}
 
The \dloner NN has a multidimensional output corresponding to the probabilities for a jet to be a
$b$-jet, a $c$-jet or a light-flavour jet.
The use of a multi-class network architecture provides the algorithm with a
smaller memory footprint than the previous ATLAS \mvtwo
algorithm~\cite{FTAG-2018-01} based on boosted decision trees (BDTs).
The topology of the network consists of a mixture of fully connected hidden layers.
The \dloner algorithm parameters, listed in Table~\ref{tab:DL1rHyperPar},
include the architecture of the NN, the number of training epochs, the learning
rates and training batch size.
Each of these parameters is optimised in order to maximise the \btag\ performance.
Batch normalisation~\cite{Ioffe:2015aa} is added by default since it is found to
improve the performance.
 
\begin{table}[htbp]
\caption{List of optimised hyperparameters used in the \dloner flavour-tagging algorithms.}
\label{tab:DL1rHyperPar}
\begin{center}
\small
\begin{tabular}{|l|c|}
\hline
Hyperparameter & Value \\
\hline
Number of input variables  & 31 \\
Number of hidden layers & 8 \\
Number of nodes [per layer] & [256, 128, 60, 48, 36, 24, 12, 6] \\
Learning rate & 0.01 \\
Training batch size & 15\,000 \\
Activation function & ReLu \\
Number of training epochs &  200\\
\hline
Free (trainable) parameters & 59\,275\\
Fixed parameters & ~~1\,140\\
Training sample size	& 22\,M jets\\
\hline
\end{tabular}
\end{center}
\end{table}
 
Training with multiple output nodes offers additional flexibility when
constructing the final output discriminant by combining the $b$-jet, $c$-jet and
light-flavour jet probabilities.
Since all flavours are treated equally during training, the trained network can
be used for both $b$-jet and $c$-jet tagging.
The final \dloner \btag discriminant is defined as:
\begin{equation*}
\Ddloner = \ln \left( \frac{\pb}{f_{c} \cdot \pc + (1-f_{c}) \cdot \pu}\right),
\end{equation*}
where $\pb$, $\pc$ and $\pu$ represent respectively
the $b$-jet, $c$-jet and light-flavour jet probabilities, and $f_{c}$ denotes
the effective $c$-jet fraction in the background hypothesis.
Using this approach, the $c$-jet fraction in the background can be chosen a
posteriori in order to optimise the performance of the algorithm at physics analysis level.
In this paper, an optimised $c$-jet fraction of 0.018 is used to evaluate the performance of
the \dloner \btag algorithm in simulated $t\bar{t}$ and $Z'\rightarrow q\bar{q}$ events.
This value is chosen as a compromise to ensure good rejection factors for both
$c$-jet and light-flavour jets in a large \btag efficiency range across a number
of analyses.
In particular, the ATLAS measurements of $VH,\, H\to b \bar b$
production~\cite{HIGG-2018-50} and $t\bar t H,\, H\to b\bar b$
production~\cite{HIGG-2020-23} were considered in this optimisation.
 
\begin{figure}[htbp]
\centering
\subfigure[]{\includegraphics[width=0.45\textwidth]{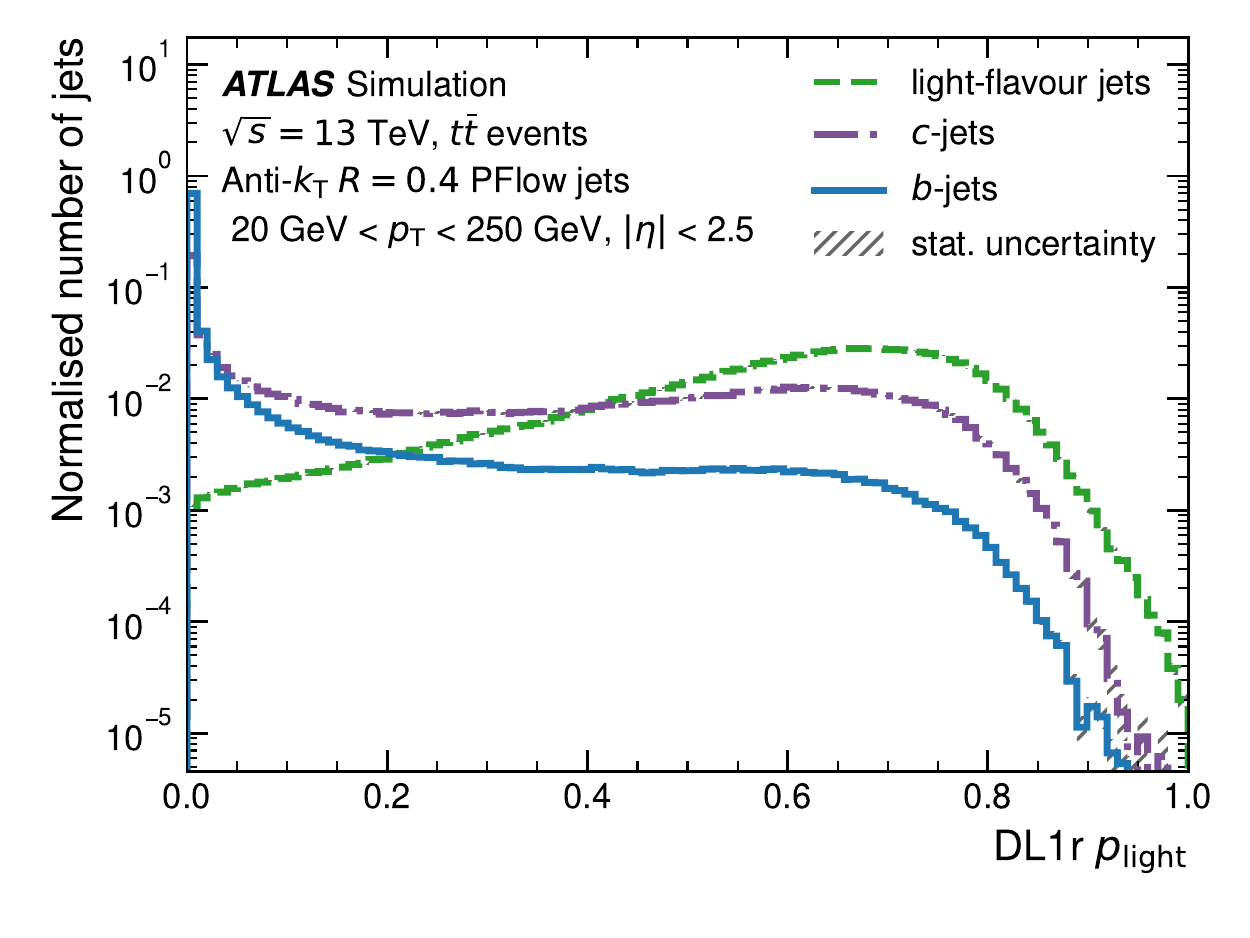}} \\
\subfigure[]{\includegraphics[width=0.45\textwidth]{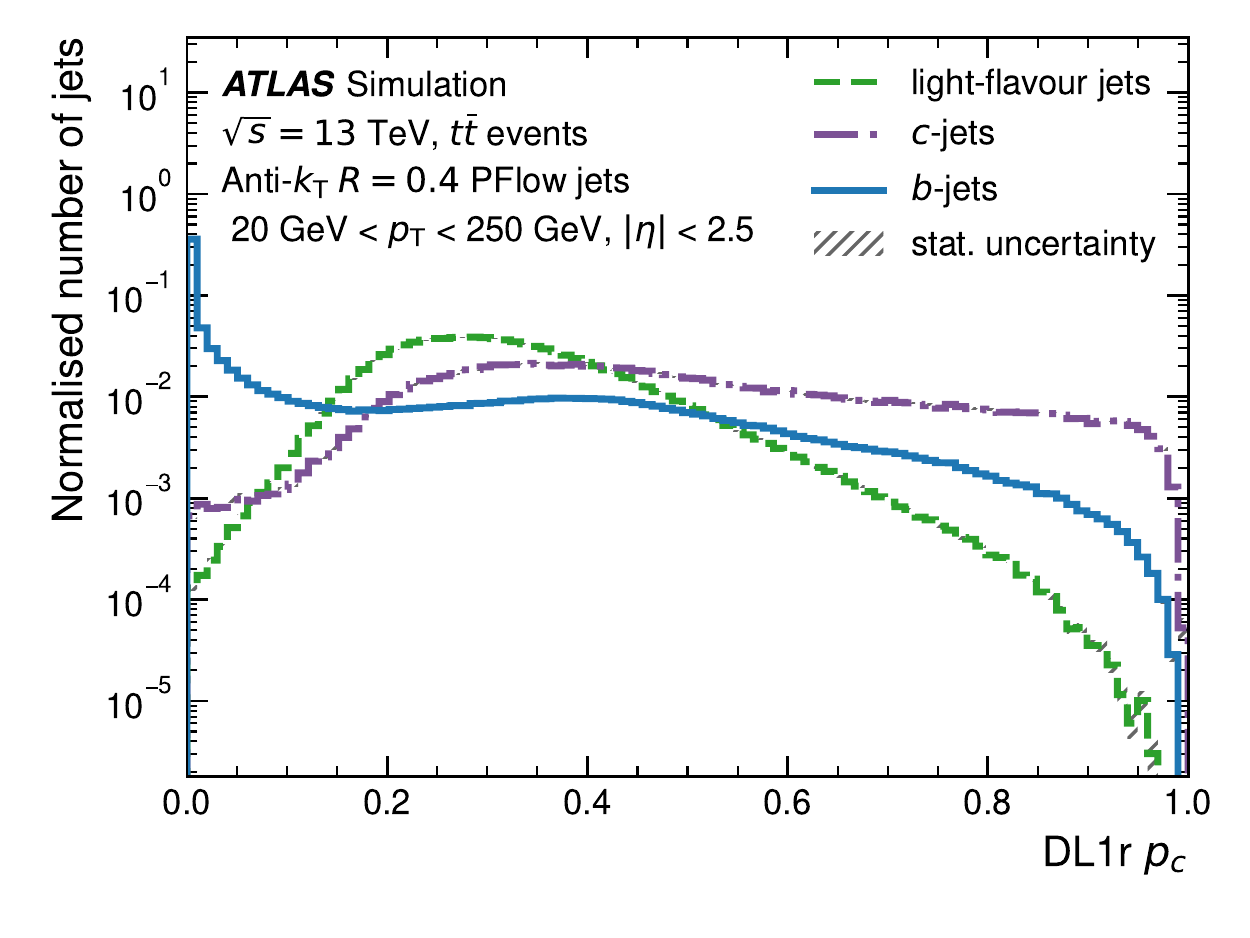}}
\subfigure[]{\includegraphics[width=0.45\textwidth]{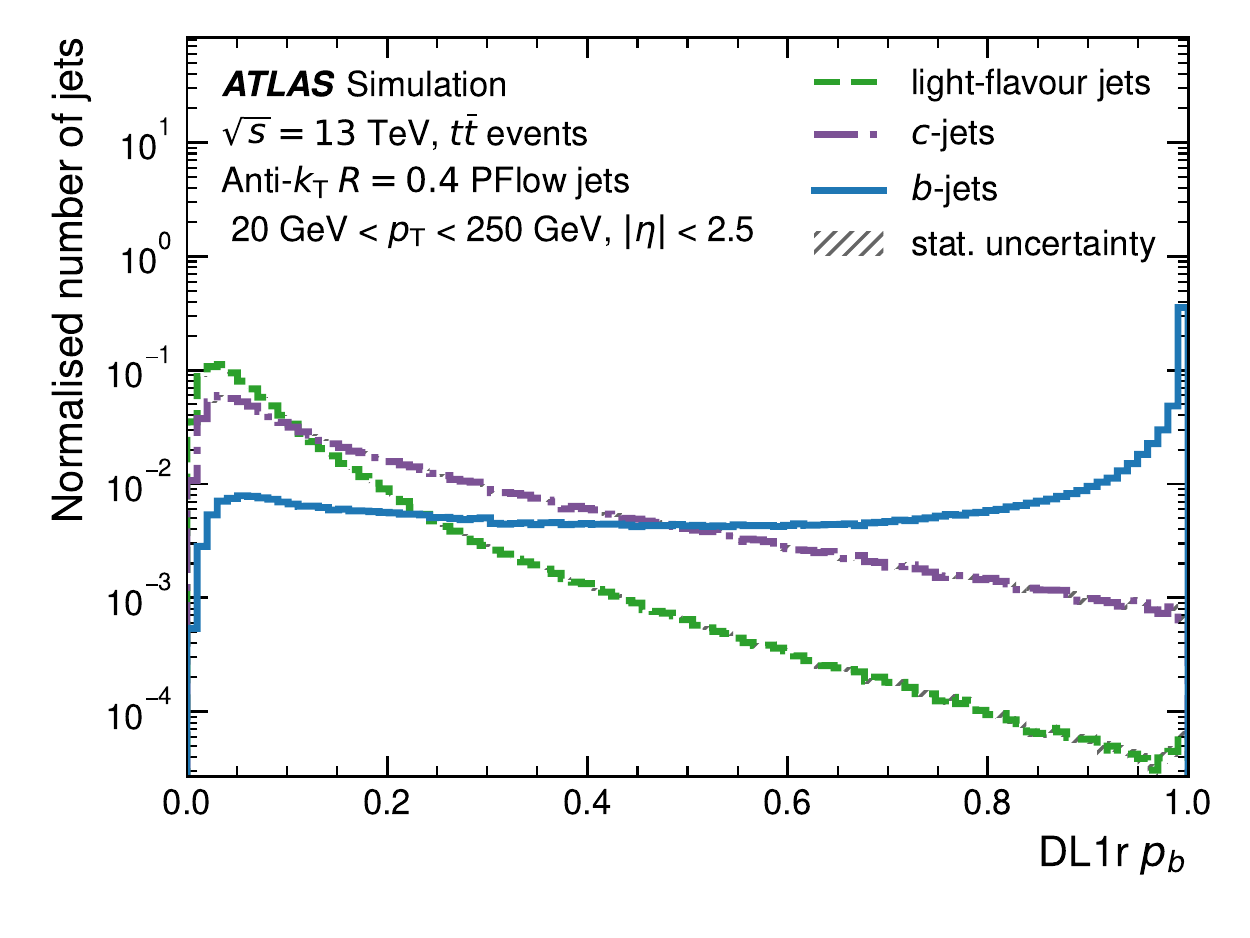}} \\
\subfigure[]{\includegraphics[width=0.45\textwidth]{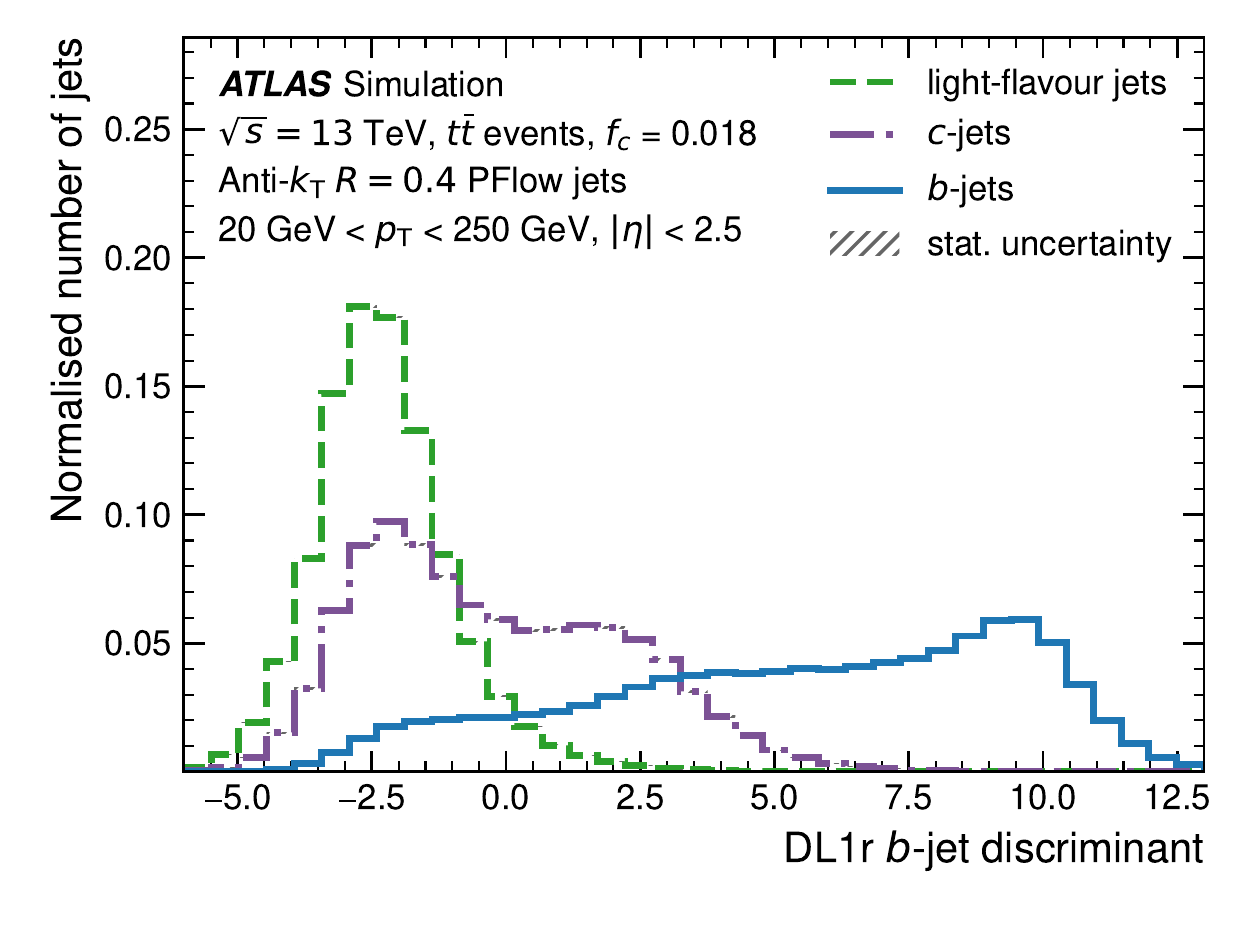}}
\subfigure[]{\includegraphics[width=0.45\textwidth]{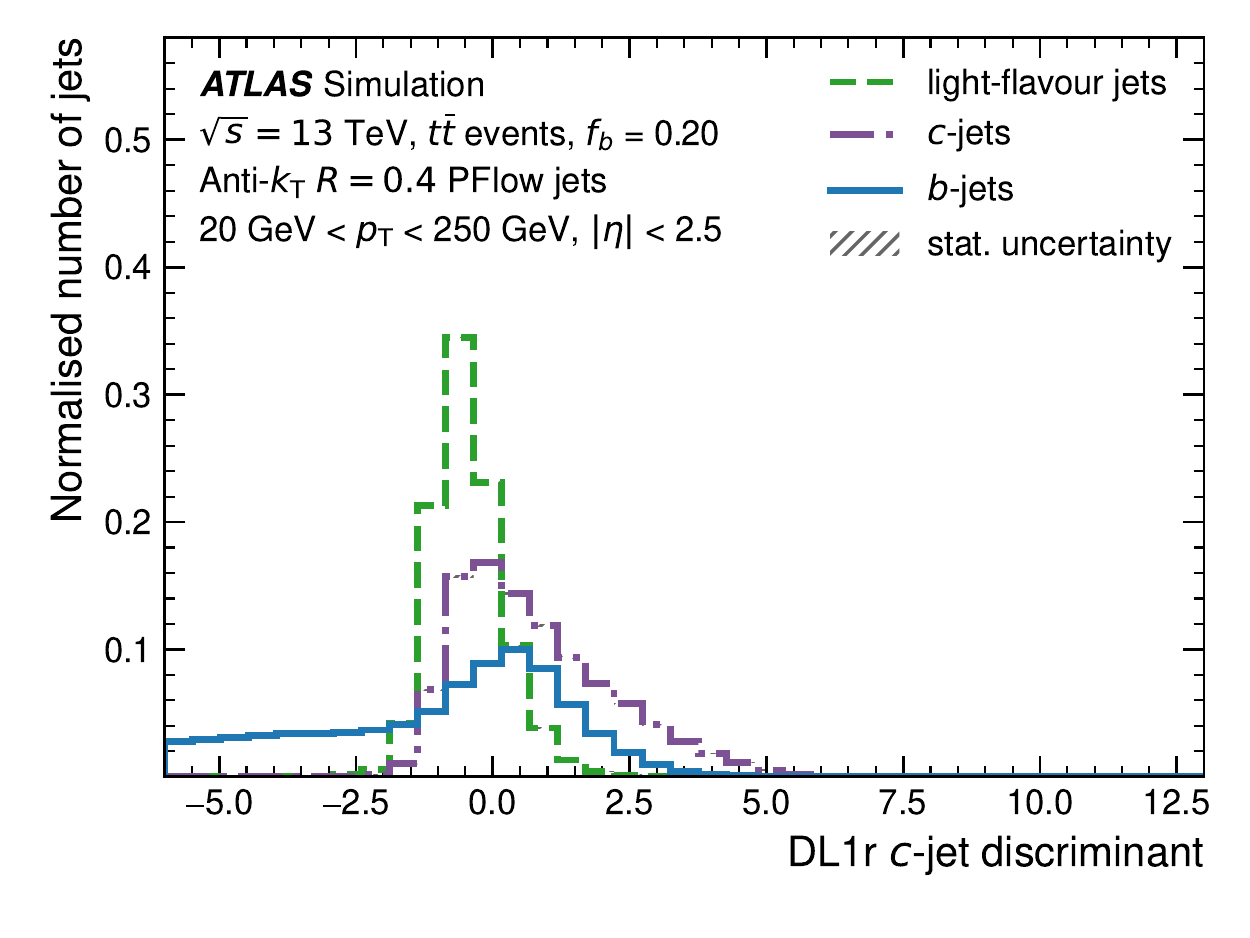}}
\caption{
Distributions of the outputs of the \dloner \btag algorithm for $b$-jets,
$c$-jets and light-flavour jets in \ttbar\ simulated events: (a) \pu, (b)
\pc, (c) \pb, (d) \Ddloner, and (e) \Ddlonerc.
}
\label{fig:DL1r}
\end{figure}

Similarly, the \dloner $c$-tagging discriminant is defined as:
\begin{equation}
\Ddlonerc = \ln \left( \frac{\pc}{f_{b} \cdot \pb + (1-f_{b}) \cdot \pu}\right),
\label{eq:dl1rc}
\end{equation}
where $f_{b}$ represents the effective $b$-jet fraction in the background
training sample.
A $b$-jet fraction of 0.2 is used to evaluate the performance of the \dloner
$c$-tagging algorithm in this paper in simulated $t\bar{t}$ and $Z'\rightarrow q\bar{q}$ events.
Larger than the $f_{c}$ fraction presented above,
the $f_{b}$ fraction is chosen here to maximise the $b$-jet rejection factor at
the given $c$-tagging efficiency rates of 20\% and 30\%.
The output probabilities of the \dloner $b$-tagging algorithms for
$b$-jets, $c$-jets and light-flavour jets in the baseline \ttbar\ simulated
events are shown in Figures~\ref{fig:DL1r}(a)--(c); the corresponding \btag and
$c$-tagging discriminants are also shown in Figures~\ref{fig:DL1r}(d) and~\ref{fig:DL1r}(e).
 
\section{Flavour-tagging performance}
\label{sec:performance}
 
The performance of a flavour-tagging algorithm is characterised by the
probability or efficiency of correctly tagging a signal jet, $\varepsilon$, and
the probability of mistakenly identifying a background jet, referred to as the
mis-tag rate.
In this paper, the performance of the algorithms is quantified in terms of
background-jet rejection factors, defined as $1 / \varepsilon$ for background jets.

\subsection{$b$-tagging performance}
 
When analysis of LHC data requires the identification of $b$-jets, the
tagging efficiency of a given requirement on the $b$-tagging discriminant is
denoted by \beff, and the charm-jet and light-flavour jet rejection factors are
$1/\ceff$ and $1/\leff$, respectively.
Many ATLAS analyses of \RunTwo LHC data use a requirement on the \dloner
discriminant, or bins built from several such requirements, that do not vary
with jet kinematics.
These are referred to as \enquote{fixed-cut operating points}, and they are
labelled according to their inclusive efficiency for the population of $b$-jets
present in the \ttbar sample used to train \dloner; for example, the \dloner
discriminant value for which 77\% of the $b$-jets in a \ttbar sample have a
higher score is called the \enquote{77\% operating point}.
 
\begin{figure}[htbp]
\centering
\includegraphics[width=0.65\textwidth]{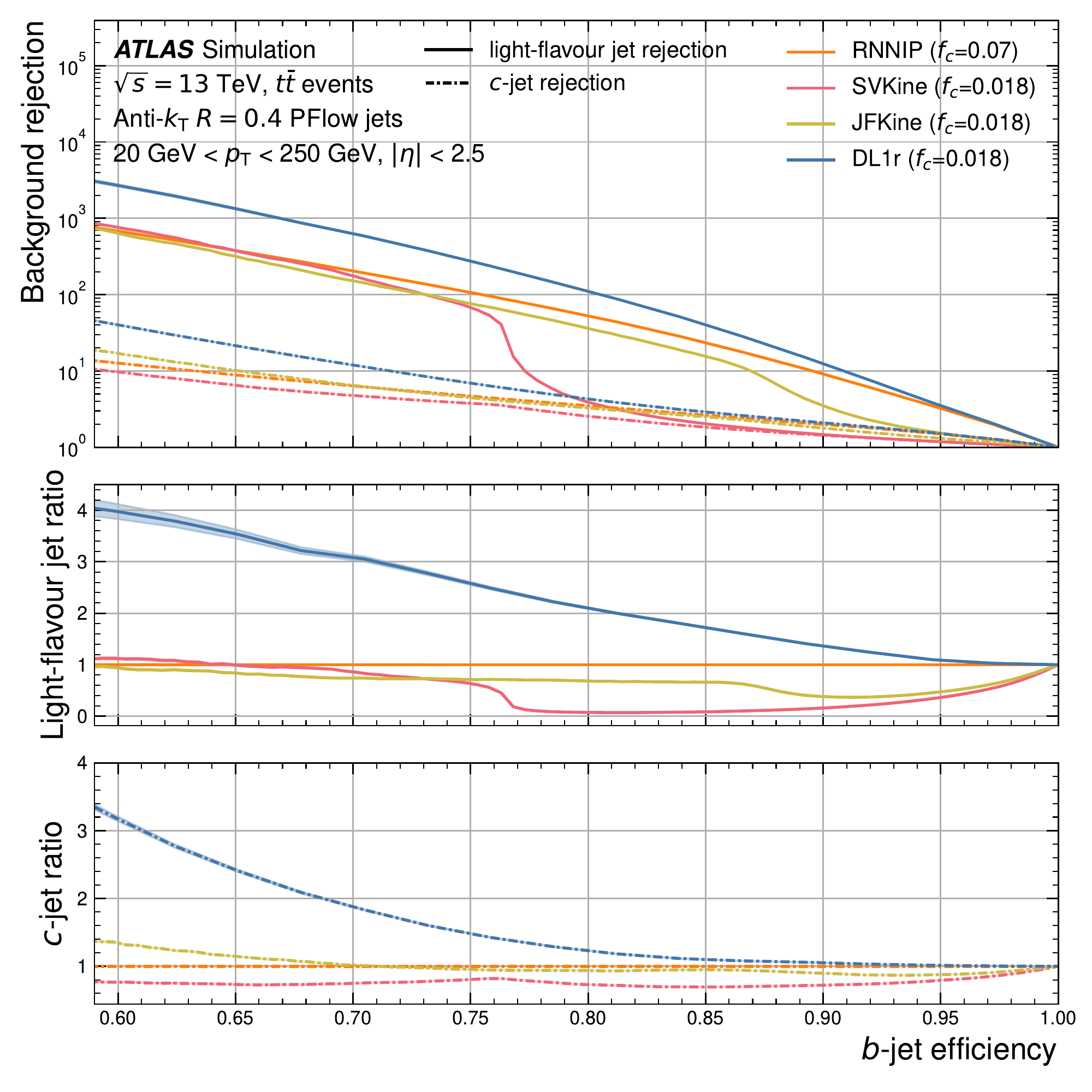}
\caption{
The light-flavour jet and $c$-jet rejection factors as a function of \beff for
\dloner as well as the low-level $b$-taggers \rnnip, \svkine , and \jfkine.
The statistical uncertainties of the rejection factors are calculated using
binomial uncertainties and are indicated as coloured bands.
The lower two panels show the ratio of the light-flavour jet rejection and the
$c$-jet rejection of the algorithms to \rnnip.
\svone and \jetfitter have secondary-vertex-finding efficiencies of
approximately 80\% and 90\%, respectively; this causes the rapid growth
in light-flavour jet rejection around these values of \beff for \svkine and \jfkine.
}
\label{fig:lowlevel-ROCs}
\end{figure}

\begin{figure}[htbp]
\centering
\includegraphics[width=0.65\textwidth]{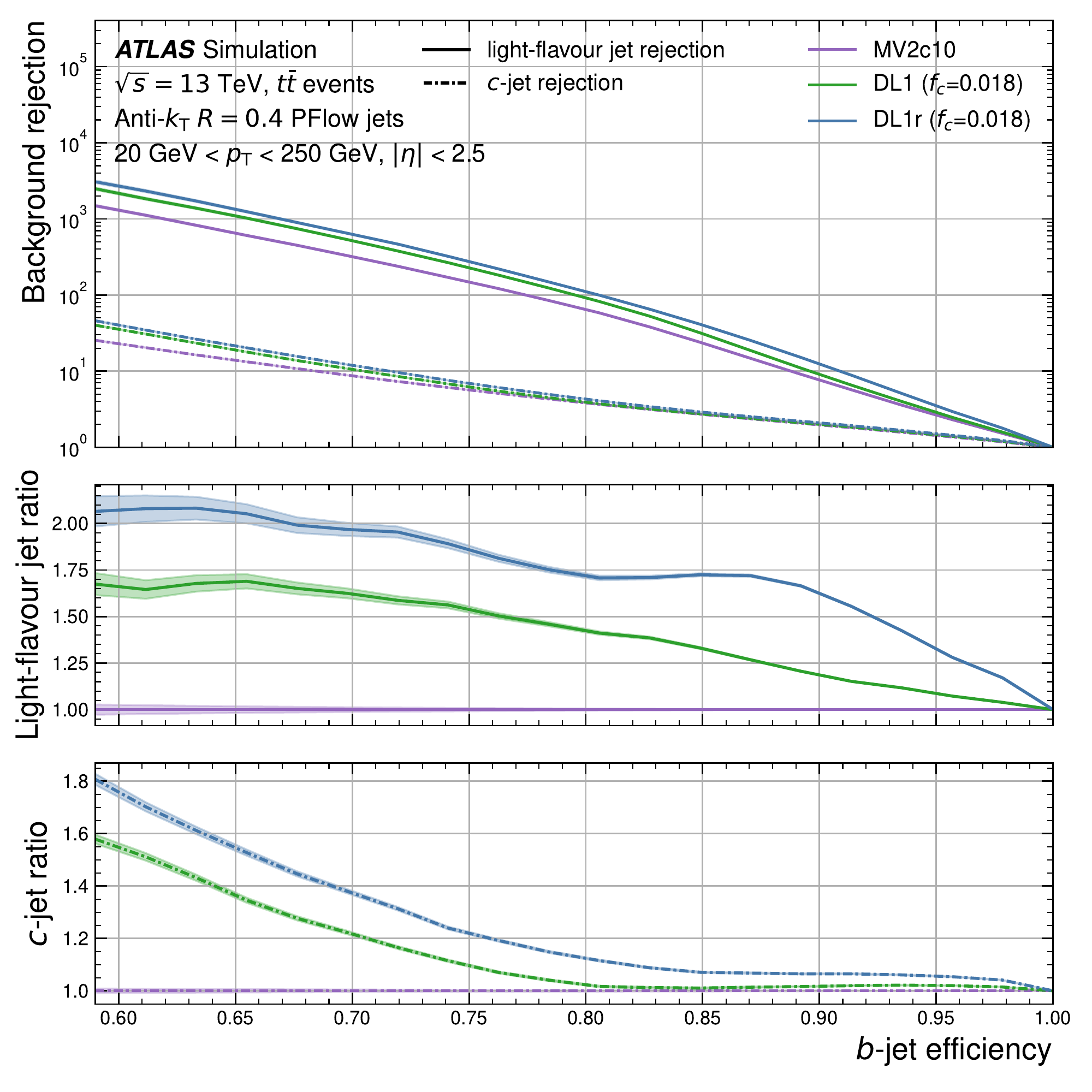}
\caption{The light-flavour jet and $c$-jet rejection factors as a function of
\beff for the high-level $b$-taggers \mvtwo, \dlone, and \dloner.
The lower two panels show the ratio of the light-flavour jet rejection and the
\(c\)-jet rejection of the algorithms to \mvtwo.
The statistical uncertainties of the rejection factors are calculated using binomial
uncertainties and are indicated as coloured bands.
\label{fig:highlevel-ROCs}
}
\end{figure}

Figures~\ref{fig:lowlevel-ROCs} and~\ref{fig:highlevel-ROCs} show the
light-flavour jet and $c$-jet rejection factors as a function of \beff for a
variety of low- and high-level $b$-taggers.
At high-efficiency operating points, \rnnip provides the best rejection among
the low-level taggers, while at low efficiency the secondary-vertex finders
\svkine and \jfkine achieve the highest background rejection.
\dloner substantially outperforms all low-level taggers across the \beff range:
the low-level algorithms exploit different jet properties, and combining these
produces large gains.
\dloner also exceeds the performance of the BDT-based \mvtwo tagger and the
\dlone tagger~\cite{FTAG-2018-01}.
 
\begin{figure}[htbp]
\centering
\subfigure[]{\includegraphics[width=0.45\textwidth]{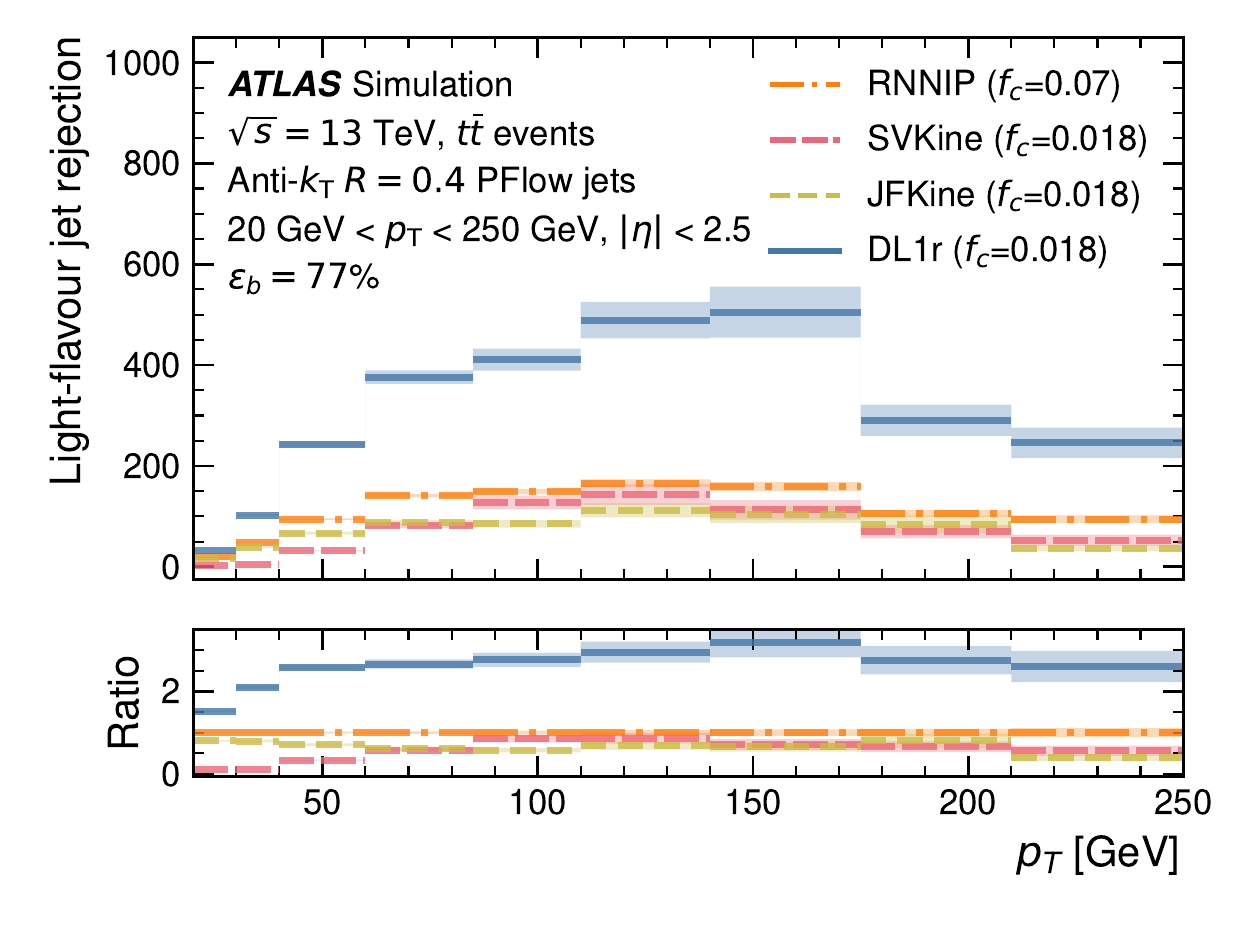}}
\subfigure[]{\includegraphics[width=0.45\textwidth]{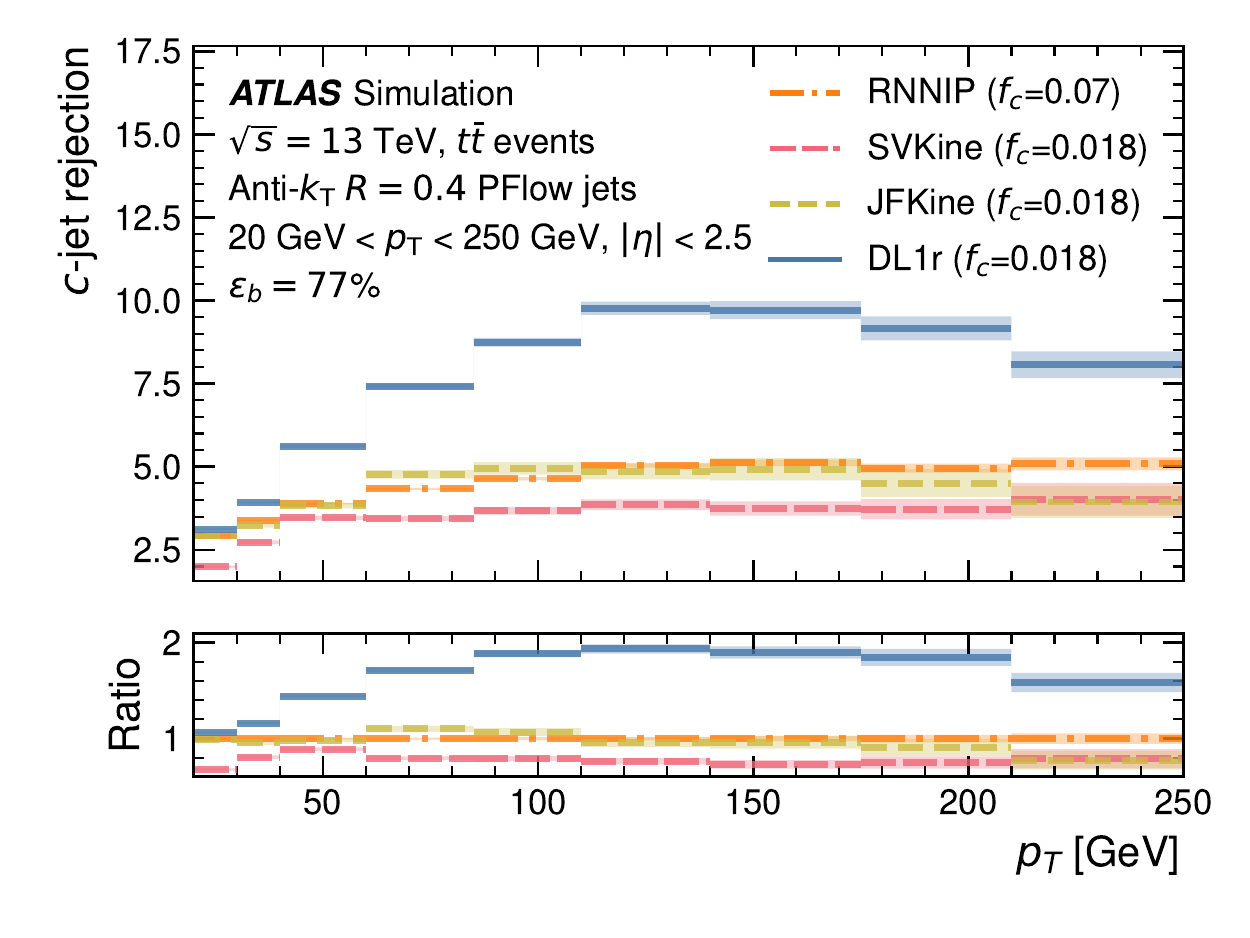}} \\
\subfigure[]{\includegraphics[width=0.45\textwidth]{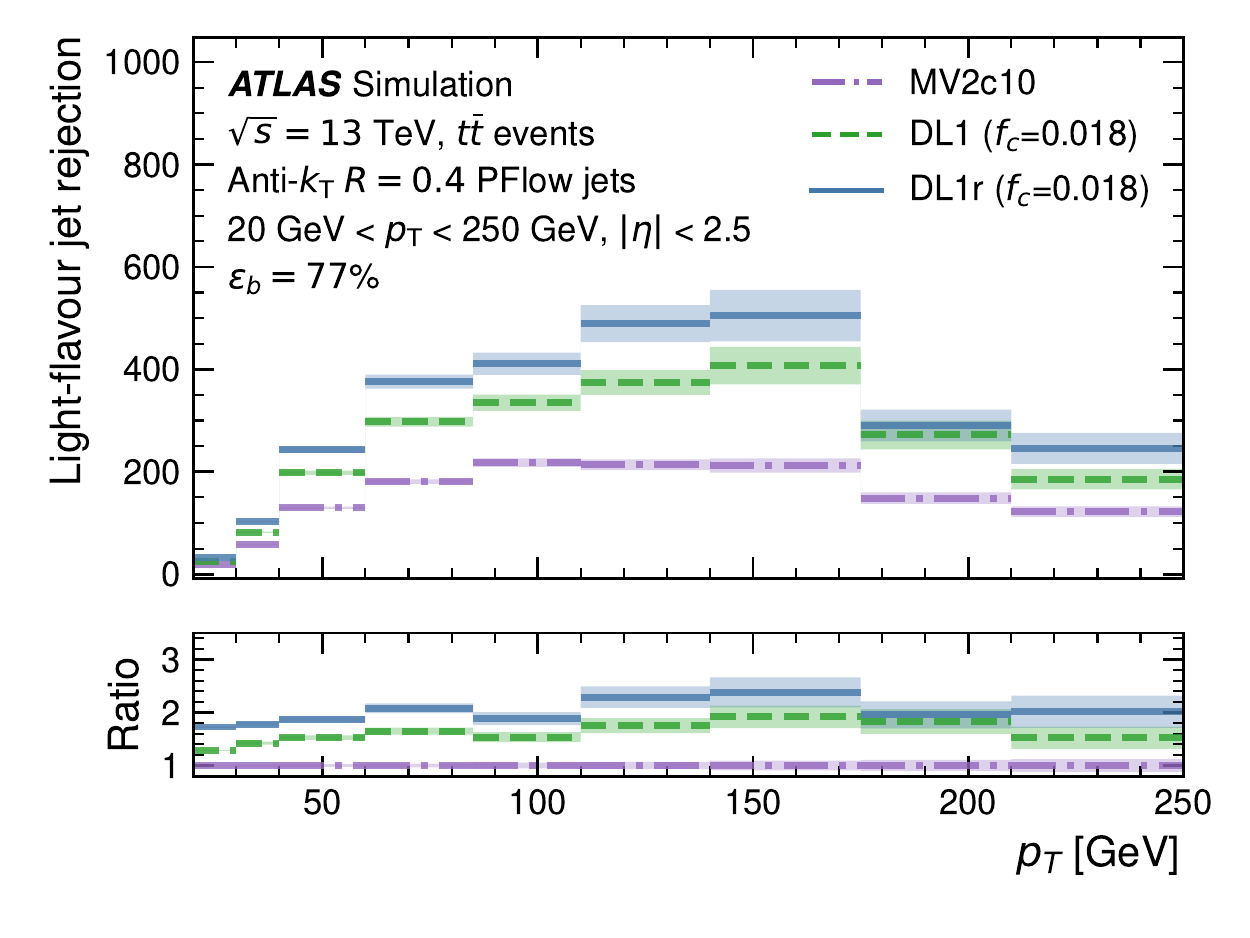}}
\subfigure[]{\includegraphics[width=0.45\textwidth]{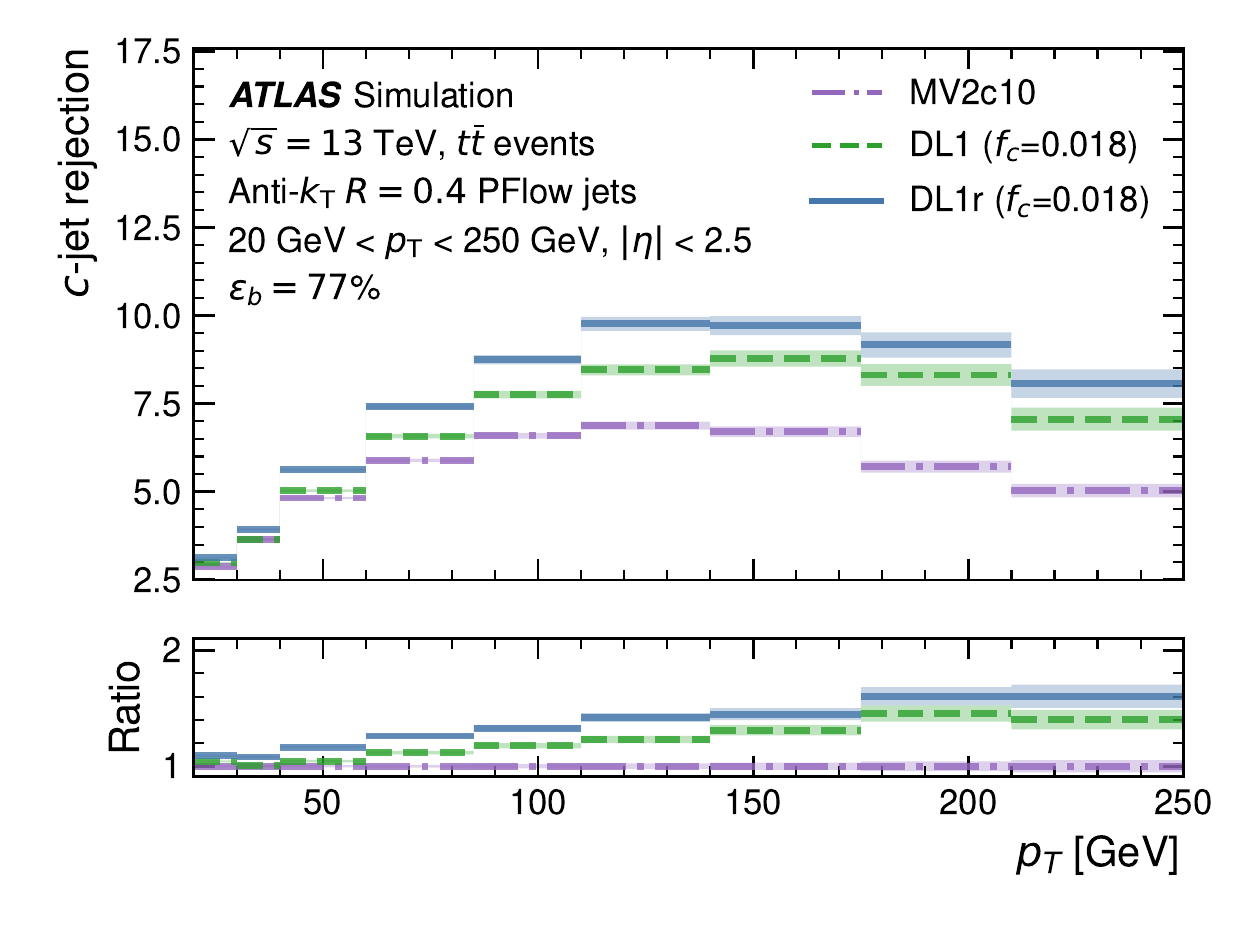}}
\caption{The light-flavour jet and $c$-jet rejection factors at a fixed $b$-tagging efficiency
\beff as a function of jet \pt.
In each bin, the $b$-tagging efficiency is set to \SI{77}{\percent}, and the
resulting background rejection is shown.
(a) shows the light-flavour jet rejection for several low-level tagging algorithms and
\dloner,
(b) shows the $c$-jet rejection for the same taggers.
(c) and (d) show light-flavour jet and $c$-jet rejection, respectively, for the \dloner, \dlone, and
\mvtwo high-level taggers.
The lower panels of (a) and (b) show the ratio of each algorithm's background
rejection to that of \rnnip ; the lower panels of (c) and (d) show the
corresponding ratios to \mvtwo .
The statistical uncertainties of the efficiencies and rejection factors are calculated
using binomial uncertainties and are indicated as coloured bands.
The drop in light-flavour jet rejection at $\pt \sim 175~\GeV \sim m_{\text{top}}$ is due to
the overlap of decay products from boosted top-quark decays.
\label{fig:pt-dependence}
}
\end{figure}

It is also important to gauge the $b$-tagging performance across a broad \pt
range, because the ATLAS physics programme relies on excellent background rejection in
a variety of situations, depending on the needs of a given analysis of the LHC
data.
In Figure~\ref{fig:pt-dependence}, the background rejection achieved at a fixed
$b$-tagging efficiency of \SI{77}{\percent} is shown in bins of jet \pt; this
fixed-efficiency is obtained by choosing the appropriate $b$-tagging
discriminant requirement in each \pt bin.
 
\begin{figure}[htbp]
\centering
\subfigure[]{\includegraphics[width=0.45\textwidth]{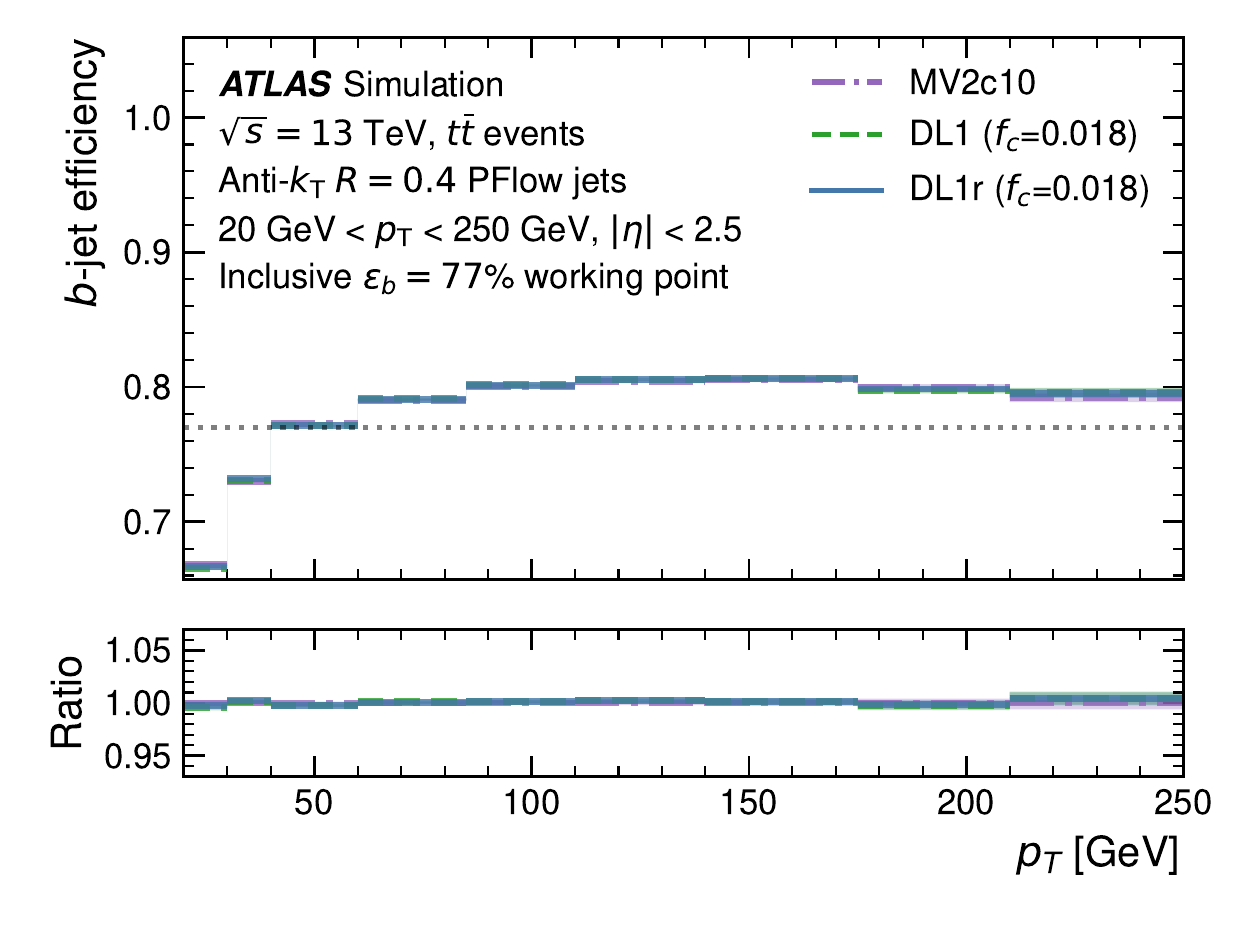}} \\
\subfigure[]{\includegraphics[width=0.45\textwidth]{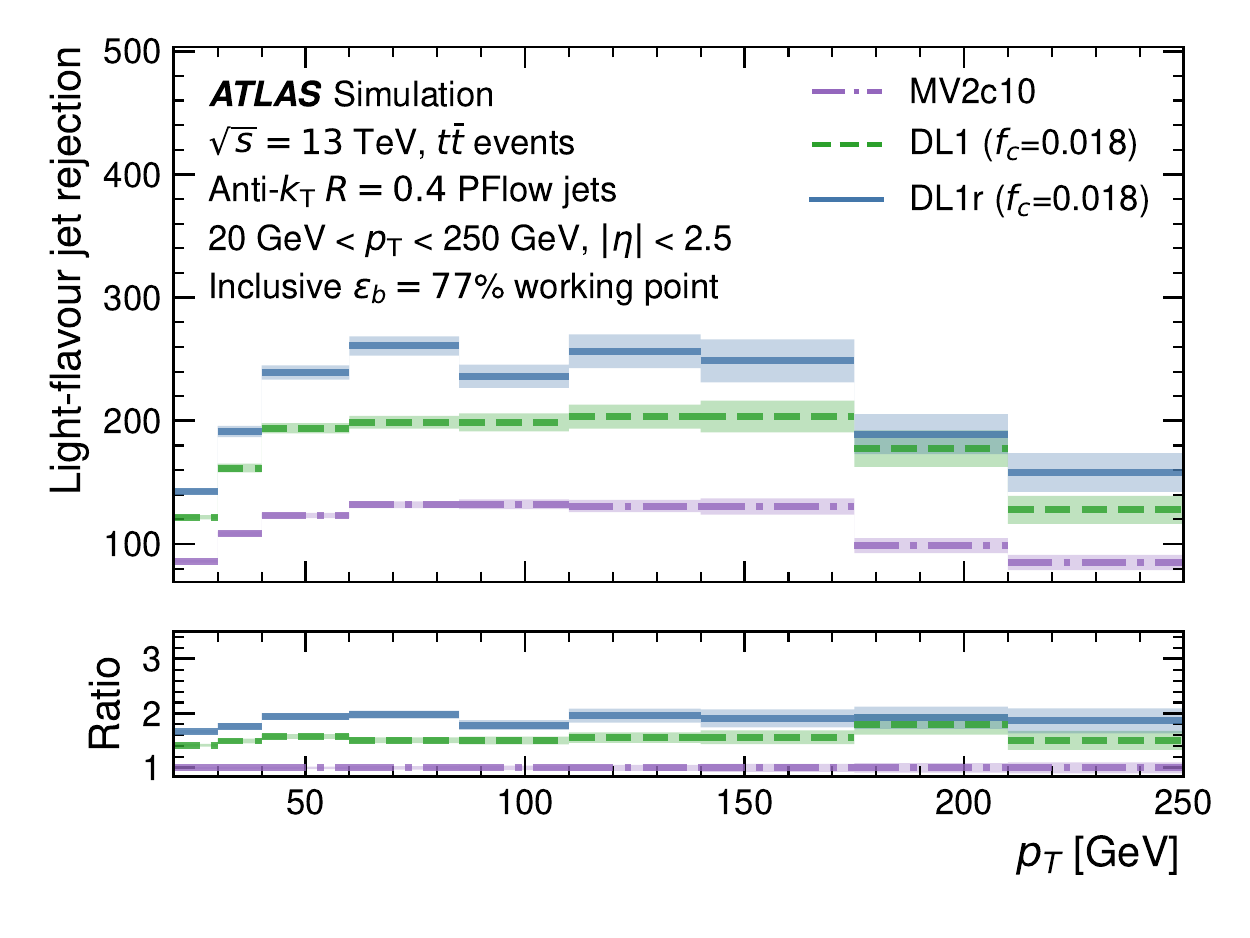}}
\subfigure[]{\includegraphics[width=0.45\textwidth]{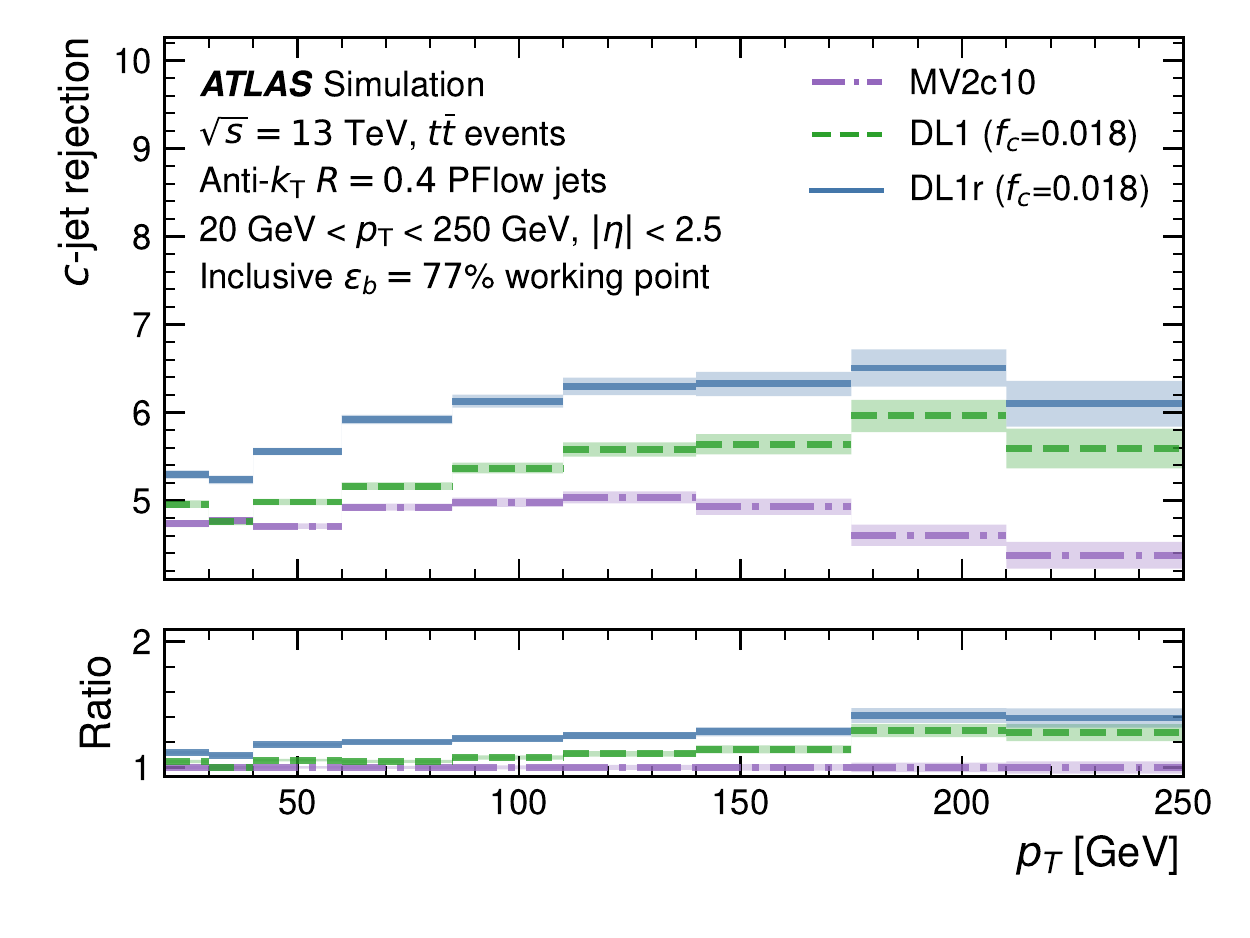}}
\caption{
The $b$-tagging efficiency, \beff, and background-jet rejection factors for
several high-level $b$-taggers, including \dloner, as a function of jet \pt,
evaluated using simulated \ttbar\ events.
(a) shows \beff, (b) shows the light-flavour jet rejection, and (c) shows
the $c$-jet rejection, all for a range of jet \pt bins at the inclusive 77\%
efficiency operating point, commonly used in ATLAS analyses of LHC \RunTwo
$pp$ collision data.
The lower panels show the ratio of each algorithm's performance to that
of \mvtwo.
The statistical uncertainties of the efficiencies and rejection factors are calculated
using binomial uncertainties and are indicated as coloured bands.
\label{fig:fixed-cut-performance-highlevel}
}
\end{figure}

\begin{figure}[htbp]
\centering
\subfigure[]{\includegraphics[width=0.45\textwidth]{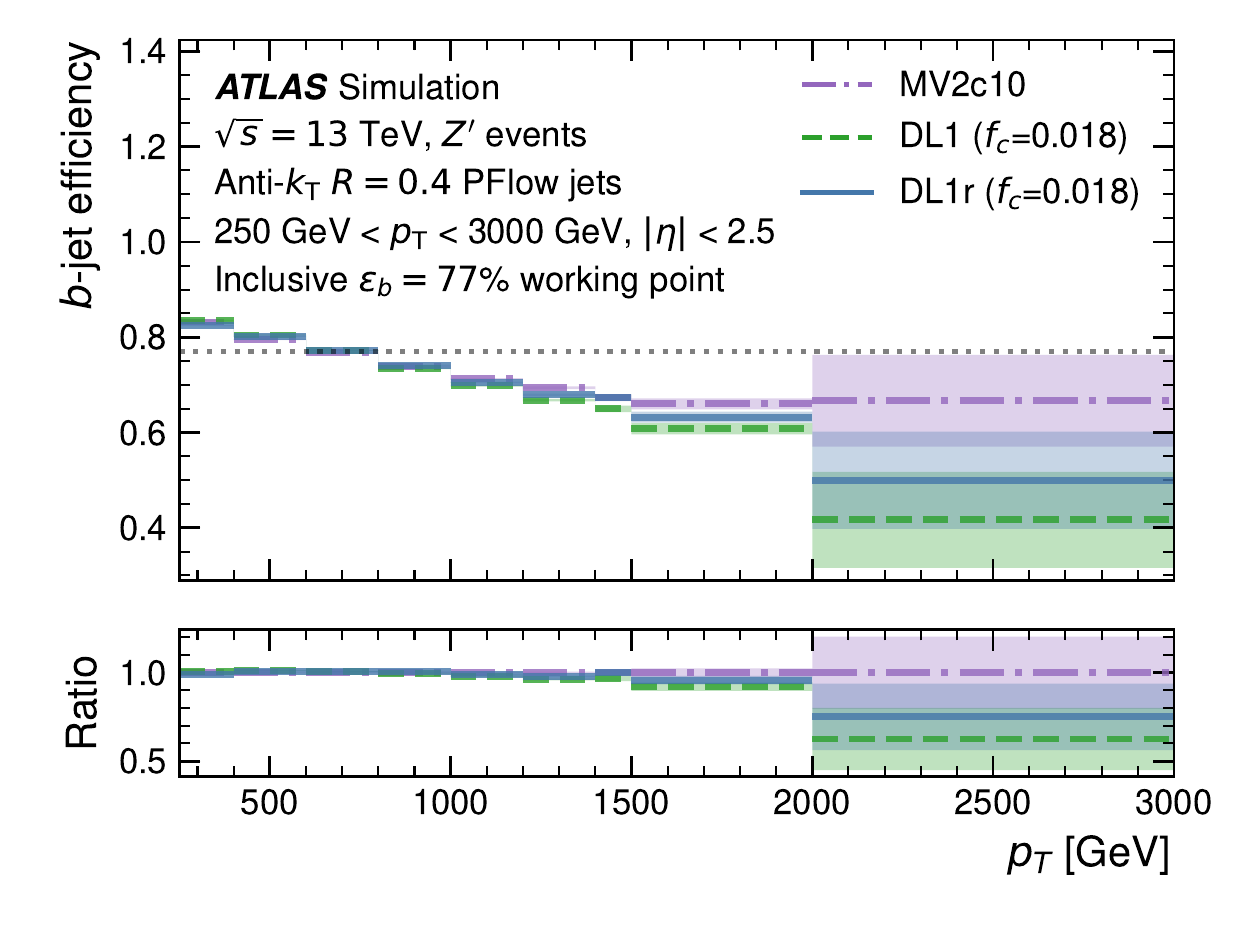}} \\
\subfigure[]{\includegraphics[width=0.45\textwidth]{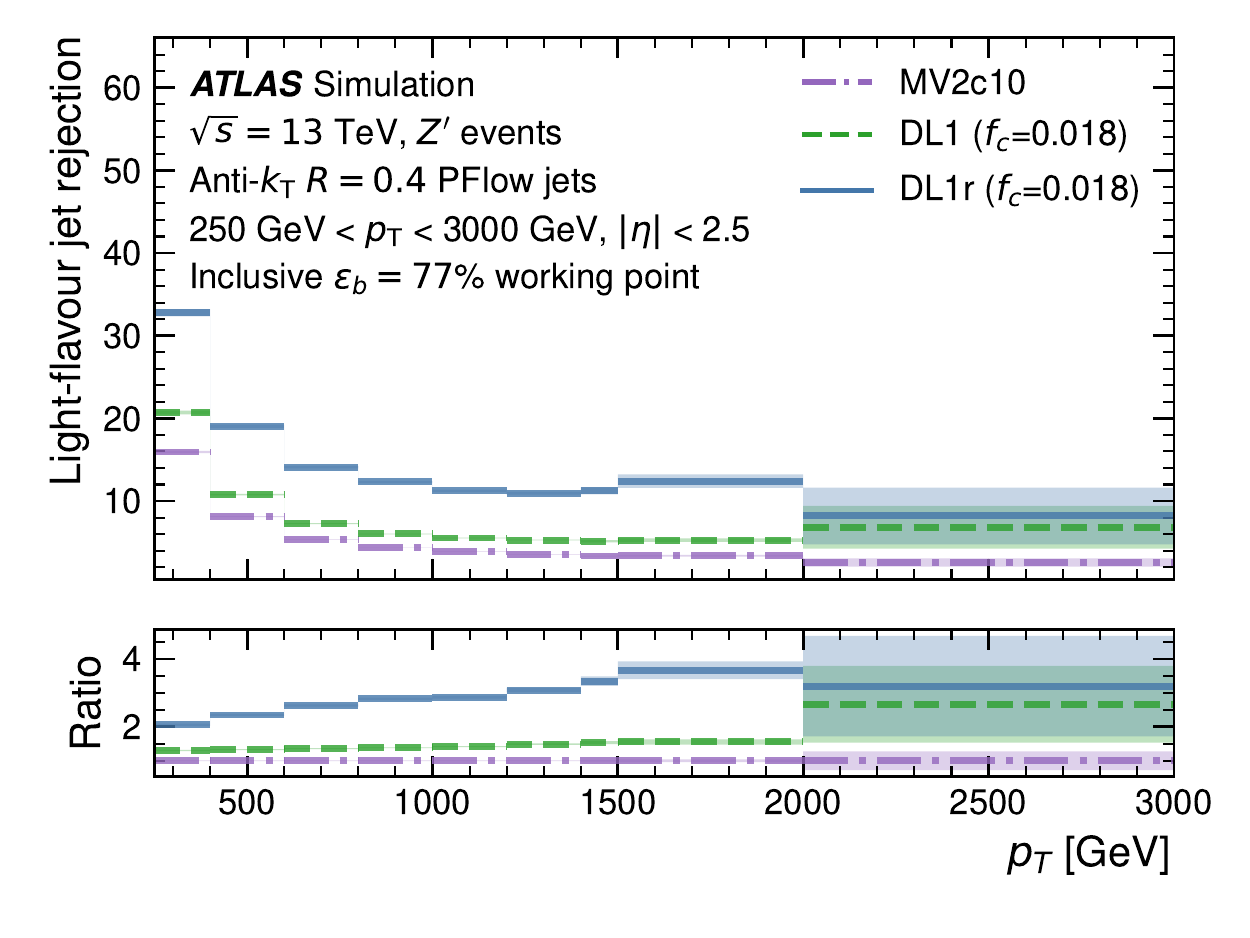}}
\subfigure[]{\includegraphics[width=0.45\textwidth]{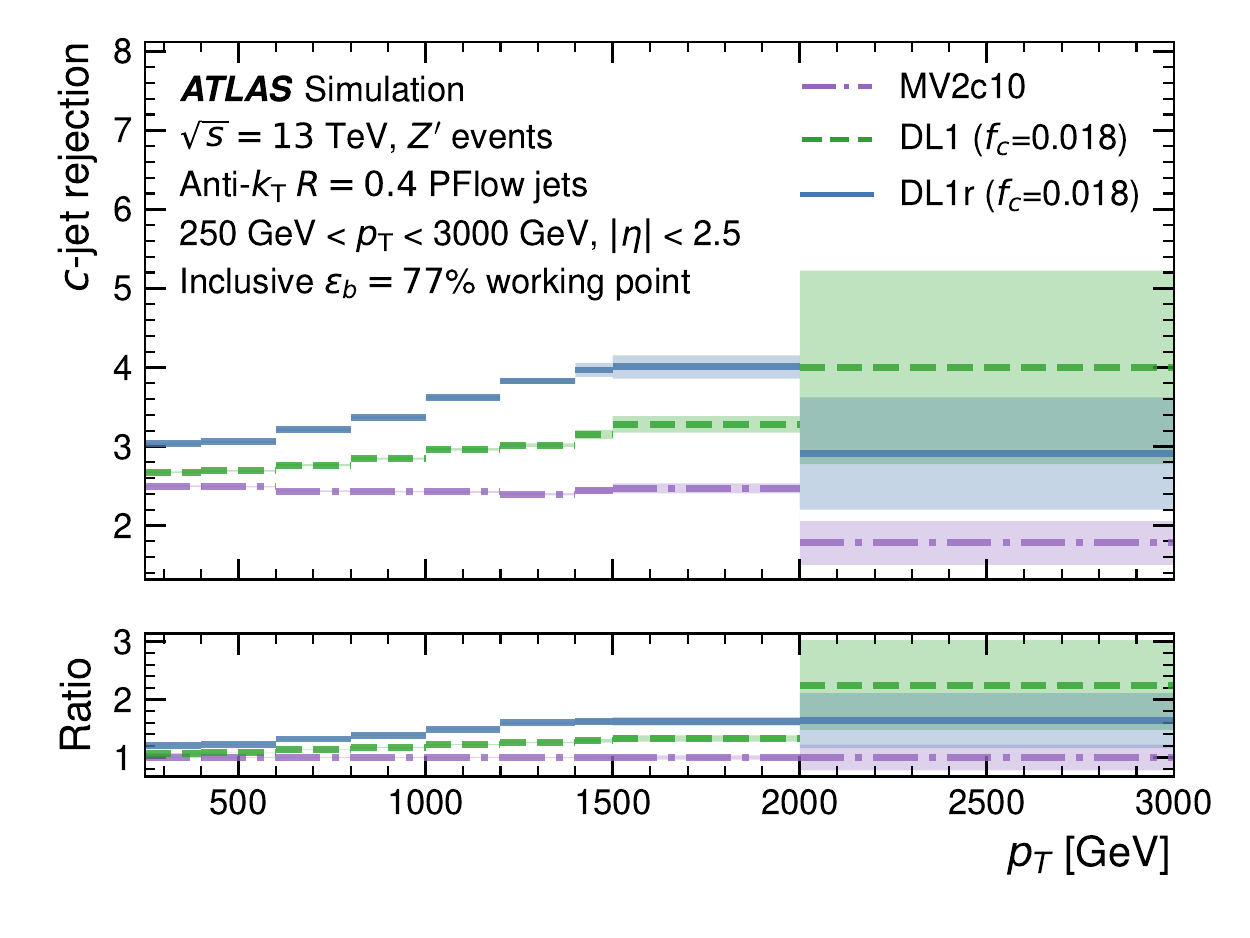}}
\caption{
The $b$-tagging efficiency, \beff, and background-jet rejection factors for several
high-level $b$-taggers, including \dloner, as a function of jet \pt, evaluated
using simulated \(Z'\) events, which mostly provide high-\pt jets.
(a) shows \beff, (b) shows the light-flavour jet rejection, and (c) shows
the $c$-jet rejection, all for a wide range of jet \pt bins at the inclusive 77\%
efficiency operating point, commonly used in ATLAS analyses of LHC \RunTwo
$pp$ collision data.
The lower panels show the ratio of each algorithm's performance to that of
\mvtwo.
The statistical uncertainties of the efficiencies and rejection factors are calculated
using binomial uncertainties and are indicated as coloured bands.
\label{fig:fixed-cut-performance-highlevel-Zprime}
}
\end{figure}

Figures~\ref{fig:fixed-cut-performance-highlevel}
and~\ref{fig:fixed-cut-performance-highlevel-Zprime}
show the \beff values and background rejection factors for jets from simulated SM \ttbar and flattened
\Zprime samples, respectively, in bins of jet \pt; several high-level
$b$-taggers are compared at the 77\% operating point.
\dloner performs significantly better than previous ATLAS $b$-taggers across a
broad range of jet \pt, although some common patterns are worth noting: (1) the
around $\pt \approx 175~\GeV$ and falls with \pt above this point, and (2) the
light-flavour jet rejection falls until about 1~\TeV, above which it is
approximately constant.
However, while \mvtwo maintains a nearly flat $c$-jet rejection versus
jet \pt, the \dlone and \dloner rejection factors improve with \pt.
The enhanced performance for highly energetic jets has yielded substantially
stronger tests of the Standard Model with the ATLAS data.
For example, recent searches for new resonances decaying into $b \bar{b}$ pairs
using the \dloner $b$-tagger achieved about a factor of 3 stronger limits on new
narrow resonances decaying into $b\bar b$ than predicted via luminosity-scaling of
previous results using \mvtwo~\cite{EXOT-2019-03}.

\begin{figure}[htbp]
\centering
\subfigure[]{\includegraphics[width=0.45\textwidth]{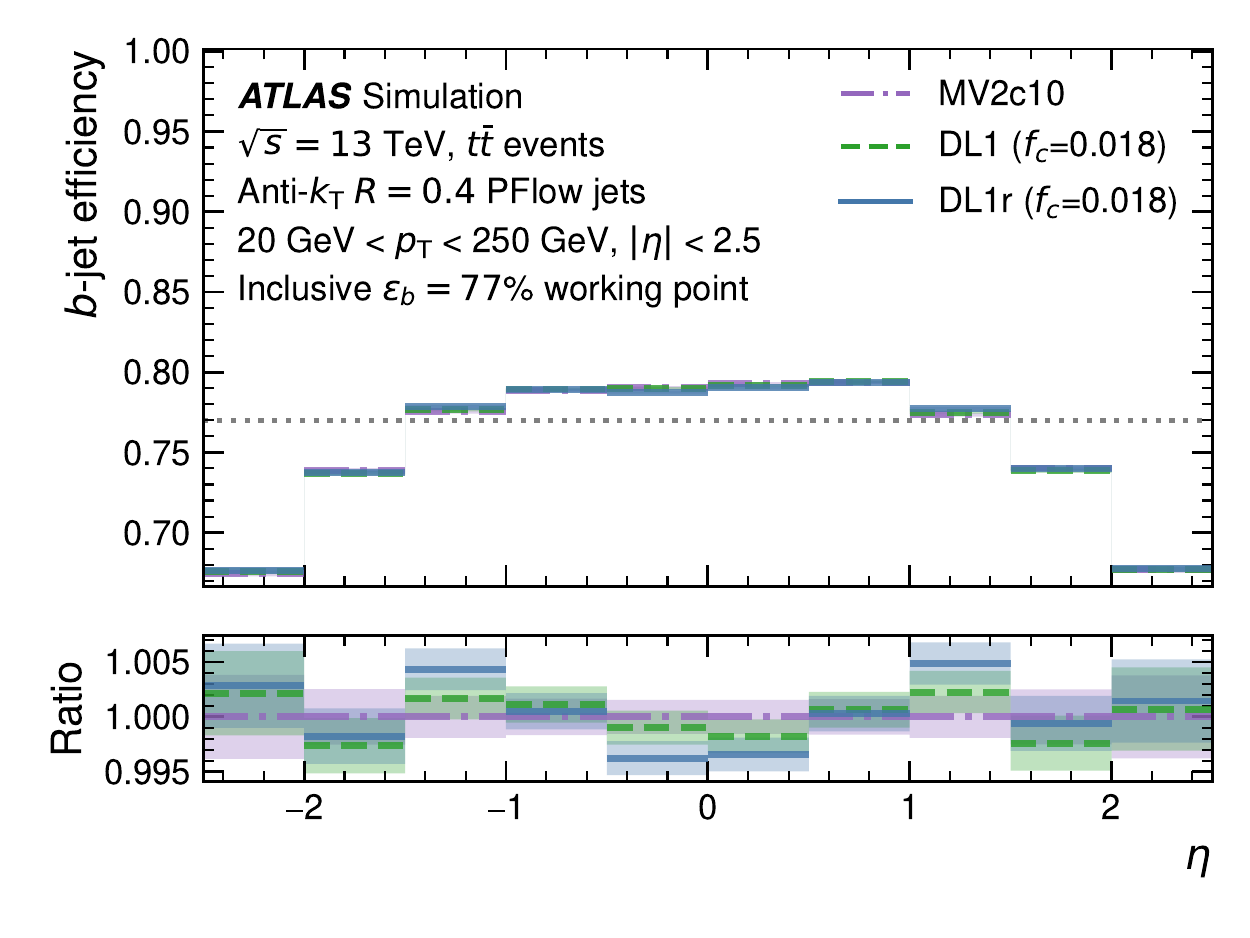}} \\
\subfigure[]{\includegraphics[width=0.45\textwidth]{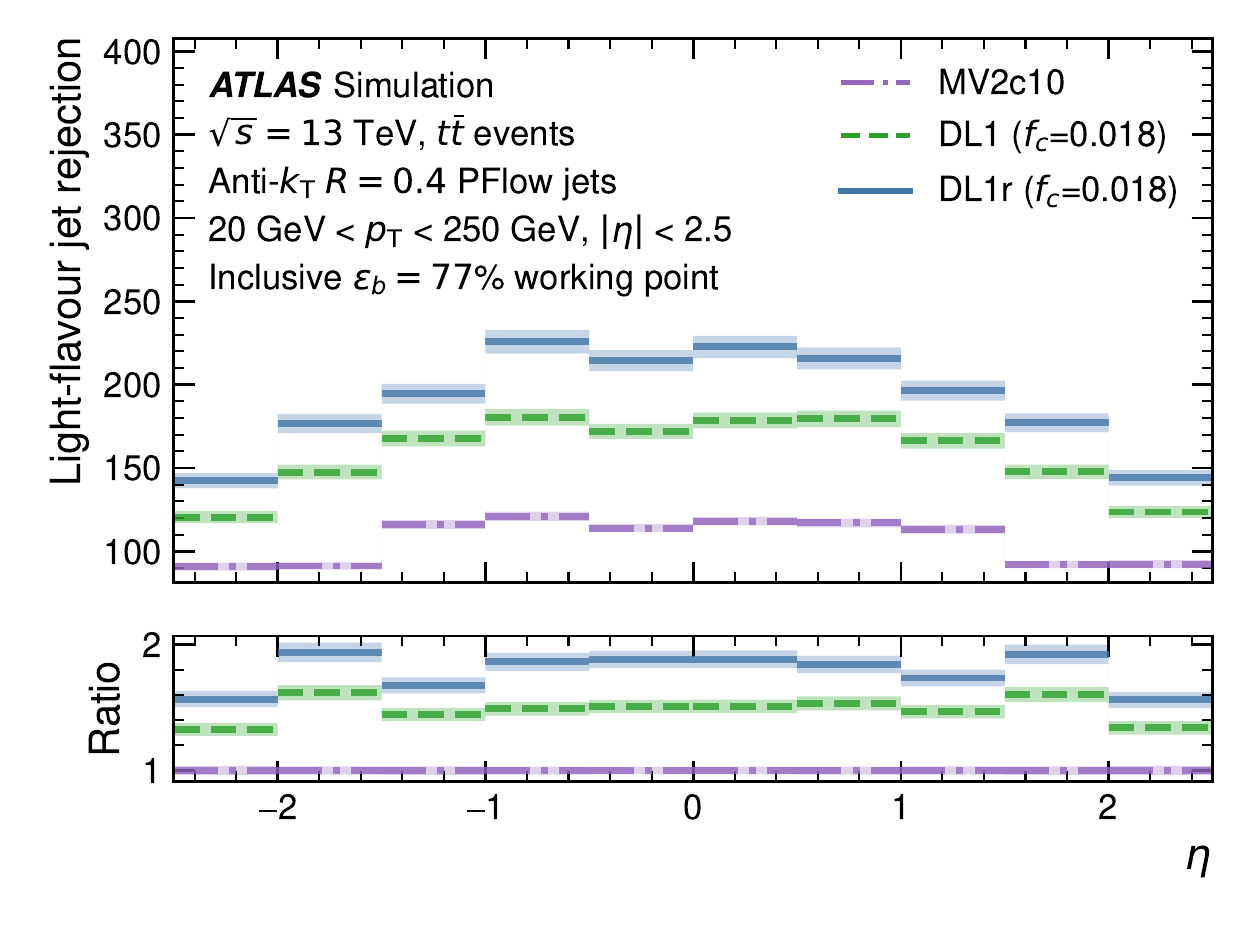}}
\subfigure[]{\includegraphics[width=0.45\textwidth]{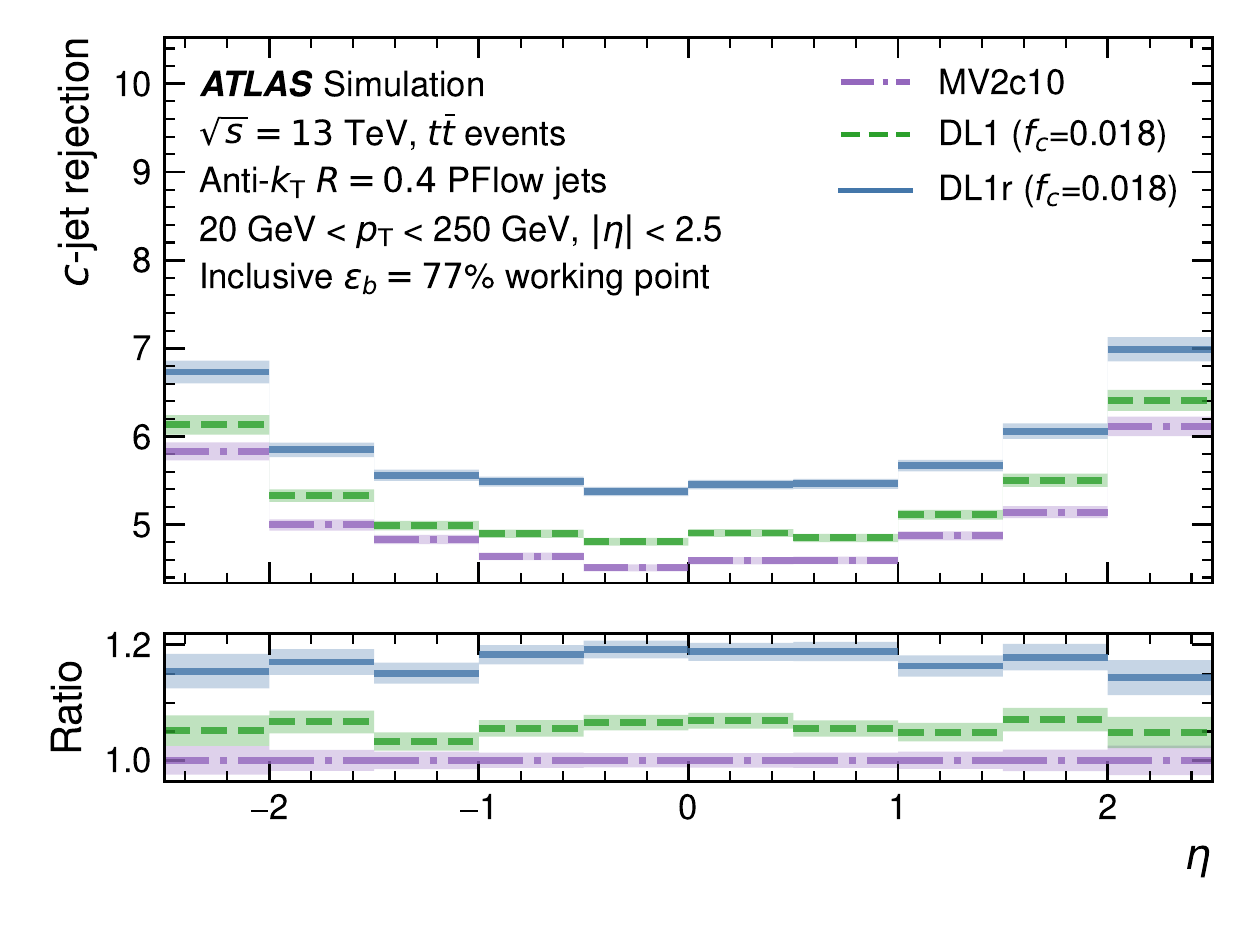}}
\caption{
The $b$-tagging efficiency, \beff, and background-jet rejection factors for several
high-level $b$-taggers, including \dloner, as a function of jet \(\eta\).
(a) shows \beff, (b) shows the light-flavour jet rejection, and (c) shows
the $c$-jet rejection, all at the inclusive 77\% efficiency operating point, commonly
used in ATLAS analyses of LHC \RunTwo $pp$ collision data.
The lower panels show the ratio of each algorithm's performance to that of
\mvtwo.
The statistical uncertainties of the efficiencies and rejection factors are calculated
using binomial uncertainties and are indicated as coloured bands.
}
\label{fig:fixed-cut-performance-highlevel-eta}
\end{figure}

Similarly, the $b$-tagging efficiencies and background-jet rejection factors
vary with the jet pseudorapidity $\eta$, in large part due to the poorer track
$d_0$ and $z_0$ resolutions at high $|\eta|$~\cite{ATL-PHYS-PUB-2015-018}.
Figure~\ref{fig:fixed-cut-performance-highlevel-eta} shows \beff and the
background-jet rejection as a function of jet $\eta$.
The $b$-tagging efficiency and $c$-jet mis-tag rates are higher for all compared
high-level taggers in the central region than at high $|\eta|$, in part due to
inefficiency of secondary-vertexing in the forward
region~\cite{ATL-PHYS-PUB-2017-011,ATL-PHYS-PUB-2018-025}.
The light-flavour jet rejection performance also deteriorates for more forward
jets.
However, \dloner consistently outperforms \dlone and \mvtwo across jet $\eta$
ranges.

\begin{figure}[htbp]
\centering
\subfigure[]{\includegraphics[width=0.45\textwidth]{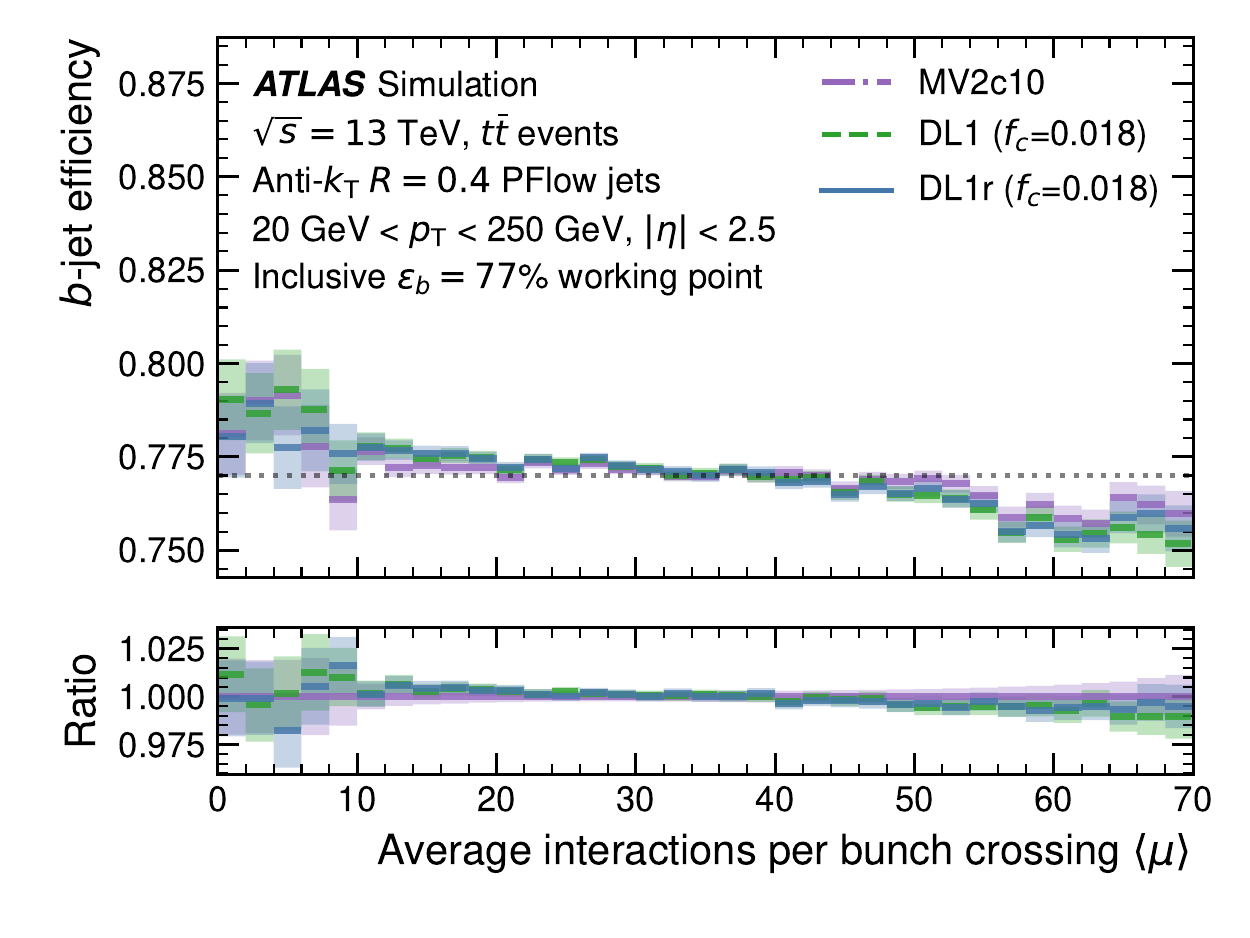}} \\
\subfigure[]{\includegraphics[width=0.45\textwidth]{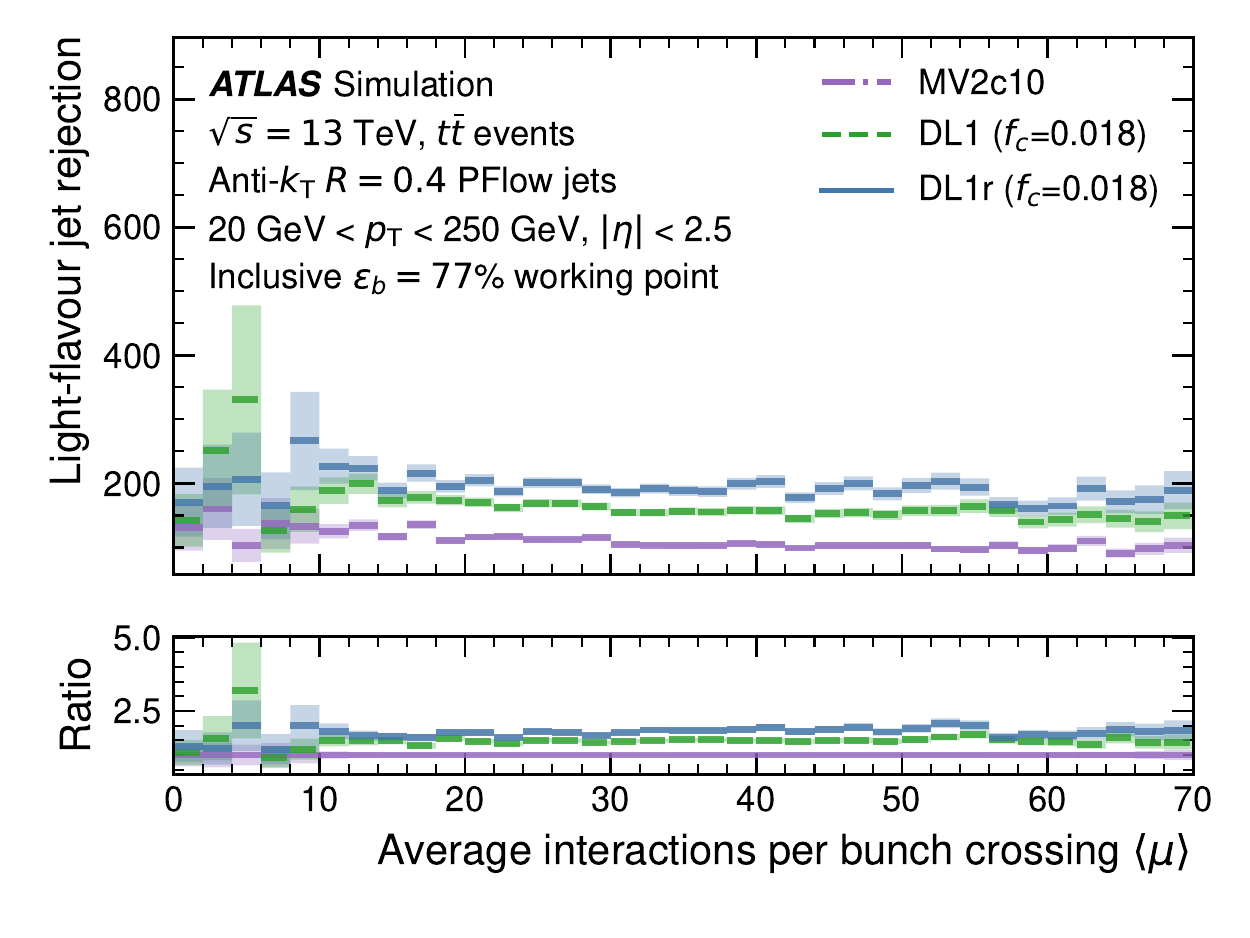}}
\subfigure[]{\includegraphics[width=0.45\textwidth]{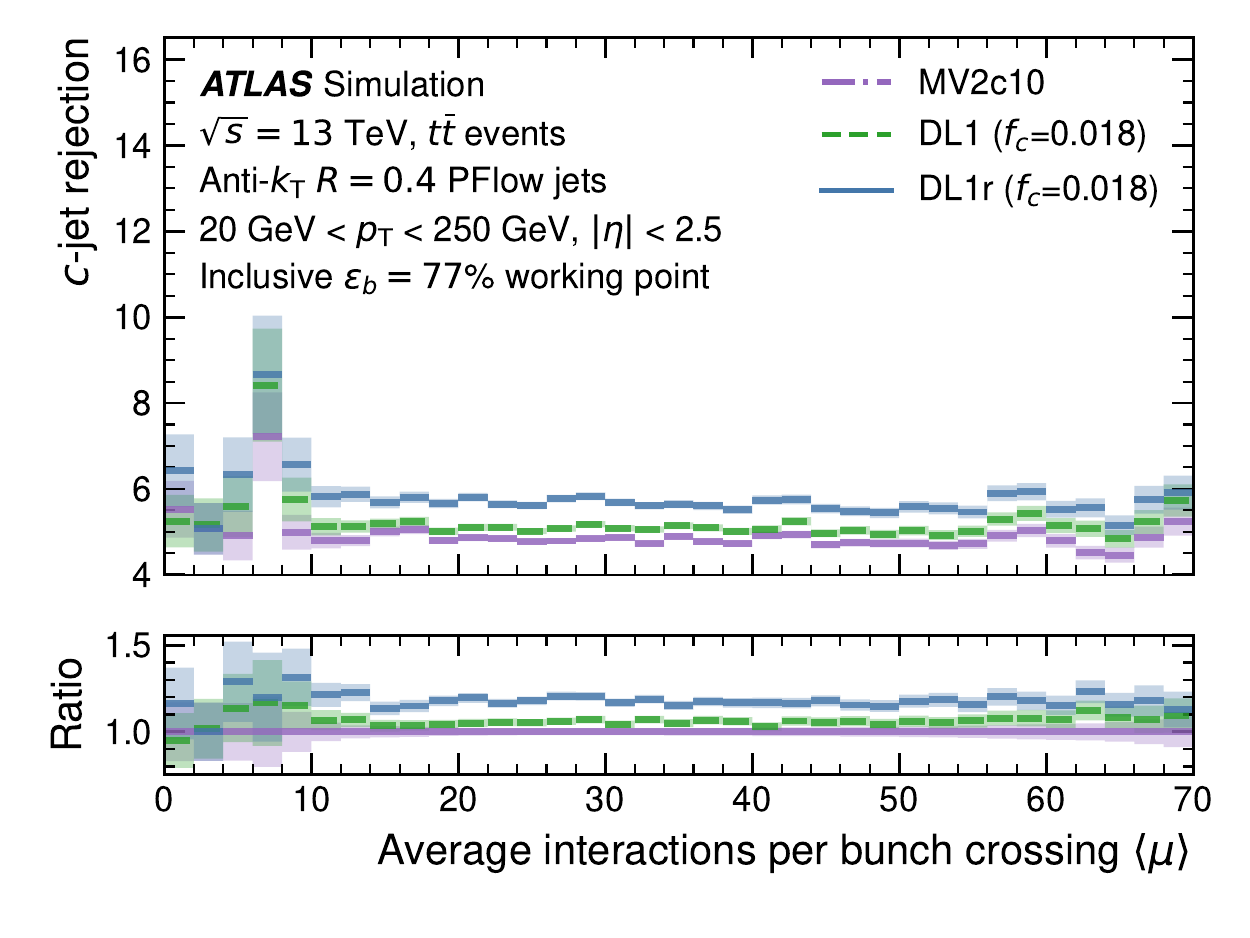}}
\caption{
The $b$-tagging efficiency, \beff, and background-jet rejection factors for several high-level
$b$-taggers, including \dloner, as a function of the average instantaneous
luminosity \(\langle \mu \rangle\).
(a) shows \beff, (b) shows the light-flavour jet rejection, and (c) shows
the $c$-jet rejection, all for a wide range of jet \pt bins at the inclusive 77\%
efficiency operating point, commonly used in ATLAS analyses of LHC \RunTwo
$pp$ collision data.
The lower panels show the ratio of each algorithm's performance to that of
\mvtwo.
The statistical uncertainties of the efficiencies and rejection factors are calculated
using binomial uncertainties and are indicated as coloured bands.
}
\label{fig:fixed-cut-performance-highlevel-pileup}
\end{figure}

The ATLAS flavour-tagging algorithms are stable versus the number of pile-up
interactions accompanying the hard-scatter collisions, as shown in
Figure~\ref{fig:fixed-cut-performance-highlevel-pileup}.
While there is a small slope in the $b$-tagging efficiency versus the average
number of interactions per bunch crossing $\langle \mu \rangle$, \beff only
changes by about 2\% over the range $10 < \langle \mu \rangle < 70$ for the
inclusive 77\% efficiency operating point.
The light-flavour jet and $c$-jet rejection factors also show little dependence on $\langle \mu
\rangle$.

\begin{figure}[htbp]
\centering
\subfigure[]{\includegraphics[width=0.45\textwidth]{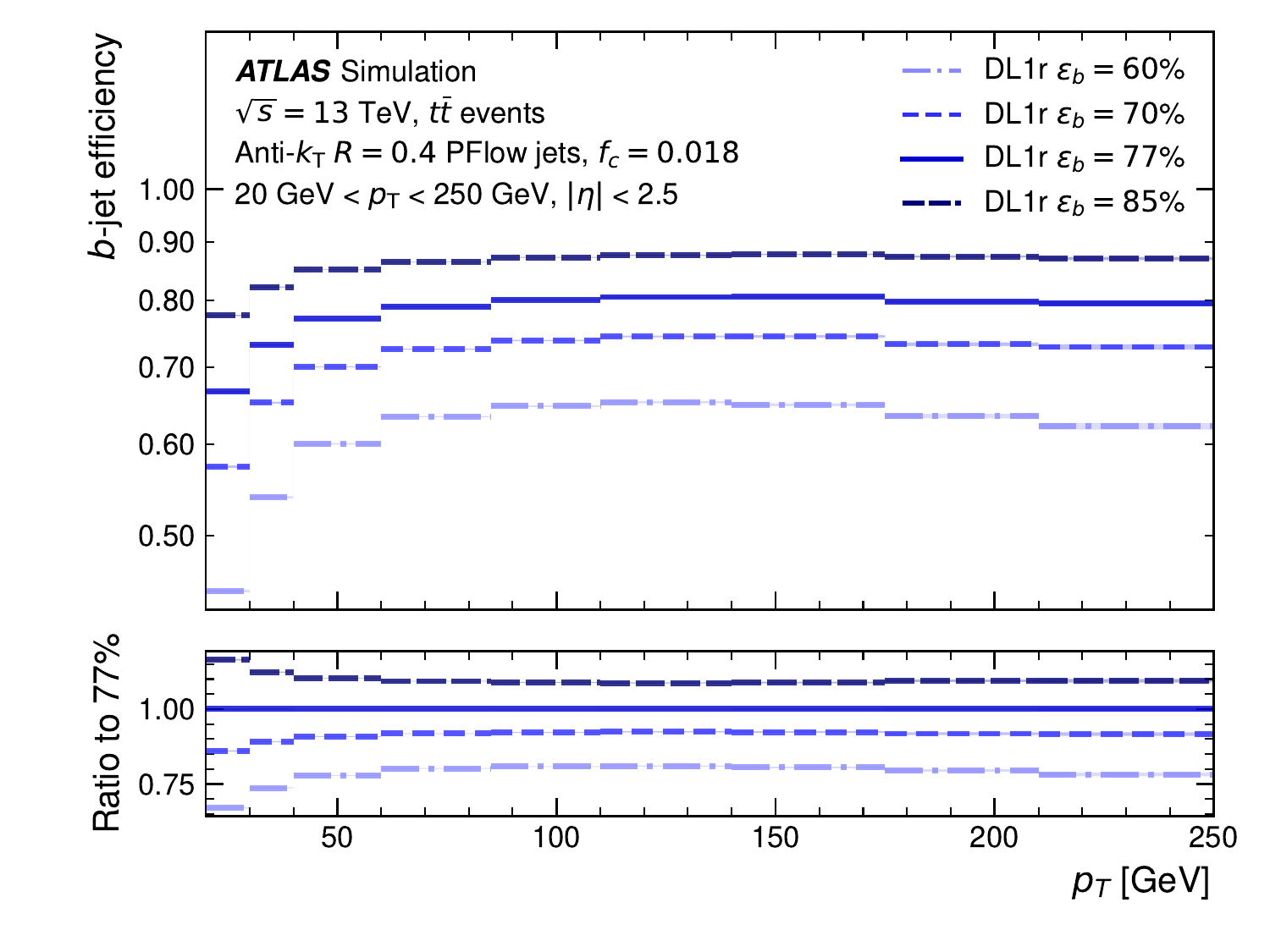}} \\
\subfigure[]{\includegraphics[width=0.45\textwidth]{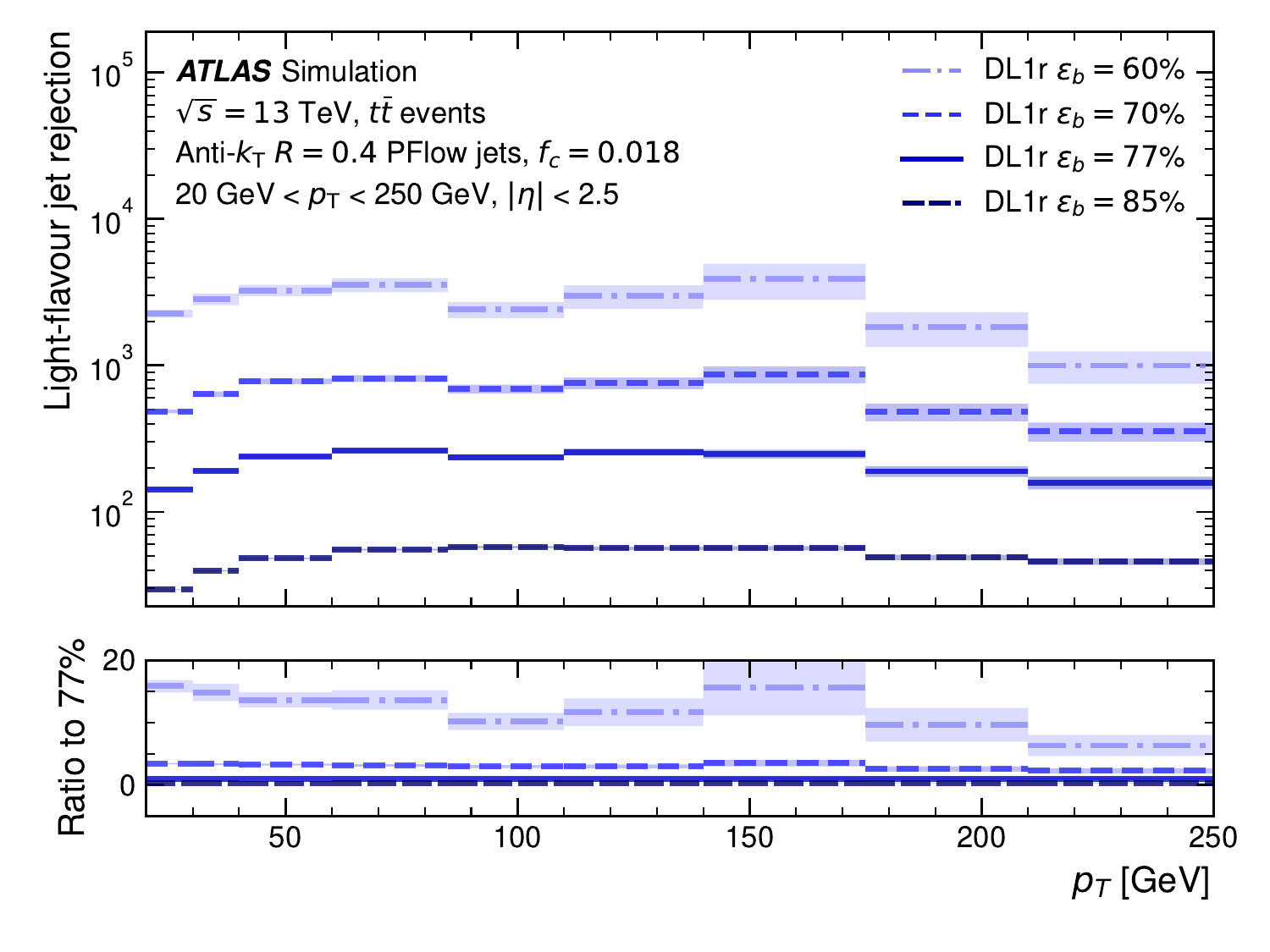}}
\subfigure[]{\includegraphics[width=0.45\textwidth]{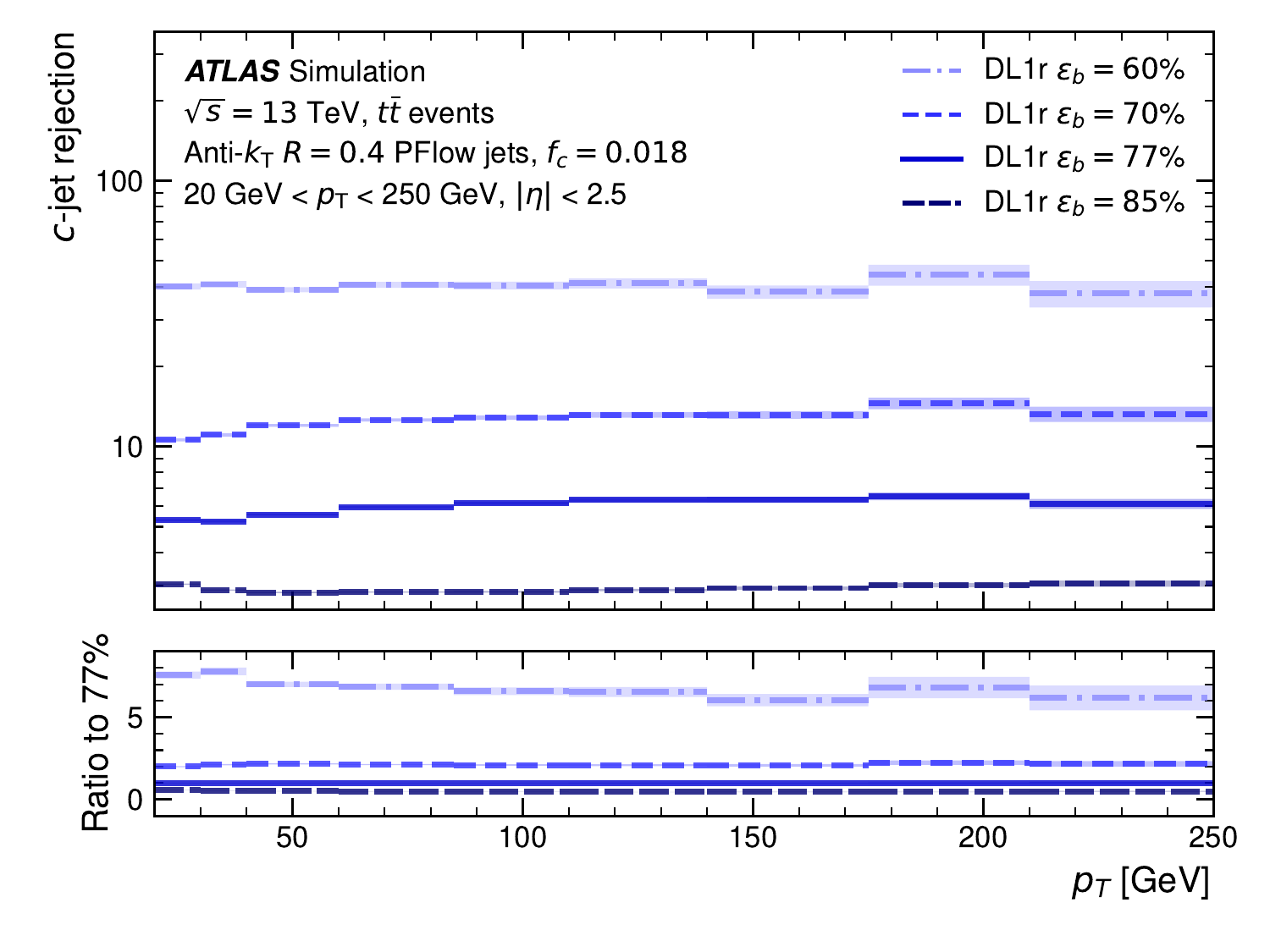}}
\caption{
The $b$-tagging efficiency \beff and background-jet rejection factors for
several operating points as a function of jet \pt.
(a) shows the \dloner \beff, (b) shows the light-flavour jet rejection, and (c) shows
the $c$-jet rejection, all for a wide range of jet \pt bins and for efficiency
operating points commonly used in ATLAS analyses of LHC \RunTwo $pp$ data.
The lower panels show the ratio of each operating point's performance to that of
the 70\% operating point.
The statistical uncertainties of the efficiency (rejection) are calculated using
binomial uncertainties and are indicated as coloured bands.
\label{fig:fixed-cut-performance-workingpoints}
}
\end{figure}

The 60\%, 70\%, 77\%, and 85\% operating points are commonly used in ATLAS
physics analyses interpreting the LHC \RunTwo $pp$ dataset~\cite{FTAG-2018-01}.
Figure~\ref{fig:fixed-cut-performance-workingpoints} shows the $b$-tagging
efficiency and background-jet rejection factors versus jet \pt for \dloner at these
commonly used operating points.
It is worth noting that relatively small changes in $b$-tagging efficiency
operating point of the order of 10\% result in very different background
rejection factors, which range from ${\sim}10$ to ${\sim}10^3$ for light-flavour jets and from
${\sim}3$ to ${\sim}40$ for charm jets.

\subsection{Charm-tagging performance}
 
A growing number of ATLAS analyses of LHC data require identification of
$c$-jets in order to probe physics processes involving final-state charm
quarks.
For example, the search for Higgs boson decays into charm quarks aims
to obtain direct evidence of the charm-quark Yukawa coupling parameter by
examining a sample of events with two $c$-tagged jets~\cite{HIGG-2021-12}.
An algorithm's performance is indicated by its $b$-jet and light-flavour jet
rejection factors, $1/\beff$ and $1/\leff$, at a given charm-tagging
efficiency, $\ceff$.
Due to charmed hadrons having shorter lifetimes and smaller masses than $b$-hadrons,
the identification of charm jets is challenging, and
the efficiency of the optimal operating point tends be much lower than for
$b$-tagging.
 
\begin{figure}[htbp]
\centering
\includegraphics[width=0.65\textwidth]{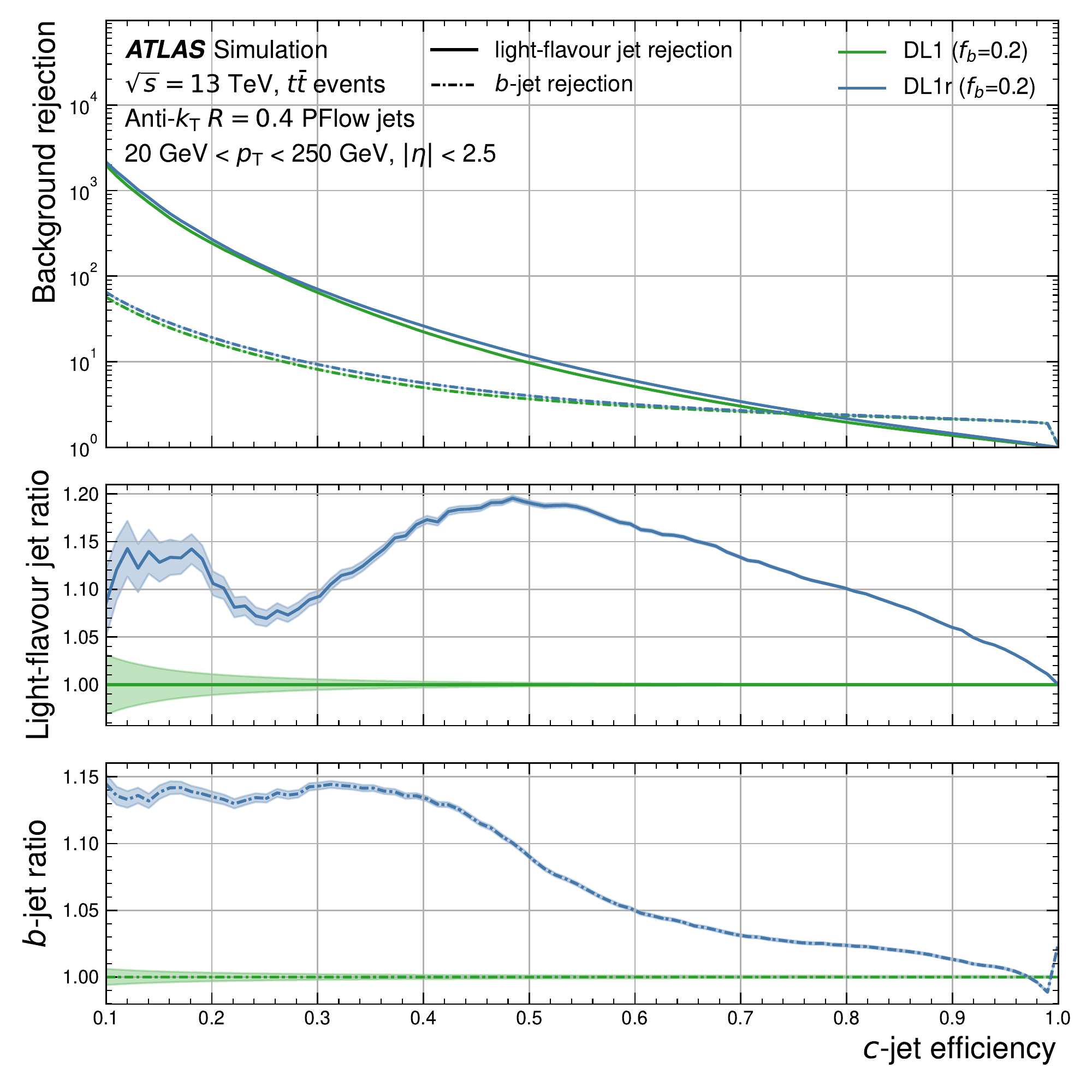}
\caption{
The light-flavour jet and $b$-jet rejection factors as a function of $c$-jet efficiency for
\dlone and \dloner.
The lower two panels show the \dloner-to-\dlone ratios of the light-flavour jet rejection and the
\(c\)-jet rejection.
The statistical uncertainties of the rejection are calculated using binomial
uncertainties and are indicated as coloured bands.
\label{fig:highlevel-ROCs_ctag}
}
\end{figure}

\begin{figure}[htbp]
\centering
\includegraphics[width=0.65\textwidth]{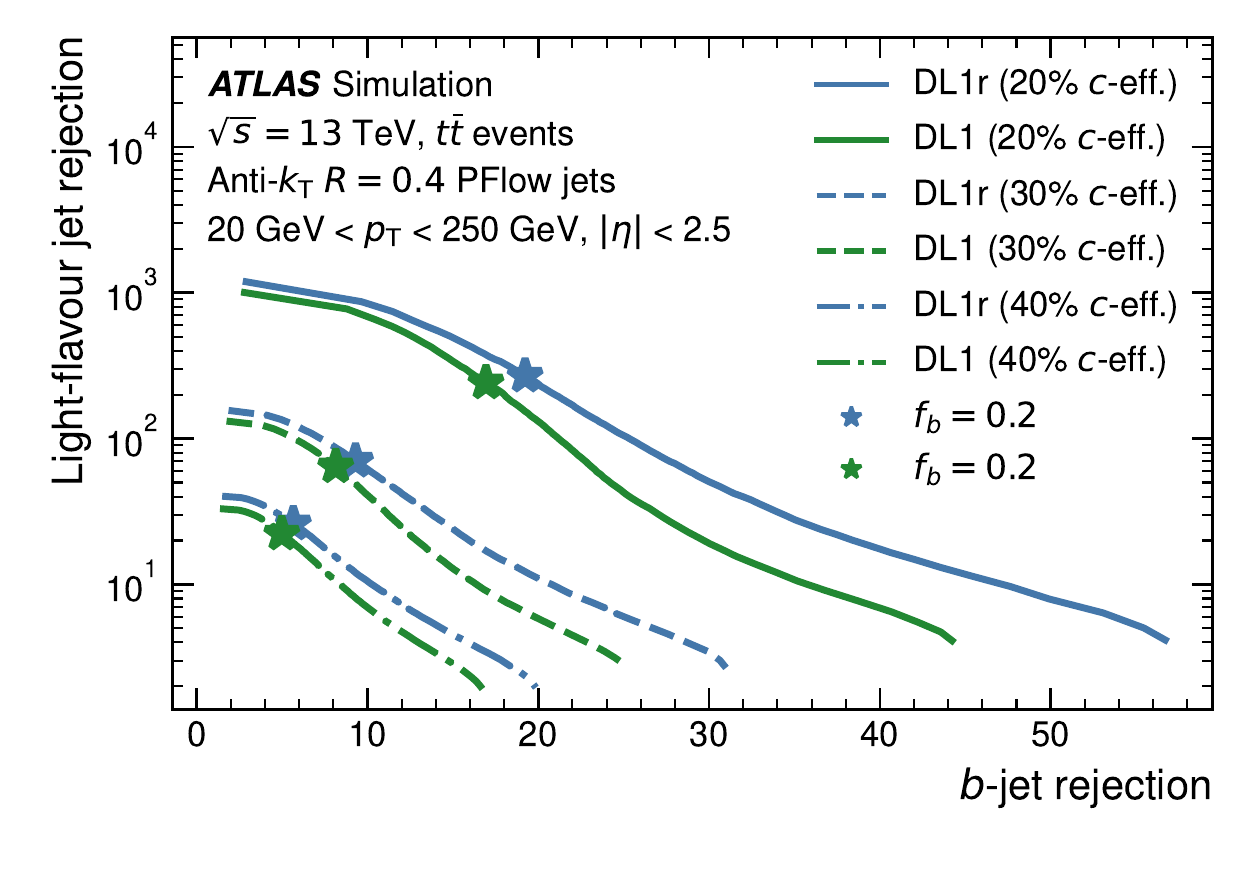}
\caption{
The light-flavour jet rejection as a function of $b$-jet rejection for
inclusive 20\%, 30\%, and 40\% $c$-jet efficiency operating points for the
\dlone and \dloner high-level $c$-taggers.
Each point on a curve corresponds to a particular choice of $f_b$, the $b$-jet
background fraction in the log-likelihood ratio that defines the tagging
discriminant; the star symbols indicate the $f_b = 0.2$ point.
\label{fig:highlevel-IsoEff}
}
\end{figure}

Figure~\ref{fig:highlevel-ROCs_ctag} presents the $b$-jet and light-flavour jet rejection factors
as a function of the $c$-tagging efficiency, evaluated in a population of jets
taken from a sample of simulated \ttbar events.
Figure~\ref{fig:highlevel-IsoEff} shows the light-flavour jet and $b$-jet rejection factors
attained by the \dlone and \dloner algorithms for a fixed charm-tagging
efficiency, again evaluated in jets from a \ttbar sample.
These \enquote{iso-efficiency} curves are obtained by varying the parameter $f_b$ used
to define the $c$-tagging discriminant introduced in Eq.~(\ref{eq:dl1rc}).

\subsection{MC generator dependence}
 
A variety of MC event generators are used in ATLAS analyses to model various
production processes at the LHC.
\Pythia~\cite{Sjostrand:2007gs}, \Herwig~\cite{Bellm:2017bvx}, and
\Sherpa~\cite{Sherpa:2019gpd} are commonly used parton-shower and hadronisation
programs that describe the structure and composition of hadronic jets.
ATLAS \Pythia and \Herwig MC samples also utilise the \EvtGen package to simulate
the decay chains of $b$- and $c$-hadrons.
The choice of parton-shower generator, hadronisation model, and hadron-decay
model affects the predicted performance of the ATLAS flavour-taggers.
Tagging efficiencies and mis-tag rates have been studied using various MC generator
versions, settings, and tuned parameter values~\cite{ATL-PHYS-PUB-2020-009}.
These effects are taken into account by applying generator-specific corrections
to the simulation.

\subsection{Overtraining checks}
 
In order to correct the tagging rates of jets in simulation, corresponding
rates are measured in data through a variety of
procedures~\cite{FTAG-2018-01,FTAG-2020-08,ATLAS-CONF-2018-055,ATLAS-CONF-2018-045,ATLAS-CONF-2018-006},
reaching a precision as good as 1\% in $b$-tagging efficiency for $b$-jets with
$\pt \sim 100$~\GeV; the most precise measurement of the $c$-jet (light-flavour jet)
mis-tag rate has uncertainties of approximately 5\% (15\%).
Efficiencies and mis-tag rates in simulation are adjusted via per-jet weights or
\enquote{scale factors} such that they reflect the performance measured in data.
However, \ttbar MC events used to train the \rnnip and \dloner algorithms are
also used in the efficiency measurements and in physics analyses utilising
flavour-tagging.
For the calibration procedure to result in a properly corrected simulation, the
\dloner performance on the \enquote{training} sample, used to optimise the tagging
algorithms, and a \enquote{validation} sample, comprising MC events
not contained in the training sample, must be consistent to within the
uncertainty associated with the corresponding data-based measurement.
 
\begin{figure}[htbp]
\centering
\includegraphics[width=0.65\textwidth]{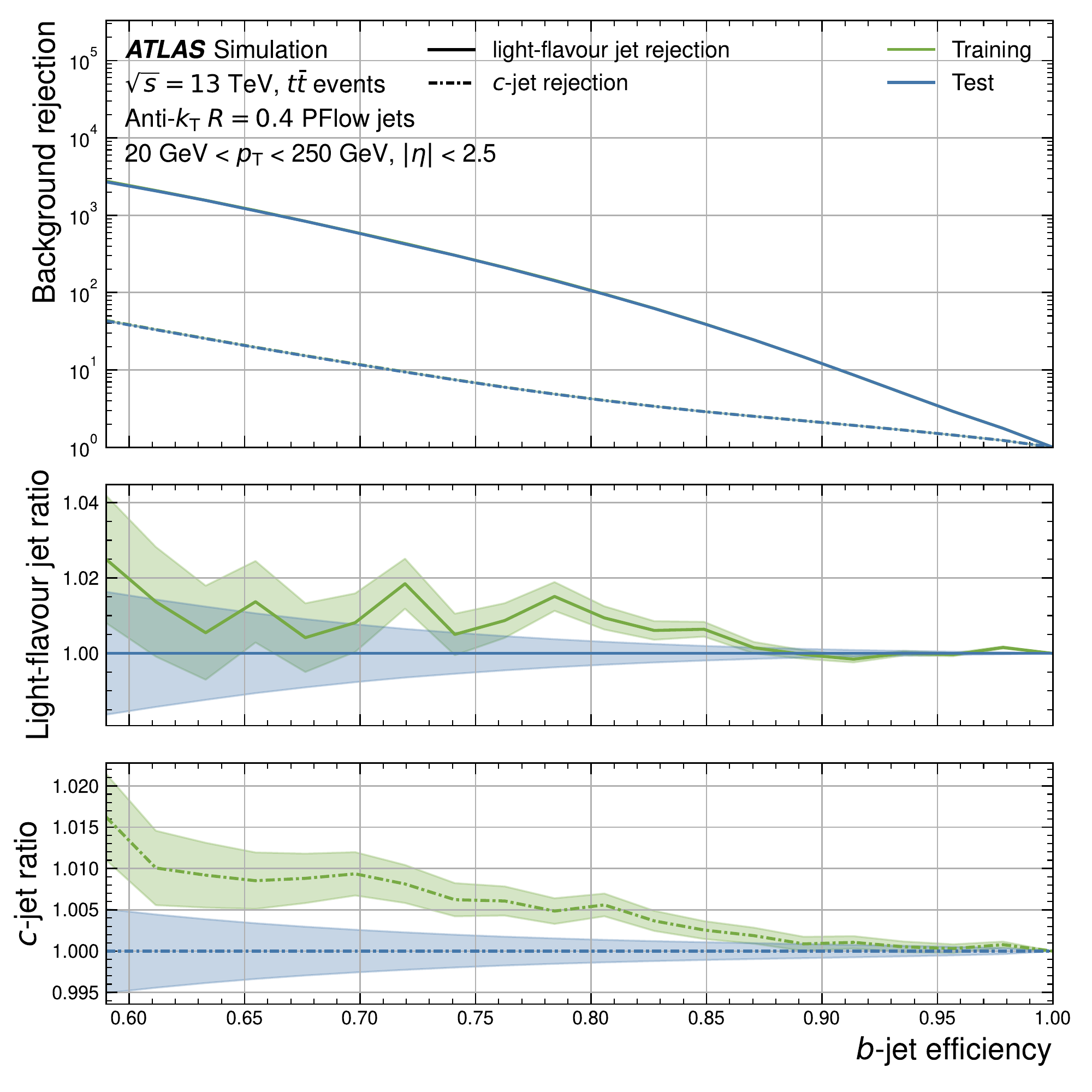}
\caption{
A comparison of the $c$-jet and light-flavour jet background rejection factors versus the
$b$-tagging efficiency for training and validation samples for the \dloner
$b$-tagging algorithm.
The training sample contains events used to optimise the \dloner
algorithm, while the validation sample comprises a statistically distinct
set of events that were not used during training.
Efficiencies and rejection factors are both derived separately for each sample.
The lower two panels show the ratios of training sample to validation
sample light-flavour jet rejection and $c$-jet rejection for \dloner.
The statistical uncertainties of the rejection are calculated using binomial
uncertainties and are indicated as coloured bands.
No difference in performance is observed above the 2\% level, which is below
the precision of the calibration measurements for these quantities.
\label{fig:trainvsvalid-roc}
}
\end{figure}

\begin{figure}[htbp]
\centering
\subfigure[]{\includegraphics[width=0.45\textwidth]{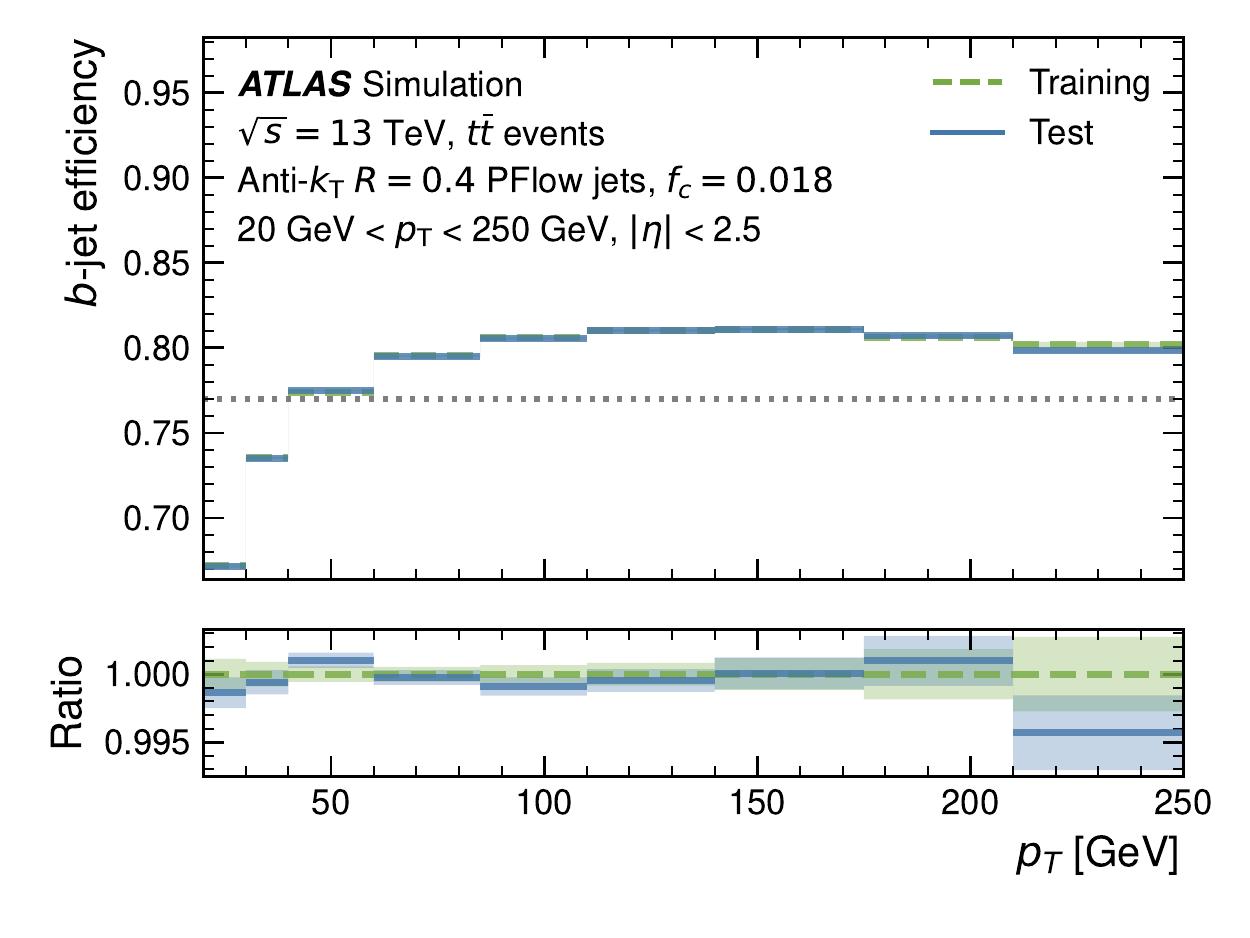}}
\subfigure[]{\includegraphics[width=0.45\textwidth]{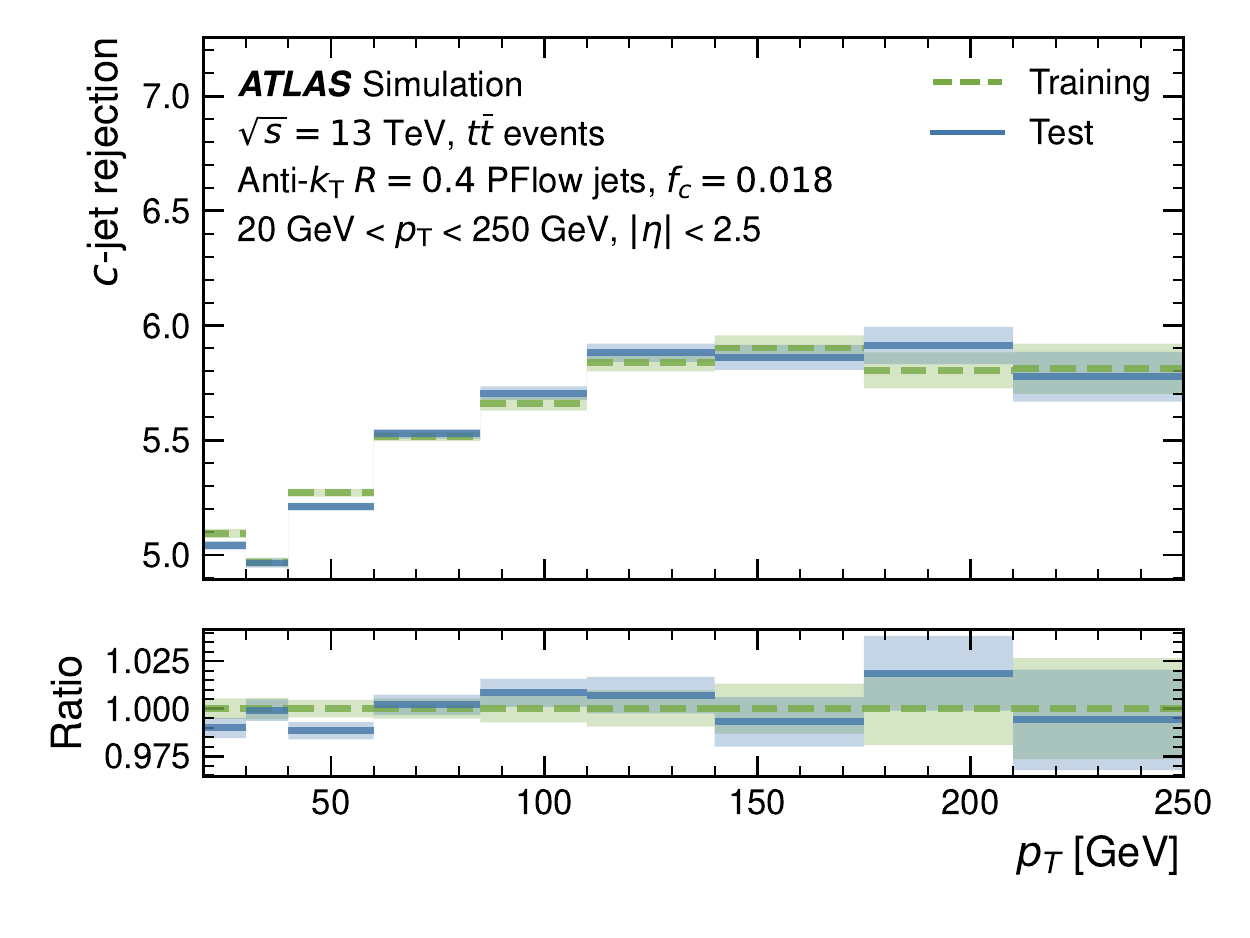}} \\
\subfigure[]{\includegraphics[width=0.45\textwidth]{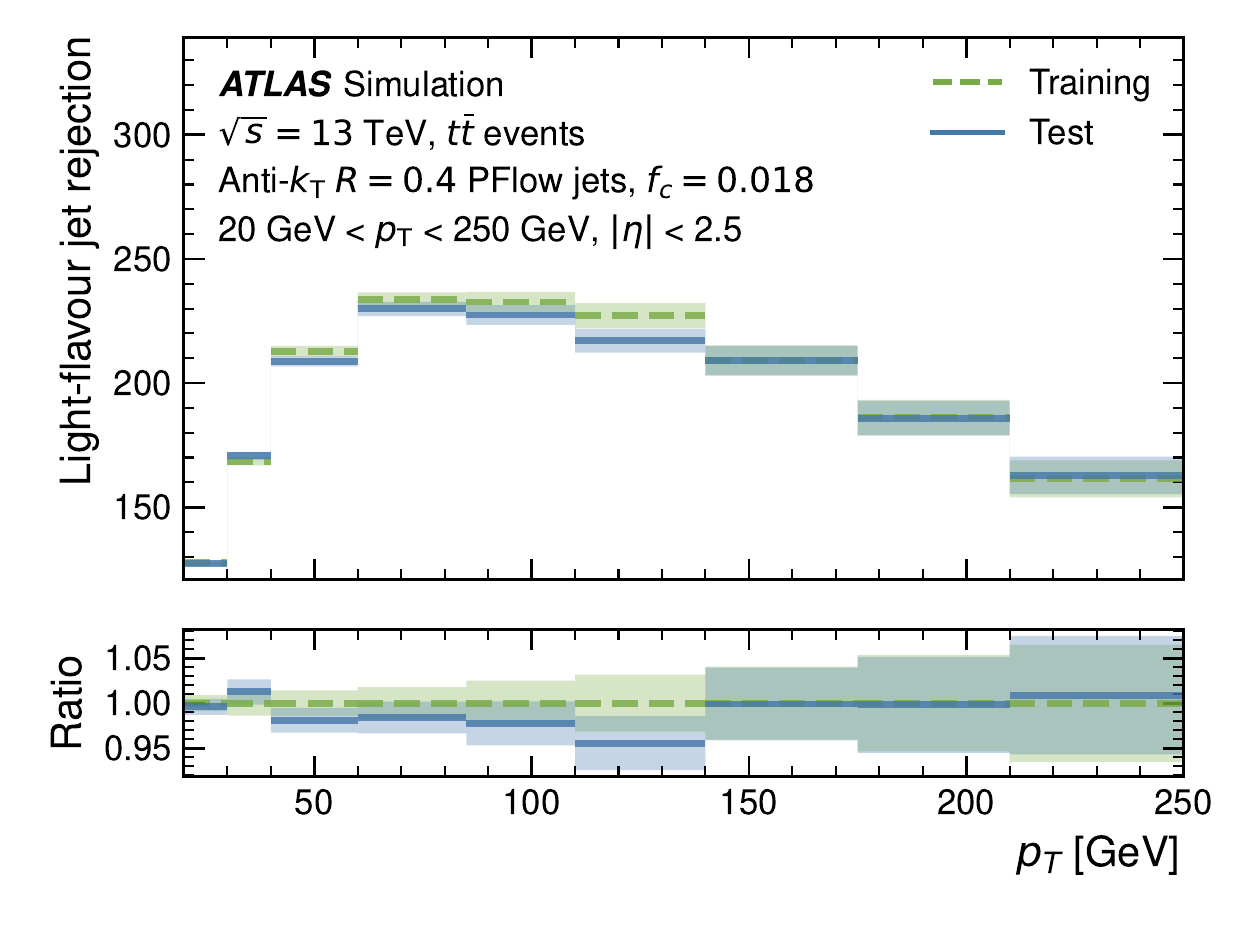}}
\caption{
A comparison of performance in jet \pt bins between the training and
validation \ttbar samples for \dloner.
The training sample contains events used to optimise the \dloner
algorithm, while the validation sample comprises a statistically distinct
set of events that were not used during training.
The (a) $b$-tagging efficiency, (b) $c$-jet rejection and (c) light-flavour jet
rejection are shown for a broad \pt range at the inclusive 77\%
efficiency operating point, commonly used in ATLAS analyses of LHC \RunTwo
$pp$ data.
This operating point is derived using the combined training and validation
samples.
The lower panels show the ratio of training sample performance to validation sample performance.
No statistically significant difference in performance is observed at the level
of the precision of data-based efficiency measurements, which is ${\sim}1$\% for
$b$-jets, ${\sim}5$\% for $c$-jets, and ${\sim}15$\% for light-flavour jets.
The statistical uncertainties of the efficiency (rejection) are calculated using
binomial uncertainties and are indicated as coloured bands.
\label{fig:trainvsvalid-pt}
}
\end{figure}

Figure~\ref{fig:trainvsvalid-roc} shows the background rejection factors versus the
$b$-tagging efficiency separately for the training and validation samples of
\ttbar events.
Figure~\ref{fig:trainvsvalid-pt} presents background rejection factors and $b$-tagging
efficiency versus the jet \pt, and the expected performance in the training and
validation samples is again compared.
No discrepancy is observed that is significant relative to the precision of
efficiency measurements, indicating that it is safe to re-use the events used in
\dloner training in physics analyses.
 
\FloatBarrier
\section{Conclusion}
 
Several flavour-tagging algorithms are used to identify jets containing
heavy-flavour hadrons in data recorded by the ATLAS experiment during \RunTwo of
the LHC.
The recent ATLAS strategy combines the results of low-level algorithms (\iptwod,
\ipthreed, \svone, \jetfitter and \rnnip) into high-level algorithms based on
the \dloner feed-forward neural network classifier.
The low-level algorithms either exploit the large impact parameters of
tracks left by heavy-flavour hadron decay products or attempt to
directly reconstruct their decay vertices.
Large increases in background-jet rejection are obtained by the \dloner
algorithms compared to each individual low-level algorithm and to previous
tagging algorithms, illustrating the high complementarity of the low-level
inputs.
In a sample of simulated Standard Model \ttbar\ events, light-flavour jet (charm-jet) rejection
factors of 170 (5) are achieved at a $b$-jet identification efficiency of 77\%;
similarly, at a $c$-jet efficiency of 30\%, the obtained light-flavour jet ($b$-jet)
rejection factor is 70 (9).


\section*{Acknowledgements}


We thank CERN for the very successful operation of the LHC, as well as the
support staff from our institutions without whom ATLAS could not be
operated efficiently.
 
We acknowledge the support of
ANPCyT, Argentina;
YerPhI, Armenia;
ARC, Australia;
BMWFW and FWF, Austria;
ANAS, Azerbaijan;
CNPq and FAPESP, Brazil;
NSERC, NRC and CFI, Canada;
CERN;
ANID, Chile;
CAS, MOST and NSFC, China;
Minciencias, Colombia;
MEYS CR, Czech Republic;
DNRF and DNSRC, Denmark;
IN2P3-CNRS and CEA-DRF/IRFU, France;
SRNSFG, Georgia;
BMBF, HGF and MPG, Germany;
GSRI, Greece;
RGC and Hong Kong SAR, China;
ISF and Benoziyo Center, Israel;
INFN, Italy;
MEXT and JSPS, Japan;
CNRST, Morocco;
NWO, Netherlands;
RCN, Norway;
MEiN, Poland;
FCT, Portugal;
MNE/IFA, Romania;
MESTD, Serbia;
MSSR, Slovakia;
ARRS and MIZ\v{S}, Slovenia;
DSI/NRF, South Africa;
MICINN, Spain;
SRC and Wallenberg Foundation, Sweden;
SERI, SNSF and Cantons of Bern and Geneva, Switzerland;
MOST, Taiwan;
TENMAK, T\"urkiye;
STFC, United Kingdom;
DOE and NSF, United States of America.
In addition, individual groups and members have received support from
BCKDF, CANARIE, Compute Canada and CRC, Canada;
PRIMUS 21/SCI/017 and UNCE SCI/013, Czech Republic;
COST, ERC, ERDF, Horizon 2020 and Marie Sk{\l}odowska-Curie Actions, European Union;
Investissements d'Avenir Labex, Investissements d'Avenir Idex and ANR, France;
DFG and AvH Foundation, Germany;
Herakleitos, Thales and Aristeia programmes co-financed by EU-ESF and the Greek NSRF, Greece;
BSF-NSF and MINERVA, Israel;
Norwegian Financial Mechanism 2014-2021, Norway;
NCN and NAWA, Poland;
La Caixa Banking Foundation, CERCA Programme Generalitat de Catalunya and PROMETEO and GenT Programmes Generalitat Valenciana, Spain;
G\"{o}ran Gustafssons Stiftelse, Sweden;
The Royal Society and Leverhulme Trust, United Kingdom.
 
The crucial computing support from all WLCG partners is acknowledged gratefully, in particular from CERN, the ATLAS Tier-1 facilities at TRIUMF (Canada), NDGF (Denmark, Norway, Sweden), CC-IN2P3 (France), KIT/GridKA (Germany), INFN-CNAF (Italy), NL-T1 (Netherlands), PIC (Spain), ASGC (Taiwan), RAL (UK) and BNL (USA), the Tier-2 facilities worldwide and large non-WLCG resource providers. Major contributors of computing resources are listed in Ref.~\cite{ATL-SOFT-PUB-2021-003}.


\clearpage

\appendix

\printbibliography
 
\clearpage
\input{atlas_authlist}

\end{document}

%% file: atlas_authlist.tex
 
\begin{flushleft}
\hypersetup{urlcolor=black}
{\Large The ATLAS Collaboration}

\bigskip

\AtlasOrcid[0000-0002-6665-4934]{G.~Aad}$^\textrm{\scriptsize 102}$,
\AtlasOrcid[0000-0002-5888-2734]{B.~Abbott}$^\textrm{\scriptsize 120}$,
\AtlasOrcid[0000-0002-7248-3203]{D.C.~Abbott}$^\textrm{\scriptsize 103}$,
\AtlasOrcid[0000-0002-1002-1652]{K.~Abeling}$^\textrm{\scriptsize 55}$,
\AtlasOrcid[0000-0002-8496-9294]{S.H.~Abidi}$^\textrm{\scriptsize 29}$,
\AtlasOrcid[0000-0002-9987-2292]{A.~Aboulhorma}$^\textrm{\scriptsize 35e}$,
\AtlasOrcid[0000-0001-5329-6640]{H.~Abramowicz}$^\textrm{\scriptsize 151}$,
\AtlasOrcid[0000-0002-1599-2896]{H.~Abreu}$^\textrm{\scriptsize 150}$,
\AtlasOrcid[0000-0003-0403-3697]{Y.~Abulaiti}$^\textrm{\scriptsize 117}$,
\AtlasOrcid[0000-0003-0762-7204]{A.C.~Abusleme~Hoffman}$^\textrm{\scriptsize 137a}$
% ...
\bigskip
\\

% ...
$^{29}$Physics Department, Brookhaven National Laboratory, Upton NY; United States of America.\\
% ...
$^{35}$$^{(a)}$Facult\'e des Sciences Ain Chock, R\'eseau Universitaire de Physique des Hautes Energies - Universit\'e Hassan II, Casablanca;$^{(b)}$Facult\'{e} des Sciences, Universit\'{e} Ibn-Tofail, K\'{e}nitra;$^{(c)}$Facult\'e des Sciences Semlalia, Universit\'e Cadi Ayyad, LPHEA-Marrakech;$^{(d)}$LPMR, Facult\'e des Sciences, Universit\'e Mohamed Premier, Oujda;$^{(e)}$Facult\'e des sciences, Universit\'e Mohammed V, Rabat;$^{(f)}$Institute of Applied Physics, Mohammed VI Polytechnic University, Ben Guerir; Morocco.\\
% ...
$^{55}$II. Physikalisches Institut, Georg-August-Universit\"{a}t G\"ottingen, G\"ottingen; Germany.\\
% ...
$^{102}$CPPM, Aix-Marseille Universit\'e, CNRS/IN2P3, Marseille; France.\\
$^{103}$Department of Physics, University of Massachusetts, Amherst MA; United States of America.\\
% ...
$^{117}$Department of Physics, New York University, New York NY; United States of America.\\
% ...
$^{120}$Homer L. Dodge Department of Physics and Astronomy, University of Oklahoma, Norman OK; United States of America.\\
% ...
$^{137}$$^{(a)}$Departamento de F\'isica, Pontificia Universidad Cat\'olica de Chile, Santiago;$^{(b)}$Millennium Institute for Subatomic physics at high energy frontier (SAPHIR), Santiago;$^{(c)}$Instituto de Investigaci\'on Multidisciplinario en Ciencia y Tecnolog\'ia, y Departamento de F\'isica, Universidad de La Serena;$^{(d)}$Universidad Andres Bello, Department of Physics, Santiago;$^{(e)}$Instituto de Alta Investigaci\'on, Universidad de Tarapac\'a, Arica;$^{(f)}$Departamento de F\'isica, Universidad T\'ecnica Federico Santa Mar\'ia, Valpara\'iso; Chile.\\
% ...
$^{150}$Department of Physics, Technion, Israel Institute of Technology, Haifa; Israel.\\
$^{151}$Raymond and Beverly Sackler School of Physics and Astronomy, Tel Aviv University, Tel Aviv; Israel.\\
% ...
$^{a}$ Also Affiliated with an institute covered by a cooperation agreement with CERN.\\
$^{b}$ Also at An-Najah National University, Nablus; Palestine.\\
$^{c}$ Also at Borough of Manhattan Community College, City University of New York, New York NY; United States of America.\\
$^{d}$ Also at Bruno Kessler Foundation, Trento; Italy.\\
$^{e}$ Also at Center for High Energy Physics, Peking University; China.\\
$^{f}$ Also at Center for Interdisciplinary Research and Innovation (CIRI-AUTH), Thessaloniki ; Greece.\\
% ...

\end{flushleft}
